\documentclass[11pt,draftclsnofoot,onecolumn,journal,letterpaper]{IEEEtran}

\usepackage{graphicx}
\usepackage{amsmath,amssymb,amsthm,mathrsfs,amsfonts,dsfont}
\usepackage{epsfig}
\usepackage[ruled,linesnumbered]{algorithm2e}
\usepackage{color}
\usepackage{subfigure}
\usepackage{cite}
\usepackage{diagbox}
\usepackage{multirow}
\usepackage{float}
\usepackage{stfloats}
\usepackage{enumitem}
\usepackage{cases}
\usepackage{setspace}

\newcommand{\mypara}[1]{{\smallskip \noindent \bf #1}\hspace{0.1in}}
\newcommand{\ssf}[1]{\textrm{$\sf{#1}$}{}}

\newtheorem{theorem}{Theorem}

\newtheorem{lemma}{Lemma}
\newtheorem{assumption}{Assumption}

\DeclareMathOperator*{\argmin}{arg\,min}

\newcommand{\vect}[1]{\mathbf{#1}}
\newcommand{\avgvect}[1]{\mathbf{\overline{#1}}}
\newcommand{\expt}{\mathbb{E}}
\newcommand{\norm}[1]{\left \| #1 \right \|}
\newcommand{\squab}[1]{\left [ #1 \right ]}
\newcommand{\dotp}[2]{\left \langle #1, #2 \right \rangle}

\ifodd 0
\newcommand{\rev}[1]{{\color{red}#1}}
\else
\newcommand{\rev}[1]{#1}
\fi

\ifodd 0
\newcommand{\revsec}[1]{{\color{red}#1}}
\else
\newcommand{\revsec}[1]{#1}
\fi

\ifodd 0

\else

\fi

\ifodd 0
\newcommand{\congc}[1]{{\color{magenta}(Cong: #1)}}
\else
\newcommand{\congc}[1]{}
\fi

\ifodd 1
\newcommand{\sihui}[1]{{\color{magenta}(Sihui: #1)}}
\else
\newcommand{\sihui}[1]{}
\fi

\ifodd 0
\newcommand{\new}[1]{{\color{magenta}#1}}
\else
\newcommand{\new}[1]{#1}
\fi

\ifodd 0
\newcommand{\crt}[1]{{\color{cyan}#1}}
\else
\newcommand{\crt}[1]{#1}
\fi

\begin{document}
\title{Design and Analysis of Uplink and Downlink Communications for Federated Learning}
\author{Sihui~Zheng \qquad  Cong~Shen  \qquad  Xiang~Chen
\thanks{S. Zheng and X. Chen are with School of Electronics and Information Technology, Sun Yat-sen University, China. {(e-mail: \texttt{zhengsh28@mail2.sysu.edu.cn}; \texttt{chenxiang@mail.sysu.edu.cn}).}}
\thanks{C. Shen is with the Charles L. Brown Department of Electrical and Computer Engineering, University of Virginia, USA. \new{(e-mail: \texttt{cong@virginia.edu}).}}
}

\maketitle

\begin{abstract}

Communication has been known to be \rev{one of the primary bottlenecks} of federated learning (FL), and yet existing studies have not addressed the efficient communication design, particularly in wireless FL where both uplink and downlink communications have to be considered. In this paper, we focus on the design and analysis of physical layer quantization and transmission methods for wireless FL. We answer the question of \textit{what} and \textit{how} to communicate between clients and the parameter server and evaluate the impact of the various quantization and transmission options of the updated model on the learning performance. We provide new convergence analysis of the well-known \textsc{FedAvg} under non-i.i.d. dataset distributions, partial clients participation, and finite-precision quantization in uplink and downlink communications. These analyses reveal that, in order to achieve an $\mathcal{O}({1}/{T})$ convergence rate with quantization, transmitting the weight requires increasing the quantization level at a \textit{logarithmic} rate, while transmitting the weight differential can keep a constant quantization level.  Comprehensive numerical evaluation on various real-world datasets reveals that the benefit of a FL-tailored uplink and downlink communication design is enormous -- a carefully designed quantization and transmission achieves more than $98\%$ of the floating-point baseline accuracy with fewer than $10\%$ of the baseline bandwidth, for majority of the experiments on both i.i.d. and non-i.i.d. datasets. In particular, 1-bit quantization ($3.1\%$ of the floating-point baseline bandwidth) achieves $99.8\%$ of the floating-point baseline accuracy at almost the same convergence rate on MNIST, representing the best known bandwidth-accuracy tradeoff to the best of the authors' knowledge. 

\end{abstract}


\section{Introduction}
\label{sec:intro}

Wireless federated learning (FL) \cite{niknam2019federated,lim2020federated} is an emerging edge artificial intelligence framework \cite{zhu2020}. FL has many attractive properties that cater to the growing trend of how data is generated and how machine learning (ML) model is trained. Empowered by the growing storage and computational capabilities of mobile devices and motivated by the increasing concern over transmitting private information to a central server, FL has become an attractive ML paradigm that trains ML models locally on each device where data never leaves the device \cite{mcmahan2017fl,konecny2016fl}. 

\rev{While FL offers many important benefits, it also faces several critical challenges including but not limited to significant communication cost, handling client heterogeity (both dataset and the  computation and communication capabilities) and the straggler problem, preserving the privacy of user data, improving robustness to adversarial attacks and failures, and ensuring fairness. A comprehensive review of these challenges can be found in \cite{li2019federated}. In particular, despite being recognized as one of the primary bottlenecks of FL \cite{kairouz2019advances,li2019federated,lim2020federated}, research on the \textit{communication} aspect in the FL pipeline has not been on par with the \textit{learning} component, particularly in a wireless environment.} Early research on communication-efficient FL largely focuses on reducing the number of communication rounds and the amount of information for communication, while assuming that the underlying communication ``tunnel'' has been established by existing wireless protocols. More recent research starts to fill this void from a communication and signal processing point of view.  In general, the principle is to balance learning performance and communication efficiency via, e.g., device selection, bandwidth allocation, and power control; see Section~\ref{sec:related} for an overview. There are also recent studies that focus on the communication system design \cite{zhu2019broadband,du2020high,shlezinger2020icassp}, but they are either system-specific (e.g., cellular networks) or with high complexity beyond the current implementation capability (e.g., very high dimensional vector quantization).

While the early studies provide a glimpse of the potential of optimizing communication for learning, the actual implementation of the communication algorithms has not been tailored to the unique characteristics of FL. In particular, it is often taken for granted that standard signal processing and communication techniques can be directly applied to FL. We show in this paper that this can be highly suboptimal because they are mostly designed for independent and identically distributed (i.i.d.) sources over time, while the communicated model update in FL represents a long-term process consisting of many progressive learning rounds that collectively determine the final learning outcome. This phenomenon is known in the machine learning literature and has been leveraged to optimize the learning hyperparameters, e.g., decaying the learning rate over time \cite{li2019convergence}, but has not been considered for the communication algorithms. To further complicate the matter, the overall FL performance is determined by both local model weight (i.e., parameters of the ML model) upload and global model weight download over multiple learning rounds, suggesting that both uplink and downlink communications have to be considered.

In this paper, we study FL-tailored communication designs for training ML models locally at mobile devices and aggregation at the base station, where the information for communication (both uplink and downlink) is the model weight (or its update) after each learning round. The design goal is to maximize the learning accuracy and convergence rate, two prime objectives in FL. We answer the questions of \textit{what} and \textit{how} to transmit the updated model in each round between clients (mobile devices) and parameter server (base station), and study practical quantization and transmission methods that leverage the inherent structure of the machine learning model. \rev{ Our main  contributions are as follows.

\begin{enumerate}[leftmargin=*] \itemsep=0pt
    \item We study practical quantization schemes for FL and show that the dynamic range of the weight needs to be taken into account, and the choice of rounding has a profound impact on the performance. For uplink, we demonstrate that transmitting only the weight differential is beneficial if the practical constraint allows, while pointing out that this differential transmission cannot be utilized for downlink when only \textit{partial} clients participate each round. We also propose an  enhancement called \textit{layered quantization} for downlink, in which the quantization gain is adjusted to match the dynamic range of the weights in different network layers.
    
    \item We rigorously prove convergence rate upper bounds of the well-known \textsc{FedAvg} \cite{mcmahan2017fl} under finite-precision quantization in both uplink and downlink communications. The theoretical analysis reveals a novel conclusion: in order to maintain the $\mathcal{O}(1/T)$ convergence rate of the floating-point \textsc{FedAvg}, the uplink or downlink quantization for direct weight transmission should increase the quantization precision at a \textit{logarithmic} rate $\mathcal{O} \left(\log(t) \right)$, while transmitting the weight differential can maintain a \textit{constant} (i.e., $\mathcal{O} \left(1 \right)$) quantization precision throughout the FL process. This result holds for non-i.i.d. dataset and partial (randomly selected) clients participation in each learning round, which is more general and matches the unique characteristics of FL \cite{mcmahan2017fl}.
    
    \item Comprehensive numerical evaluation on \revsec{four} widely adopted datasets with increasing learning difficulties, \textit{MNIST}, \revsec{\textit{F-EMNIST}}, \textit{CIFAR-10} and \textit{Shakespeare} are done. We design a series of experiments to show the impact of each step in the quantization including quantization gain, rounding method and the relationship between the quantization and hyperparameters like batch size, local epoch, etc. In particular, we corroborate the theoretical conclusion that the quantization precision needs to increase at a logarithmic rate for direct weight transmission via numerical experiments. The results also reveal that the benefit of a FL-tailored uplink and downlink communication design is significant. In majority of the experiments, we see that a carefully designed quantization and transmission achieves more than $98\%$ of the floating-point baseline accuracy with fewer than $10\%$ of the baseline bandwidth, for both i.i.d. and non-i.i.d. datasets. As a final exclamation point, a 1-bit quantization ($3.1\%$ of the floating-point baseline bandwidth) achieves $99.8\%$ of the floating-point baseline accuracy at almost the same convergence rate in the MNIST experiment, representing the best known bandwidth-accuracy tradeoff to the best of the authors' knowledge.
\end{enumerate} }

The rest of this paper is organized as follows. A brief overview of the related literature is provided in Section~\ref{sec:related}. Section~\ref{sec:mod} describes the wireless FL system model. Uplink and downlink communication designs, together with the theoretical convergence analyses, are presented in Section~\ref{sec:comm_ul} and \ref{sec:comm_dl}, respectively. Experimental results are reported in Section~\ref{sec:exp}. Section~\ref{sec:conclusions} concludes the paper, and the technical proofs of the main theorems are provided in the Appendices.

\section{Related Works}
\label{sec:related}

Federated learning \cite{mcmahan2017fl} is an emerging distributed machine learning \cite{verbraeken2020} paradigm that addresses several new features created by modern ML applications.  It has been extensively studied in recent years in the machine learning community, which aims to address various questions around improving machine learning efficiency and effectiveness \cite{stich2018local,wang2018cooperative,haddadpour2019local,li2019convergence}, preserving the privacy of user data \cite{bhowmick2018protection,truex2019hybrid,niu2019secure}, robustness to attacks and failures \cite{xie2019zeno++,xie2019practical}, and ensuring fairness and addressing sources of bias \cite{bonawitz2019towards,li2019fair}. However, these works mostly focus on the machine learning aspect of FL and largely consider over-simplified communication models.

Recently, researchers have started looking into the communication design of FL, particularly the communication algorithms, protocols, and systems.  Joint radio and computation resource management is another active research topic. Existing works \cite{mo2020energy} study the inherent trade-off between local model update and global model aggregation, to optimize over transmission power/rate and training time. To enable FL at scale and address the straggler problem, client selection is essential. In this regard, various joint radio resource allocation and client selection policies \cite{zeng2019energy,chen2019joint,shi2019device,yang2019energy,chen2020convergence} have been proposed to minimize the learning loss or the training time.

Communication-efficient design has been another active research topic in FL \cite{li2019federated}, where the attempts have largely focused on either reducing the total number of communication rounds, or reducing the size of the exchanged messages in each round. One of the representative approaches for reducing the communication rounds is \textsc{FedAvg} \cite{mcmahan2017fl}, which allows periodic model aggregation and local model updates and thus enables flexible communication-computation tradeoff \cite{li2018federated}.  Theoretical understanding of this tradeoff has been an active research area and, depending on the underlying assumptions (e.g., i.i.d. or non-i.i.d. local datasets, convex or non-convex loss functions, gradient descent or stochastic gradient descent), rigorous analysis of the convergence behavior has been carried out \cite{stich2018local,wang2018cooperative,wang2019adaptive,li2019convergence}.  For model compression, general discussions on sparsification, subsampling, and quantization are given in \cite{konecny2016fl}. Particularly, sparsification methods reduce the number of non-zero entries in the stochastic gradient \cite{alistarh2017qsgd}. Structured and sketched updates are proposed in \cite{konevcny2016federated} to reduce the size of model updates, which are further extended by lossy compression and federated dropout \cite{caldas2018expanding}. \rev{There have been recent efforts in developing quantization  and source coding to reduce the communication cost  \cite{alistarh2017qsgd,bernstein2018signsgd,reisizadeh2019tsp,shlezinger2020icassp,zhu2020one,reisizadeh2019fedpaq, amiri2020federated,amiri2020quantized}. However, most of the quantizers studied in these papers do not consider practical constraints and are not widely used in practice. Reference \cite{reisizadeh2019fedpaq} only considers i.i.d. datasets and uplink quantization, and  \cite{amiri2020quantized} focuses on downlink quantization of the model differential and does not apply to partial clients participation, which is an important feature of FL. }

\section{System Model}
\label{sec:mod}

\begin{figure*}[ht]
    \hsize=\textwidth
    \centering
    \includegraphics[width=\textwidth]{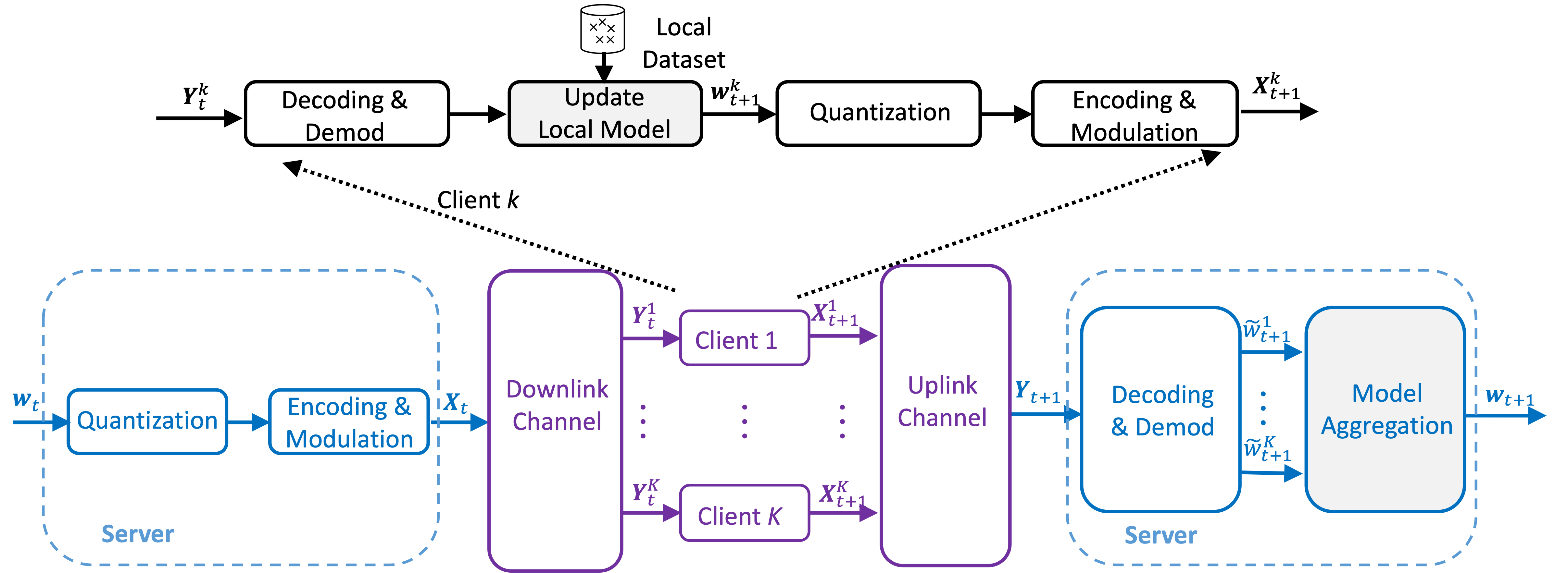}
    \caption{Wireless FL system model. The $t$th to the $(t+1)$th round of operations at both clients (mobile devices, shown in black) and server (BS, shown in blue) are illustrated. The shaded boxes correspond to the learning operations and others refer to communications.}
    \label{fig:system_model}
\end{figure*}
 
The wireless federated learning system is illustrated in Fig.~\ref{fig:system_model}. We assume a federated learning task of collaboratively training a ML model (e.g., logistic regression or deep neural network (DNN)) as in \cite{mcmahan2017fl}. In particular, there is a central parameter server (e.g., base station) and a set of at most $N$ clients (e.g., mobile devices).  Client \rev{$k$} stores a (disjoint) local dataset $\mathcal{D}_k = \{\vect{z}_i\}_{i=1}^{D_k}$, with its size denoted by $D_k$, that never leaves the device. Datasets across devices are assumed to be non-i.i.d., which is an important feature of FL \cite{mcmahan2017fl,kairouz2019advances}. The maximum data size when all devices participate in FL is $D = \sum_{k=1}^N D_k$. Each data sample $\vect{z}$ is given as an input-output pair $\{\vect{x}, y\}$. The loss function $f(\vect{w}, \vect{z})$ measures how well a ML model with parameter $\vect{w} \in \mathbb{R}^d$ fits one particular data sample $\vect{z}$. For the $k$th device, its local loss function $F_k(\cdot)$ is defined by
\begin{equation*}
    F_k(\vect{w}) \triangleq \frac{1}{D_k} \sum_{\vect{z}\in\mathcal{D}_k} f(\vect{w}, \vect{z}).
\end{equation*}
Then, the global optimization objective over all $N$ clients is given by
\begin{equation} \label{eqn:global}
    F(\vect{w}) \triangleq \sum_{k=1}^N \frac{D_k}{D} F_n (\vect{w}) = \frac{1}{D}\sum_{k=1}^N \sum_{\vect{z}\in\mathcal{D}_k} f(\vect{w}, \vect{z}).
\end{equation}
The global loss function measures how well the model fits the entire corpus of data on average. As a result, the objective is to find the best model parameter $\vect{w}^*$ that minimizes the global loss function: $$\vect{w}^* = \argmin_\vect{w} F(\vect{w}).$$ Let $F^*$ and $F_k^*$ be the minimum value of $F$ and $F_k$, respectively. Then, $\Gamma = F^* - \frac{1}{N}\sum_{k=1}^N F_k^*$ quantifies the degree of non-i.i.d. \cite{li2019convergence}. \rev{We note that using $\Gamma$ to measure the degree of non-i.i.d. is more meaningful when the dataset size is large, in which the minimum loss function values of $F_k$ approach the true expected minimum loss function values with respect to the individual dataset distributions.}

This work considers a generic FL framework where partial client participation and  non-i.i.d. local datasets, two critical features that separate FL from distributed SGD, are explicitly captured. More specifically, the FL pipeline works by iteratively executing the following steps at the $t$th learning round.

\begin{enumerate}[leftmargin=*] \itemsep=0pt
    \item \textbf{Downlink communication for model download.}  The centralized server broadcasts the current global ML model, which is described by the latest weight vector $\vect{w}_t$, to a set of randomly selected clients denoted as $\mathcal{S}_t$ with $\vert \mathcal{S}_t \vert = K$. The detailed communication mechanism for this phase will be described in Section~\ref{sec:comm_dl}.
    
    \item \textbf{Local computation.} Each client uses its local data to train a local ML model improved upon the received global ML model. In this work, we assume that mini-batch stochastic gradient descent (SGD) is used in training, where the weight $\vect{w}_t^k$ is updated iteratively (for $E$ steps in the current learning round) at device $k$ as:
    \begin{align*}
        \vect{w}_{t,0}^k &= \vect{w}_t^k, \\
        \vect{w}_{t,\tau}^k &= \vect{w}_{t,\tau-1}^k - \eta_t \nabla F(\vect{w}_{t,\tau-1}^k, \xi_{\tau}^k),~ \forall \tau=1, \cdots, E, \\
        \vect{w}_{t+1}^k &=\vect{w}_{t,E}^k,
    \end{align*}
    where $\xi_{\tau}^k$ is a mini batch of data points that are independently sampled uniformly at random from the local dataset of client $k$.

    \item \textbf{Uplink communication for model upload.} The selected $K$ devices upload their latest local models to the server synchronously. The communication mechanism for this phase will be described in Section~\ref{sec:comm_ul}.
    
    \item \textbf{Global aggregation.} The server aggregates the received local models to generate a new global ML model:
        \begin{align}
            \vect{w}_{t+1} = \sum_{k \in \mathcal{S}_t } \frac{D_k}{\sum_{i \in \mathcal{S}_t} D_i}\vect{w}_{t+1}^k.
        \label{eqn:agg}
        \end{align}
    The server then moves on to the $(t+1)$th round. The process completes after $T$ rounds.
\end{enumerate}

By and large, the above process is followed by majority of the existing FL formulations. There are some variants, such as adapting the client selection \cite{xu2020client}, allowing for varying number of local updates \cite{li2019federated}, or improving the model learning by distributed primal-dual methods \cite{smith2017cocoa}. Our work nevertheless focuses on the communication aspect (both uplink and downlink) of FL and can incorporate these enhancements.

\section{Uplink Communication Design}
\label{sec:comm_ul}

The task of a particular uplink communication round (e.g. $t$th round) is to deliver $\vect{w}_t^k$ for client $k$ to the BS as accurately and efficiently as possible. However, since the FL process involves $T$ rounds of model update/communication operations, which are inherently correlated over time, there exist opportunities to improve the communication design. In particular, uplink design involves answering two questions from the communication perspective: \textit{what} to transmit, and \textit{how} to transmit.

\subsection{What to Transmit: Weight versus Weight Differential}
\label{ssec:what}

If we treat the design of the $t$th uplink communication round at client $k$ as an isolated task, i.e., we ignore the operations in the past both at client $k$ and at the server, we can directly transmit the latest local weight vector $\vect{w}_t^k$. A different choice, which leverages the past information, is to transmit the \textit{weight update} (also called \textit{weight differential} in this paper) $\vect{d}_t^k = \vect{w}_t^k -  \vect{w}_{t-1}^k$ as opposed to the weight itself. 

From a pure learning perspective, there is no difference whether the updated model itself ($\vect{w}_t^k$) or its differential ($\vect{d}_t^k$) is communicated from clients to the server. As long as the server can reconstruct $\vect{w}_t^k$, this aspect does not impact the learning performance \cite{mcmahan2017fl}. Thus, it seems that the choice is insignificant and boils down to other practical considerations. For example, transmitting weight differential relies on the server keeping the previous global model $\vect{w}_{t-1}$, so that the new local models can be reconstructed from the differential. This however may not always be true if the server deletes intermediate model aggregation for privacy preservation \cite{bonawitz2019towards}, which makes reconstruction from the model differential infeasible. As another example, transmitting weight differential implicitly assumes $\vect{w}_{t-1}^k=\vect{w}_{t-1}$, i.e., the previously received global model is accurate. This may not be true in a practical communication system where decoding error is inevitable. In both examples, transmitting the weight vector itself is more preferable. 

However, in addition to these considerations, we show in Section~\ref{ssec:conv_all} that the choice between weight or weight differential in the uplink communication phase has a more profound impact to the learning performance (in particular the convergence), when imperfect reconstruction due to quantization is captured.

\subsection{How to Transmit: Quantization Designs}
\label{ssec:quantization}

In order to transmit the $d$-dimensional source message (either weight or weight differential) to the server, we first quantize the source vector into discrete values, and then apply coding (both source and channel) and modulation in the baseband, as shown in Fig.~\ref{fig:system_model}. The coding and modulation operation can leverage existing wireless system designs \cite{TV:05} and is not the focus of this paper. The quantization method, however, bears some consideration as discussed below.

In ML models such as deep neural networks (DNN), weights are usually represented in floating point format\footnote{A 32-bit representation is a common choice in practice.}.  A \emph{quantizer} is designed to reduce the  necessary bit-width for each weight and hence decrease the message size for communications. \rev{It is worth noting that the quantization design in FL is very different to DNN model compression \cite{gupta2015deep,lin2016fixed,han2016iclr}, which focuses on reducing model storage and simplifying inference computation. Also for DNN model compression, the impact of quantization is reflected in the final model after the training is complete. Quantization design for FL, on the other hand, aims at reducing the communication bandwidth, and has to be carried out in every round such that all the quantizations collectively affect the performance of FL.}

\begin{figure}
    \centering
    \includegraphics[width=\columnwidth]{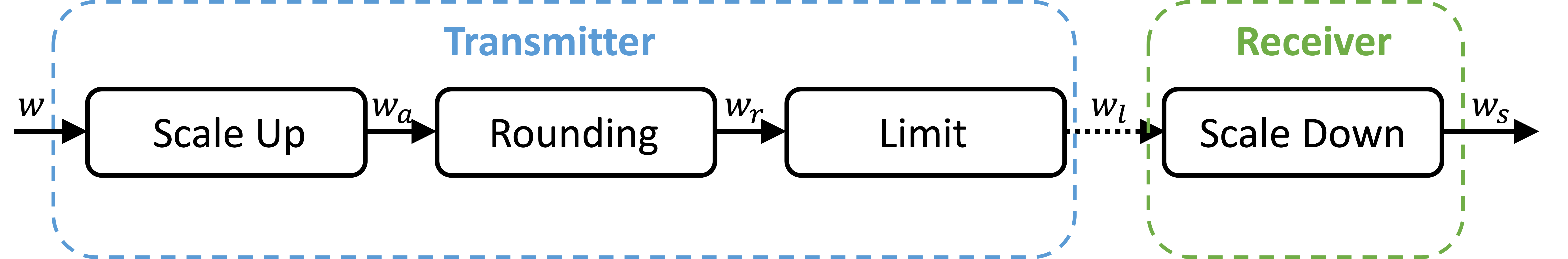}
    \caption{Illustration of the adopted quantization structure.}
    \label{fig:quan_struct}
\end{figure}

We focus on a {practical} quantizer design that is suitable for communication system implementations. For this reason, we do not consider vector quantization \cite{shlezinger2020icassp} which is highly complex and not well used in practice despite its theoretical advantages. The adopted quantizer design is illustrated in Fig.~\ref{fig:quan_struct}. We note that this diagram is different from the existing literature \cite{alistarh2017qsgd,reisizadeh2019fedpaq,du2020high,reisizadeh2019tsp}, which only has the \textit{rounding} and \textit{limit} operations. \rev{Inspired by the classical quantization methods in communication system \cite{montorsi2001design}, we add the \textit{scaling up} and \textit{scaling down} steps.} This is because the dynamic range of the weights may not match a pre-determined rounding strategy, and a proper quantization gain control\footnote{This can be implemented by the automatic gain control (AGC) module in the wireless transmitter, which is usually enforced before the analog-to-digital converter (ADC) so that the input signal can match the dynamic range of the ADC.} is often applied to handle this issue. Specifically, for a full-precision weight $w$, the quantizer output $Q(w)$ can be obtained via the following steps:

\begin{enumerate}[leftmargin=*]\itemsep=0pt
    \item \textbf{Scale Up.} $w$ is first amplified with a scaling factor called \emph{quantization gain}. Denoting the quantization gain as $G$, the amplified value is $w_a = w  G$. 
    
    \item \textbf{Rounding.} $w_a$ is truncated to only retain its integer part:  $w_r = R(w_a)$ where $R(\cdot)$ denotes the rounding function.
    
    \item \textbf{Limit.} The range of integer $w_r$ is further limited to $B$ bits:
    \begin{equation*}
        w_l = \begin{cases}
            2^{B-1}-1 & \text{if}~w_r > 2^{B-1}-1, \\
            w_r & \text{if}~w_r \in [-2^{B-1}, 2^{B-1}-1] , \\
            -2^{B-1} & \text{if}~w_r < -2^{B-1}.
        \end{cases}
    \end{equation*}
    
    \item \textbf{Scale Down.} The receiver output $w_s$ is obtained by scaling down  $w_l$: $w_s = w_l/G$.
\end{enumerate}

We now detail how different components are designed for a $B$-bit quantizer as in Fig.~\ref{fig:quan_struct}.

\mypara{Quantization Gain.}
A large $G$ preserves more decimal digits of $w$ and hence improves the representation accuracy, but it also increases the percentage of overflow in the subsequent limit operation, which introduces quantization errors in a different way. It is worth mentioning that $G$ is typically set as power of 2, which simplifies the implementation to bit shifting.

\rev{
\mypara{Quantizer Structure.} 
For comparison, we consider two quantizer structures in this work. In \emph{Native Quantization} (NQ), the scaling up is limited to $G=2^{B-1}$ (1 bit for sign and the rest for decimal), and thus the scaling down step can be done at the transmitter, which means the receiver does not need to know $G$. An alternative structure  \emph{Tuned Quantization} (TQ) allows for fine-tuning $G$ to a more suitable value (usually greater than $2^{B-1}$) but requires that the scaling down step be done at the receiver.
}

\mypara{Rounding method.} 
Two rounding functions are considered. The basic one is \emph{nearest rounding} (NR):
\begin{equation*}
    R(x) = \begin{cases}
        \left \lfloor x \right \rfloor & \text{if}~x - \left \lfloor x \right \rfloor < 0.5 \\
        \left \lfloor x \right \rfloor + 1 & \text{otherwise}
    \end{cases}
\end{equation*}
where $\left \lfloor x \right \rfloor$ is the floor of $x$. The second method is \emph{stochastic rounding} (SR) \cite{gupta2015deep}, which rounds $x$ to $\left \lfloor x \right \rfloor$ with a probability proportional to the proximity of $x$ to $\left \lfloor x \right \rfloor$ (w.p. is short for `with probability'):
\begin{equation*}
    R(x) = \begin{cases}
        \left \lfloor x \right \rfloor & \text{w.p.}~1-(x-\left \lfloor x \right \rfloor) \\
        \left \lfloor x \right \rfloor + 1 & \text{w.p.}~x-\left \lfloor x \right \rfloor.
    \end{cases}
\end{equation*}

\mypara{Enhanced 1-bit \rev{quantizer}.}
For the special case of a 1-bit \rev{quantizer}, the quantization operation can be \rev{simplified as following two steps, without following the scale up -- rounding and limit-- scale down operations. In particular, we first round $w$ with either NR or SR as follows.}
\begin{itemize}[leftmargin=*]\itemsep=0pt
    \item Nearest Rounding:
        \begin{equation*}
            R(w) = \begin{cases}
                +1, & \text{if}~w \geq 0, \\
                -1, & \text{if}~w < 0.
            \end{cases}
        \end{equation*}    
    \item Stochastic Rounding:
        \begin{equation*}
            R(w) = \begin{cases}
                +1, & \text{w.p.}~\text{Pr}, \\
                -1, & \text{w.p.}~(1-\text{Pr}),
            \end{cases}
        \end{equation*}
where $\text{Pr}=\min(1, \max(0, \frac{w+1/G}{2/G}))$.
\end{itemize}
\rev{Then, the receiver performs scale down to get $Q(w)=R(w)/G$.}

\subsection{Convergence Analysis for \textsc{FedAvg} with Uplink Quantization}
\label{ssec:conv_all}

As stated in Section~\ref{ssec:what}, both the weight itself $\vect{w}_t^k$ and weight differential $\vect{d}_t^k$ can be used for uplink communication. However, this section shows that the two options have very different convergence behaviors, which lead to different requirements on quantization.

\subsubsection{Analysis for weight transmission}
\label{ssec:conv_qwt}
We first analyze directly transmitting weight $\vect{w}_t^k$ in the uplink of \textsc{FedAvg} with quantization. The main theoretical result is that this configuration converges to the global optimum at a rate of $\mathcal{O}(\frac{1}{T})$, which is the same scaling behavior of the vanilla \textsc{FedAvg}, \textit{if we gradually increase the quantization precision over $t$}.

To simplify the analysis, we assume in the remainder of the paper that the local dataset sizes at all devices are the same: $D_i = D_j$, $\forall i, j \in [N]$, and focus on the general case of randomly selected $K$ out of $N$ clients participating in the server aggregation with non-i.i.d. dataset\footnote{\rev{As will become clear after Section~\ref{sec:comm_dl}, the case of unbalanced datasets can be easily incorporated in the analysis of both uplink and downlink communications when there is \textit{full} clients participation in FL.  However, when combined with partial clients participation, the analysis of unbalanced dataset becomes nontrivial. In this case, the coefficients in Eqn.~\eqref{eqn:agg} vary in each round, which makes the random sampling of clients no longer an unbiased estimation of the full participation case. Also, the error of SGD and quantization becomes much more complex, since it not only depends on how many clients are selected but also on which clients are selected. We leave this case for future research.}}.  Let the set $\mathcal{S}_t \subset [N]$ denote the $K$ randomly selected clients in the $t$th round. With quantization, these devices transmit $\{ Q(\vect{w}_t^k) \}_{k=1}^K$ in the uplink, and the server performs aggregation as
\begin{equation}
\label{eqn:globalw}
    \vect{w}_{t} = \frac{1}{K} \sum_{k \in \mathcal{S}_{t}} Q(\vect{w}_{t}^k).
\end{equation}

The following assumptions are made for the convergence analysis.  Assumption \ref{as:F} is fairly standard and has been widely used in the convergence analysis of \textsc{FedAvg}; see \cite{li2019convergence,stich2018local,reisizadeh2019fedpaq,haddadpour2019local}. Assumption \ref{as:w_bounded} simply upper bounds the largest value of the weight so that the error in quantization is bounded. In practice, this assumption almost always holds because of the limited bit-width of weights in storage and computation.
\begin{assumption}\label{as:F}
\begin{enumerate}[leftmargin=12pt,topsep=0pt, itemsep=0pt,parsep=0pt]
    \item[1)] \textbf{$L$-smooth:} $\forall~\vect{v}$ and $\vect{w}$, $F_k(\vect{v}) \leq F_k(\vect{w}) + (\vect{v} - \vect{w})^T \nabla F_k(\vect{w}) + \frac{L}{2} \norm{\vect{v} - \vect{w}}^2$.
    
    \item[2)] \textbf{$\mu$-strongly convex:} $\forall~\vect{v}$ and $\vect{w}$, $F_k(\vect{v}) \geq F_k(\vect{w}) + (\vect{v} - \vect{w})^T \nabla F_k(\vect{w}) + \frac{\mu}{2} \norm{\vect{v} - \vect{w}}^2$.
    
    \item[3)] \textbf{Bounded variance for mini-batch SGD:}  
    The variance of stochastic gradients satisfies: $$\expt \norm{\nabla F_k (\vect{w}_t^k, \xi_t^k) - \nabla F_k (\vect{w}_t^k)}^2 \leq \sigma_k^2$$ for $k=1, \dots, N.$ 
    
    \item[4)] \textbf{Uniformly bounded gradient:} $\expt \norm{\nabla F_k (\vect{w}_t^k, \xi_t^k)}^2 \leq H^2$ for all $k=1, \dots, N$.
\end{enumerate}
\end{assumption}

\begin{assumption}
\label{as:w_bounded}
$\max_{k \in [N], t \in [T]} \norm{\vect{w}_t^k}_{\infty} \leq M$, for constant $M \geq 0$.
\end{assumption}

\begin{theorem}
\label{thm:1}
    Define $\kappa = \frac{L}{\mu}$, $\gamma = \max\{8\kappa, E\}$. Choose learning rate $\eta_t=\frac{2}{\mu(\gamma+t)}$ and quantization level $B_t = \log_2{ \squab{\frac{\mu(\gamma+t-1)}{2} + 1}}.$ Then, under Assumptions \ref{as:F} and \ref{as:w_bounded} and using {stochastic rounding} with quantization gain $G=\frac{2^{B_t-1}}{M}$ on weight $\vect{w}_t^k$, the convergence of \textsc{FedAvg} with non-i.i.d. local datasets and partial clients participation satisfies
    \small{
        \begin{equation}
        \label{eqn:conv_res}
            \expt \squab{F(\vect{w}_T)} - F^* \leq \frac{2\kappa}{\gamma+T} \squab{\frac{D}{\mu} + \left ( 2L+\frac{E\mu}{4} \right ) \norm{\vect{w}_0 - \vect{w}^*}^2},
        \end{equation}
    }
    where the constant $D$ is
    \small{
        \begin{equation}
        \label{eqn:conv_d}
            D = \rev{\sum_{k=1}^{N}\frac{\sigma_k^2}{N^2}} + 6L \Gamma + 8(E-1)^2H^2 + \frac{N-K}{N-1}\frac{4}{K} E^2H^2 + \frac{dM^2}{K}.
        \end{equation}
        }
\end{theorem}
The complete proof of {Theorem \ref{thm:1}} can be found in Appendix \ref{app:proof_thm1}. We note that the expectation in Eqn.~\eqref{eqn:conv_res} is with respect to three random events: (a) stochastic gradient when updating the model; (b) stochastic rounding in quantization; and (c) random sampling when selecting clients in each round.

\subsubsection{Analysis for weight differential transmission}
\label{ssec:conv_qdt}
We call the communication design for weight differential as \emph{Differential Transmission} (DT). With quantized weight differentials, the $K$ randomly selected devices transmit $\{ Q(\vect{d}_t^k) \}_{k=1}^K$ in the uplink, and the server performs aggregation as
\begin{equation}
\label{eqn:globalwd}
    \vect{w}_{t} = \vect{w}_{t-1} + \frac{1}{K} \sum_{k \in \mathcal{S}_{t}} Q(\vect{d}_{t}^k).
\end{equation}
Intuitively, the global aggregation in Eqn.~\eqref{eqn:globalwd} may be better than Eqn.~\eqref{eqn:globalw} for a given quantizer design as in Fig.~\ref{fig:quan_struct}.  As stated previously, we have to strike a balance between representation range and accuracy when selecting the value of $G$. The range of the weight differential $\vect{d}_{t}^k$ is typically smaller than the raw weight $\vect{w}_{t}^k$, in particular towards convergence. Hence, a larger $G$ can be used for DT to achieve higher quantization precision while avoiding excessive overflow. This can also be interpreted as not wasting bits on the constant part of the weights, which improves communication efficiency.

In addition to the advantage in quantization precision, {Theorem~\ref{thm:2}} shows that DT can converge to the global optimum at rate of $\mathcal{O}(\frac{1}{T})$ \textit{without requiring an increasing quantization level}, which is better than {Theorem~\ref{thm:1}}. The complete proof of {Theorem \ref{thm:2}} can be found in Appendix \ref{app:proof_thm2}.

\begin{theorem}
\label{thm:2}
    Let {Assumption} \ref{as:F} hold and $\kappa, \gamma, \eta_t$ be defined in Theorem \ref{thm:1}. Let $B$ be a fixed quantization level. Using {stochastic rounding} on weight differential $\vect{d}_t^k$, the convergence bound in Eqn.~\eqref{eqn:conv_res} for \textsc{FedAvg} with non-i.i.d. local datasets, partial clients participation, and uplink quantization still holds, with $D$ being replaced by
    \begin{equation*}
        D =  \rev{\sum_{k=1}^{N}\frac{\sigma_k^2}{N^2}} + 6L \Gamma + 8(E-1)^2H^2 
             + \frac{N-K}{N-1}\frac{4}{K}E^2 H^2 + \frac{4d}{K(2^B-1)^2} E^2 H^2.
    \end{equation*}
\end{theorem}

\rev{Compared to the known convergence results without quantization \cite{stich2018local,li2019convergence,wang2018cooperative}, Theorems~\ref{thm:1} and \ref{thm:2} state that the same convergence rate can be largely preserved if the quantization is carefully designed. Intuitively, errors introduced by uplink quantizations may be accumulated and reflected in the new global model, which is then used by clients for the next round of training. This could potentially lead to error propagation over rounds and affect the convergence of FL. The core idea behind these theorems, especially reflected in their proofs, is that the quantizer design should ensure the errors introduced by quantization are well controlled at a lower level comparing to the noise of SGD, such that the overall ``noise level'' is not increased and thus the convergence of SGD is not violated.}

\rev{We can also see from Theorems~\ref{thm:1} and \ref{thm:2} that the convergence bounds have certain monotonic relationships with several hyperparameters. The bounds increase with $E$, which is consistent with the result of \cite{zhao2018federated}. Larger $B$ and $K$ reduce the bounds, which is intuitive. Finally, we note that the effect of non-i.i.d. datasets, which is captured by $\Gamma$, is reflected in both Theorem~\ref{thm:1} and \ref{thm:2}. Furthermore, the convergence upper bounds of these theorems are monotonically decreasing when $\Gamma$ reduces. When $\Gamma$ goes to zero as the dataset size increases asymptotically, we have the i.i.d. dataset and the convergence upper bounds have the best results.}

\section{Downlink Communication Design}
\label{sec:comm_dl}

The task of downlink communication is to broadcast the latest global model $\vect{w}_t$ to the selected clients at the beginning of each learning round. It is clear that the quantization design described in Section \ref{ssec:quantization} can still be adopted in downlink. However, the differential transmission scheme in uplink is no longer feasible in the downlink for \textit{partial} clients participation, which is a key feature of FL particularly when massive amount of clients exist \cite{mcmahan2017fl,bonawitz2019towards}. This is because participating clients differ from round to round, and a newly participating client does not have the ``base'' model of the previous round to reconstruct the new global model based on weight differential. Thus, we only focus on transmitting the global model $\vect{w}_t$ at round $t$, and develop an enhanced method call \emph{Layered Quantization} (LQ) for downlink communication.

\subsection{Layered Quantization}
\label{sec:layered_dl}

\begin{figure}
    \centering
    \subfigure[Empirical cumulative distribution function]{
        \label{fig:dl_cdf}
        \includegraphics[width=0.45\columnwidth]{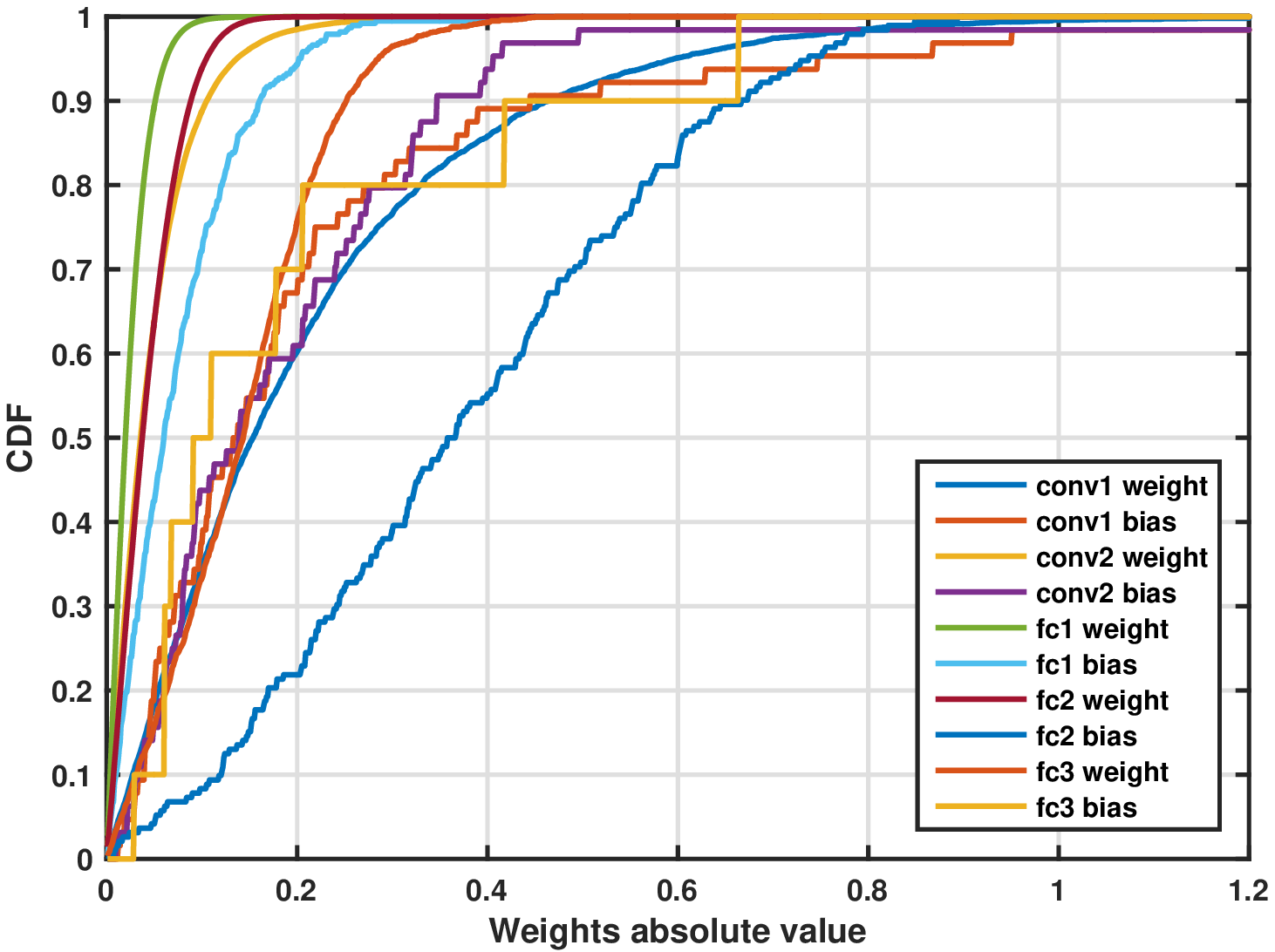}}
    \subfigure[Mean and variance]{
        \label{fig:dl_mean}
        \includegraphics[width=0.43\columnwidth]{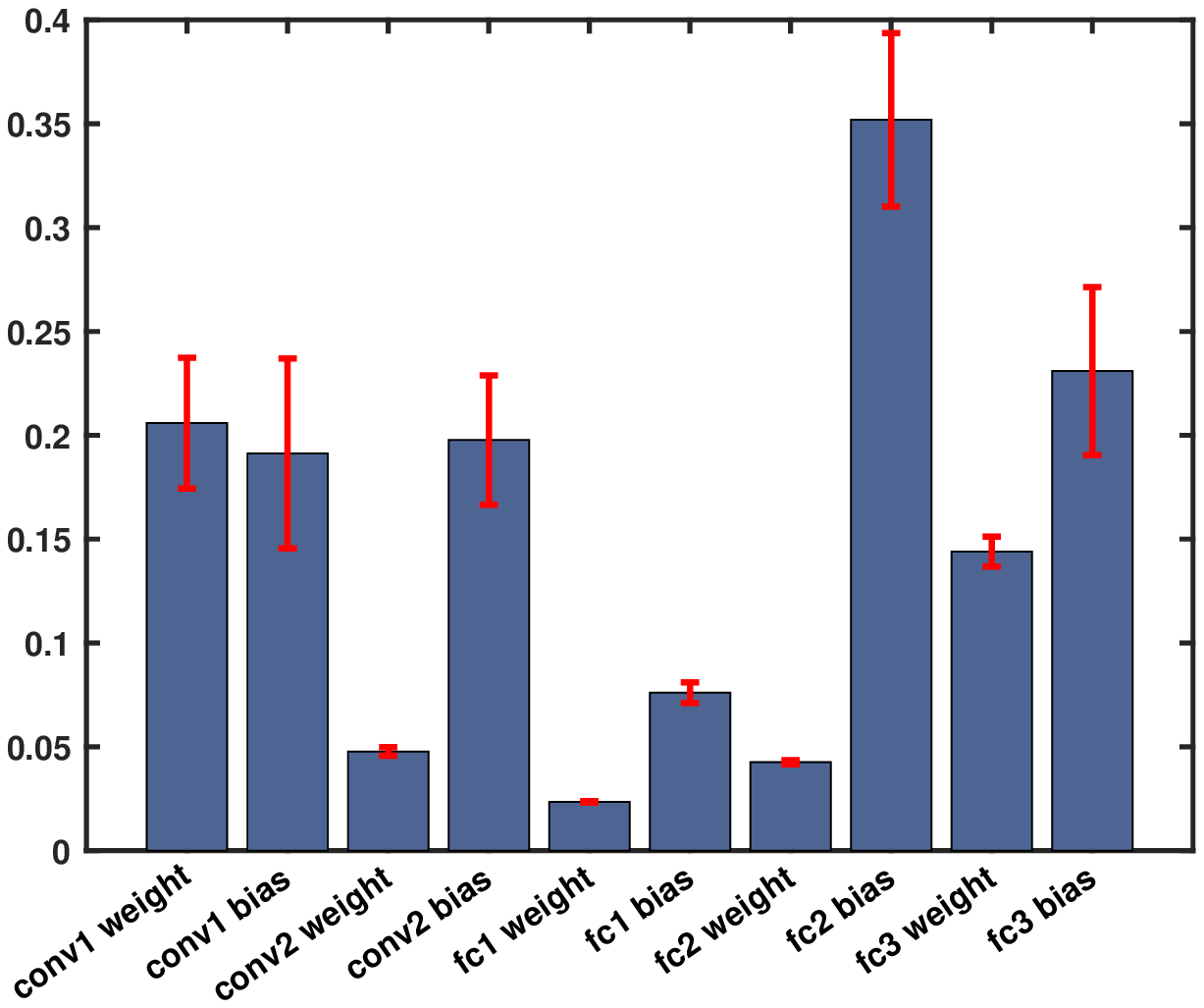}}
    \caption{Comparison of the dynamic range of weights in different layers of a typical CNN model for CIFAR-10 dataset.}
    \label{fig:dl_range}
\end{figure}

Layered quantization is an enhancement that builds on the quantization design in Section \ref{ssec:quantization}. We have emphasized the importance of selecting an appropriate quantization gain $G$ to match the dynamic range of the weight or the weight differential, depending on the specific transmission method. To empirically see this, Fig.~\ref{fig:dl_range} plots the statistics of different layers of a typical CNN model trained for CIFAR-10 dataset\footnote{More details can be found in Section~\ref{sec:exp}.}. We can see that the dynamic ranges of different layers are indeed very different. \rev{This phenomenon of varying weight distributions across layers of the DNN model has been reported in the literature \cite{han2016iclr}, and an automatic clip ranging tuning method has been proposed for the secure aggregation FL protocol \cite{bonawitz2019federated}.} Intuitively, if we apply different quantization gains to different layers, the overall performance can be improved over applying one global quantization gain.

To elaborate this approach, we first denote a quantization operation on weight $w$ with gain $G$ as $Q(w; G)$. Then, the quantization gain control on a particular layer can be written as $G=G_b G_e$, where $G_b$ represents the \textit{base quantization gain} that remains the same across different DNN layers, and $G_e$ represents the \textit{layer-specific quantization gain}. More specifically, $G_b$ is determined by the overall available quantization bit-width $B$, and $G_e$ is then applied to adjust the position of the remained digits for the specific layer. \rev{Then, for each training round, the server can implement LQ on the global model $\vect{w}_{t}$ to be broadcasted according to the following steps:}
\begin{enumerate}[leftmargin=12pt,topsep=0pt, itemsep=0pt,parsep=0pt]
    \item \textbf{Determine the base gain.} Set $G_b = 2^{B-1}$ for all layers.
    
    \rev{\item \textbf{Determine the layer-specific gain.} For each layer of $\vect{w}_t$, calculate the empirical cumulative distribution function (CDF) of this layer, and then take the 90-percentile value $\alpha$. Set $G_e = 2^\rho$, where $\rho = \lfloor \log_2 ({1/\alpha}) \rfloor$.}
    
    \item \textbf{Quantization.} Quantize the weights in this layer with $Q(w;G)$ where $G=G_b G_e$.
\end{enumerate}

\rev{The LQ design described above is ``dynamic'' in the sense that the layer-specific gain $G_e$ is updated in every round of FL. As a result, the server needs to broadcast the current $G_e$ for every layer to all participating clients, in conjunction with the latest (quantized) global model, so that the clients can properly scale down  the receiver output. We note that this additional communication of broadcasting $G_e$ for all layers is insignificant comparing to broadcasting the global model, and the overall communication overhead is not significantly increased. Furthermore, we can also adopt a ``static'' LQ design where the layer-specific gains $G_e$ are determined in advance on a pre-trained model. Then, $G_e$ can be fixed throughout the FL process (although still different across layers). This approach has the advantage of reduced computation (no need to compute the latest CDF and update $G_e$ in each round) and reduced communication (no need to communicating the latest $G_e$ in each round), at the expense of not tracking the dynamic range of weights in real time.}

\subsection{Convergence Analysis for \textsc{FedAvg} with Downlink Quantization}

We now analyze the convergence behavior of \textsc{FedAvg} with quantized downlink communication. In round $t$, the server first aggregates the uploaded weight update as
$
    \vect{w}_{t} = \frac{1}{K} \sum_{k \in \mathcal{S}_{t}} \vect{w}_{t}^k
$
and then broadcasts a quantized version $Q(\vect{w}_{t})$ for round $t+1$, as illustrated in Fig.~\ref{fig:system_model}. Suppose that we use the quantization scheme with tuned quantization and stochastic rounding as described in Section \ref{ssec:quantization}. The convergence behavior for quantized downlink communication is characterized in {Theorem~\ref{thm:3}}. \rev{The proof, which is quite different from the uplink case, can be found in Appendix \ref{app:proof_thm3}.}

\begin{theorem}
\label{thm:3}
    Reuse the definitions of $\kappa, \gamma, \eta_t$ in Theorem \ref{thm:1} and let
    \begin{equation}
        \label{eqn:B_dl}
        B_{t} = \log_2{\left ( 1 + \frac{\sqrt{1-\eta_t \mu}}{\eta_t} \right )}
    \end{equation}
     be the quantization level for the $t$-th iteration. With {stochastic rounding} on global weight $\vect{w}_t$ and under Assumptions \ref{as:F} and \ref{as:w_bounded}, the convergence bound in Eqn.~\eqref{eqn:conv_res} holds for non-i.i.d. local datasets, partial clients participation, and downlink quantization, with $D$ being replaced by
    \begin{equation*}
        D = \rev{\sum_{k=1}^{N}\frac{\sigma_k^2}{N^2}} + 6L \Gamma + 8(E-1)^2H^2 + \frac{N-K}{N-1}\frac{4}{K}E^2 H^2 + dM^2.
    \end{equation*}
\end{theorem}

\rev{Most of the dependencies on the hyperparameters still apply to the results in Theorem~\ref{thm:3}.} As a final comment, we note that the quantization precision in Eqn.~\eqref{eqn:B_dl} suggests that $B_{t} = \mathcal{O}\left( \log(t) \right)$, which matches the uplink analysis when the weight is directly transmitted. Since the downlink communication cannot adopt weight differential, it remains to be seen whether the $\mathcal{O}\left( \log(t) \right)$ requirement for quantization can be improved for the FL downlink.

\section{Experiments}
\label{sec:exp}

We validate the uplink and downlink communication design and compare the performance against the floating-point baseline, which represents a natural performance upper bound. \revsec{Following the setup in \cite{mcmahan2017fl, reddi2020adaptive}, we have carried out FL experiments on four datasets: MNIST \cite{lecun1998gradient}, CIFAR-10 \cite{krizhevsky2009learning}, Shakespeare \cite{shakespeare} and F-EMNIST \cite{caldas2018leaf}}. Details of the setup are given in Section~\ref{sec:exp_set}. Then, in Section \ref{sec:res_ul}, we focus on the performance of uplink communication and study the impact of parameters such as quantization gain and rounding. For downlink, we show in Section \ref{sec:res_dl} that a well-designed quantization scheme is critical to achieving good performance for downlink communication, and further demonstrate the performance improvement from layered quantization. Lastly, we combine both uplink and downlink designs and report the results in Section~\ref{sec:res_both}, which demonstrates that the proposed methods are capable of substantially improving the communication efficiency and, as a result, boosting the learning performance.

\subsection{Experiment Setup}
\label{sec:exp_set}

\subsubsection{MNIST}
\label{exp:mnist}
The training sets are evenly partitioned over $N=2000$ clients each containing 30 examples and we set $K=20$ per round (except in Fig.\ref{fig:upload2}, where we analysis the impact of $K$) . For the \textbf{i.i.d.} case, the data is shuffled and randomly assigned to each client while for the \textbf{non-i.i.d.} case, the data is sorted by labels, divided into 4000 shards, and each client is then assigned 2 shards randomly with 1 or 2 labels. The CNN model has two $5 \times 5$ convolution layers, a fully connected layer with 512 units and $\ssf{ReLU}$ activation, and a final output layer with softmax. The first convolution layer has 32 channels while the second one has 64 channels, and both are followed by $2 \times 2$ max pooling. The following parameters are used for training: local batch size $BS=5$, \rev{the number of local epochs $E=1$ for i.i.d. and $E=5$ for non-i.i.d., and learning rate $\eta = 0.065$.}

\subsubsection{CIFAR-10}
\label{exp:cifar}

The data partition is similar to the MNIST experiment for both i.i.d. and non-i.i.d. cases. We set $N=100$ and $K=10$ (except in Fig.\ref{fig:upload2}) for i.i.d while $N=K=10$ for non-i.i.d. We train a CNN model with two $5 \times 5$ convolution layers (both with 64 channels), two fully connected layers (384 and 192 units respectively) with $\ssf{ReLU}$ activation and a final output layer with softmax. The two convolution layers are both followed by $2 \times 2$ max pooling and a local response norm layer. The training parameters are: (a) \textbf{i.i.d.}: $BS=50$, $E=5$, learning rate
initially sets to $\eta=0.15$ and decays every 10 rounds with rate 0.99; (b) \rev{ \textbf{non-i.i.d.}: $BS=100$, $E=2$, $\eta=0.1$ and decay every round with rate 0.992.}

\subsubsection{Shakespeare}
\label{exp:shake}
This dataset is built from \textit{The Complete Works of William Shakespeare} and each speaking role is viewed as a device. Hence, the dataset is naturally unbalanced and \textbf{non-i.i.d.} since the number of lines and speaking habits of each role vary significantly. There are totally 1,129 roles in the dataset \cite{caldas2018leaf}. We randomly pick 300 of them and build a dataset with 794,659 training examples and 198,807 test examples. We also construct an i.i.d. dataset by shuffling the data and redistribute evenly to 300 roles and set $K=10$. The task is the next-character prediction, and we use a classifier with an 8D embedding layer, two LSTM layers (each with 256 hidden units) and a softmax output layer with 86 nodes. The training parameters are: $BS=20$, $E=1$, learning rate
initially sets to $\eta=0.8$ and decays every 10 rounds with rate 0.99.

\revsec{
\subsubsection{F-EMNIST}
\label{exp:femnist}
We use the federated version of the EMNIST dataset (F-EMNIST) \cite{caldas2018leaf} in this experiment. There are 3,400 clients with total 704,017 training examples and 79,952 test examples. It should be noted that F-EMNIST partitions the images of digits or English characters by their authors, thus the dataset is \emph{naturally non-i.i.d.} since the writing style varies from person to person. We use the model recommended by \cite{reddi2020adaptive}, which is a CNN model with two convolutional layers, max pooling, and dropout, followed by a 128-unit linear layer. We set $K=10, BS=10, E=1$ and $\eta=0.03$ for training.
}

\subsection{Results for Uplink Communication}
\label{sec:res_ul}

\begin{figure}
    \centering
    \subfigure{
        \includegraphics[width=0.48\columnwidth]{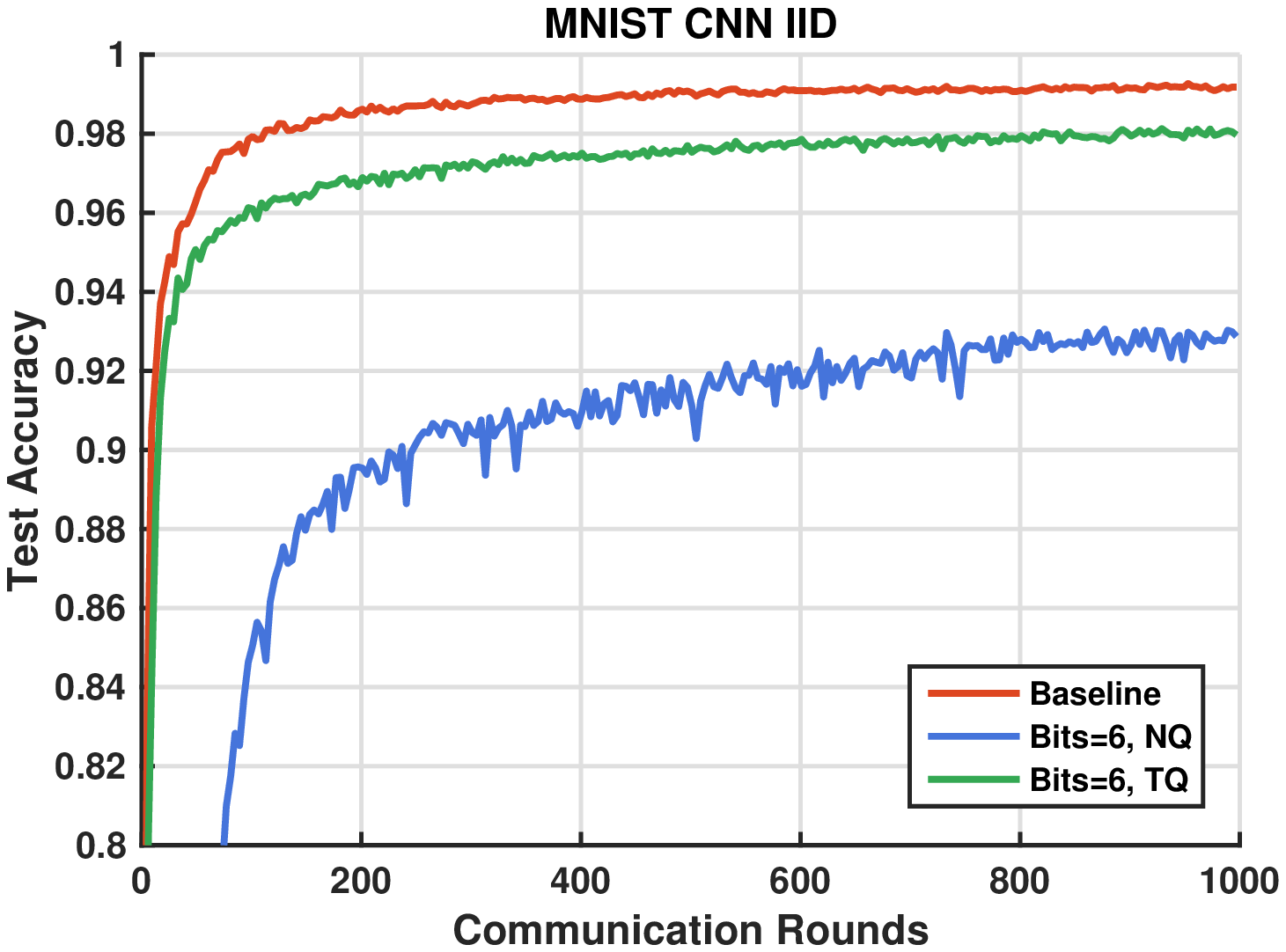}}
    \subfigure{
        \includegraphics[width=0.48\columnwidth]{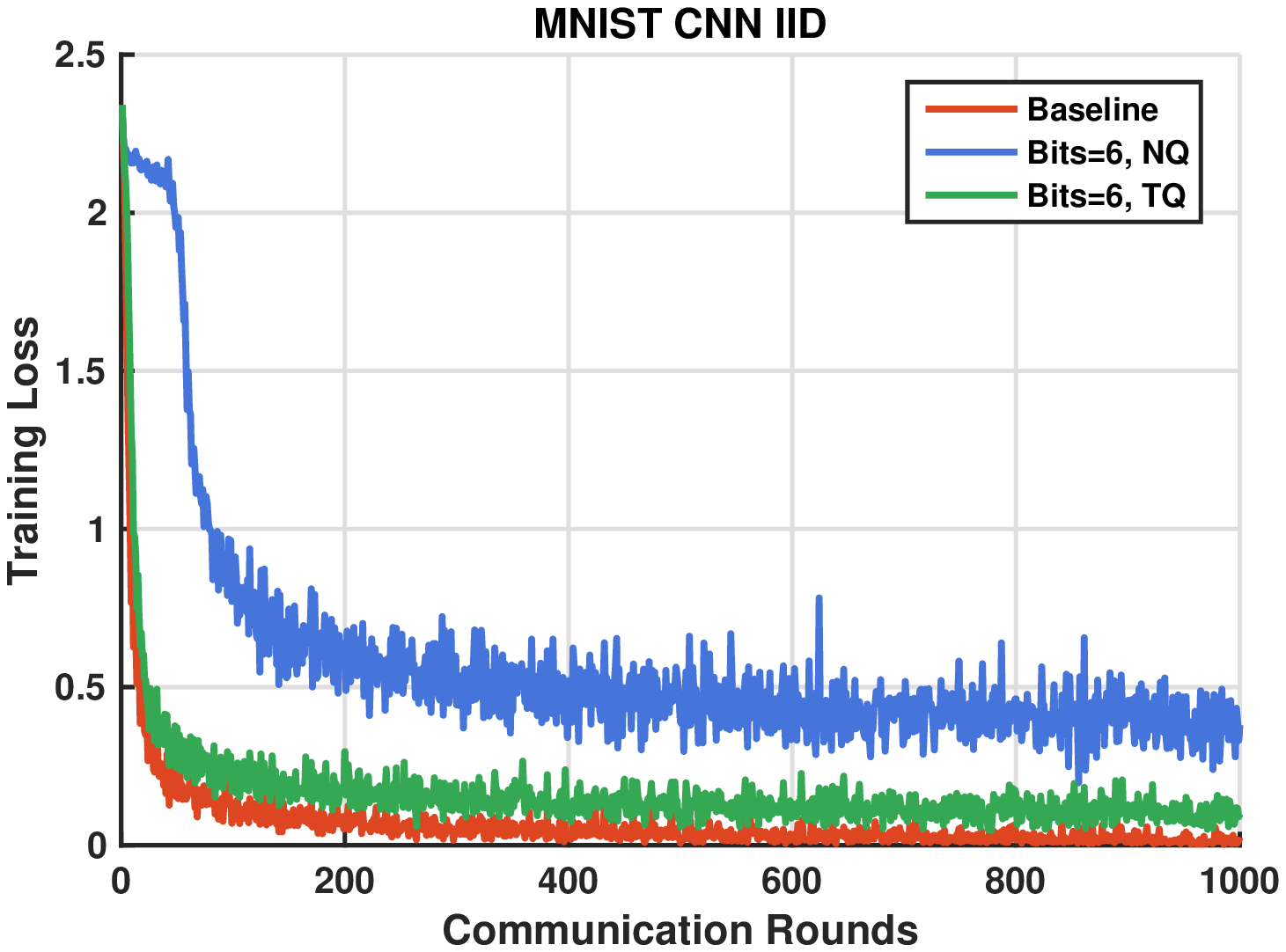}}
    \caption{Comparing the performance of Native Quantization (NQ) and Tuned Quantization (TQ).}
    \label{fig:upload1}
\end{figure}

\mypara{Native quantization versus tuned quantization.}
In Fig.~\ref{fig:upload1}, we compare the performance of Native Quantization (NQ) and Tuned Quantization (TQ), two different structures described in Section \ref{ssec:quantization}, on the MNIST dataset\footnote{Both model accuracy on the test set and the training loss are plotted for the remainder of this paper for all experiments.}. The quantization gain for NQ is set to $G=2^{6-1}=32$ (the maximum value for 6-bit), which has a 6.38\% degradation in the test accuracy compared to the baseline. TQ on the other hand allows a larger and more suitable $G$ (256 in this case) and achieves significantly better performance. This demonstrates the advantage of TQ. 

\begin{figure*}
    \hsize=\textwidth
    \centering
    \subfigure{
        \includegraphics[width=0.48\textwidth]{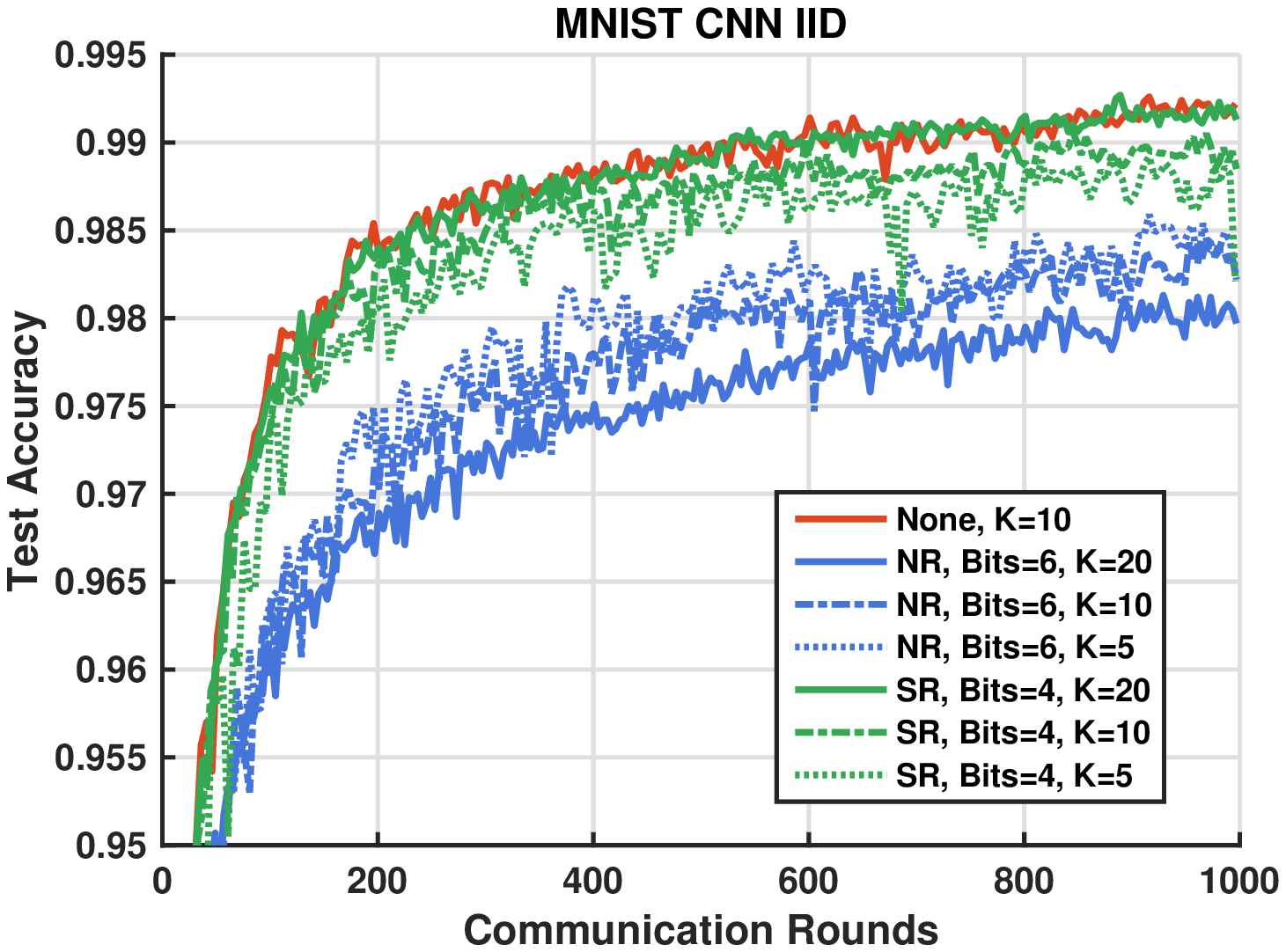}}
    \subfigure{
        \includegraphics[width=0.48\textwidth]{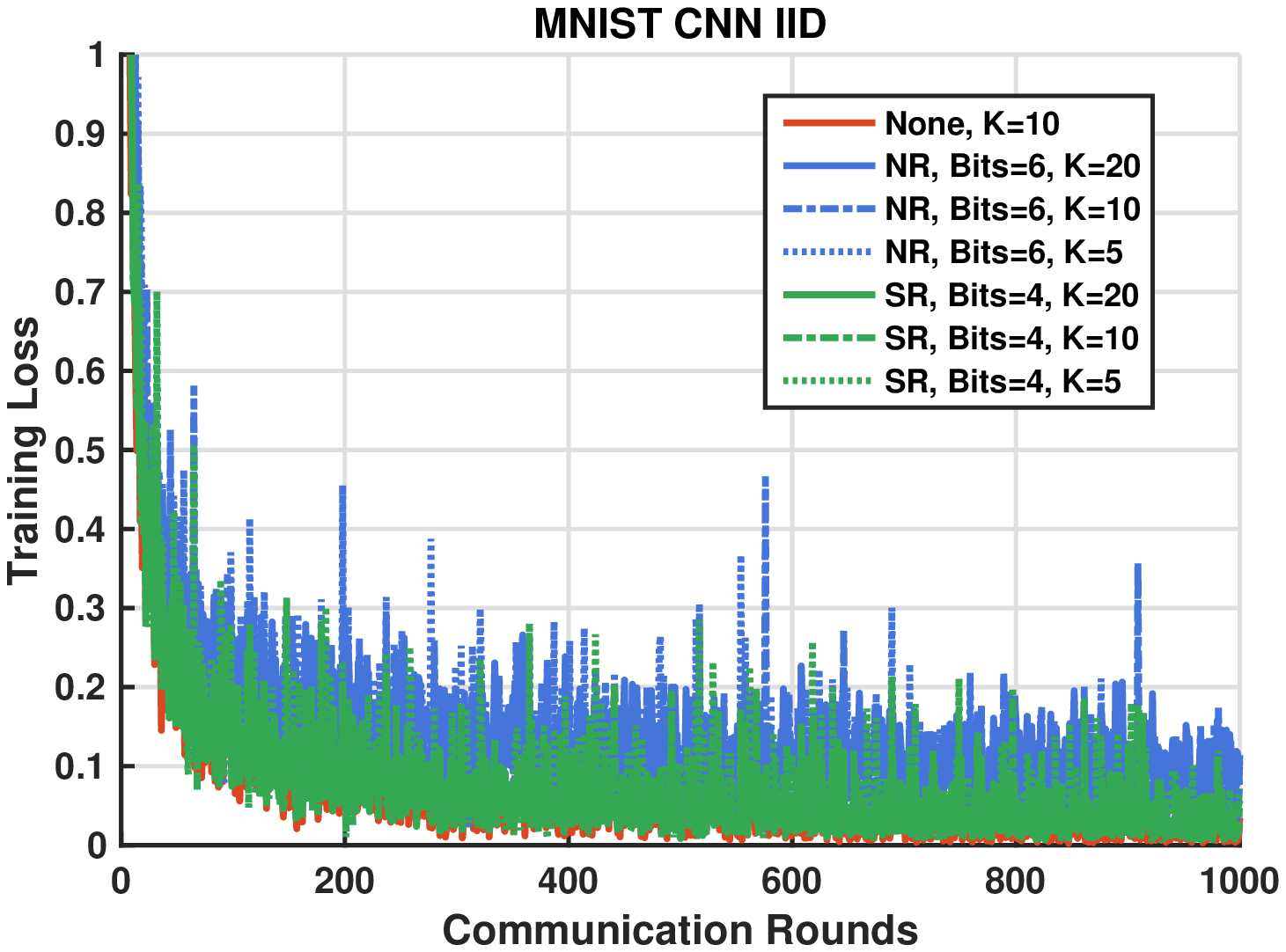}}
    \subfigure{
        \includegraphics[width=0.48\textwidth]{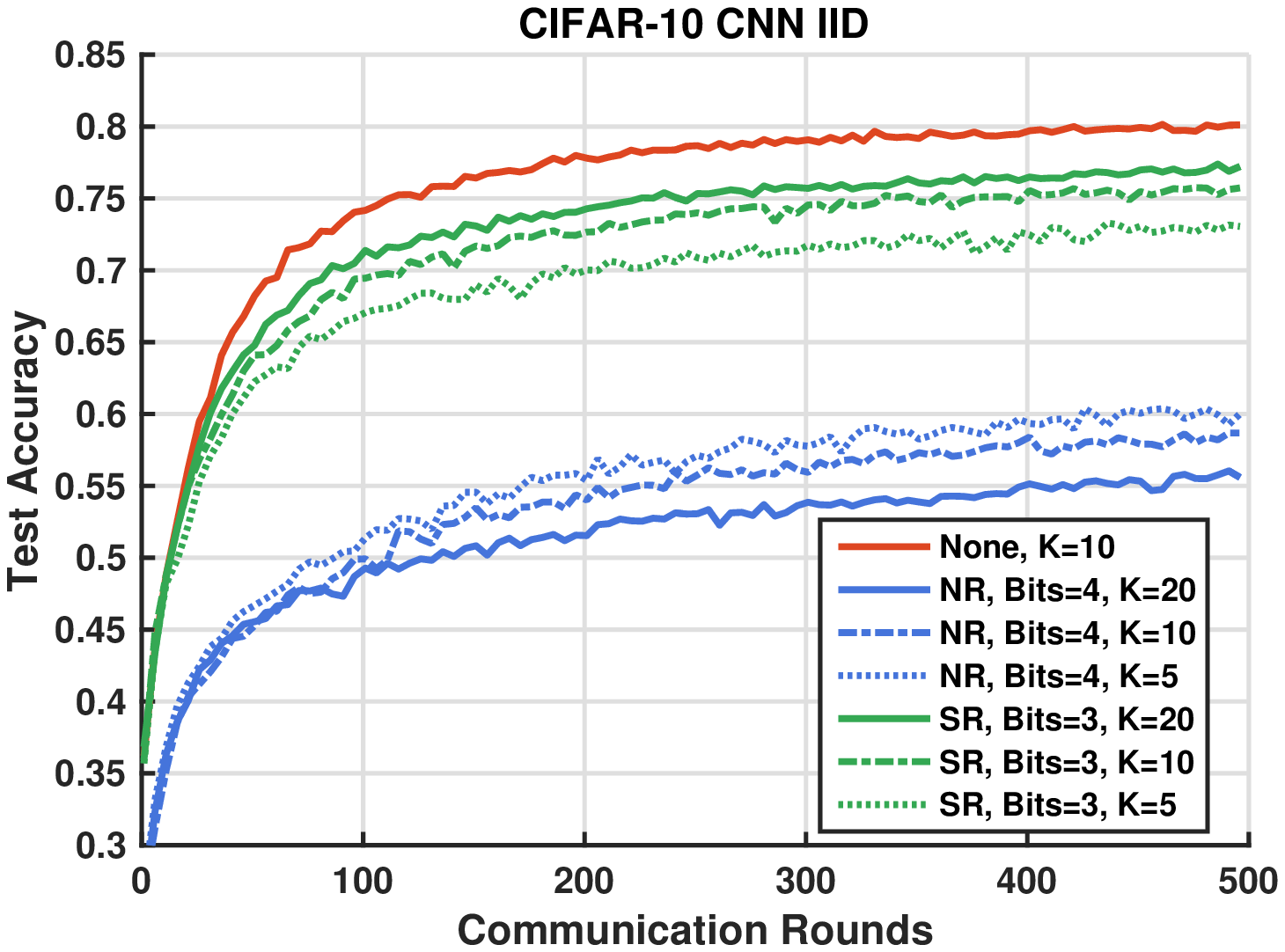}}
    \subfigure{
        \includegraphics[width=0.48\textwidth]{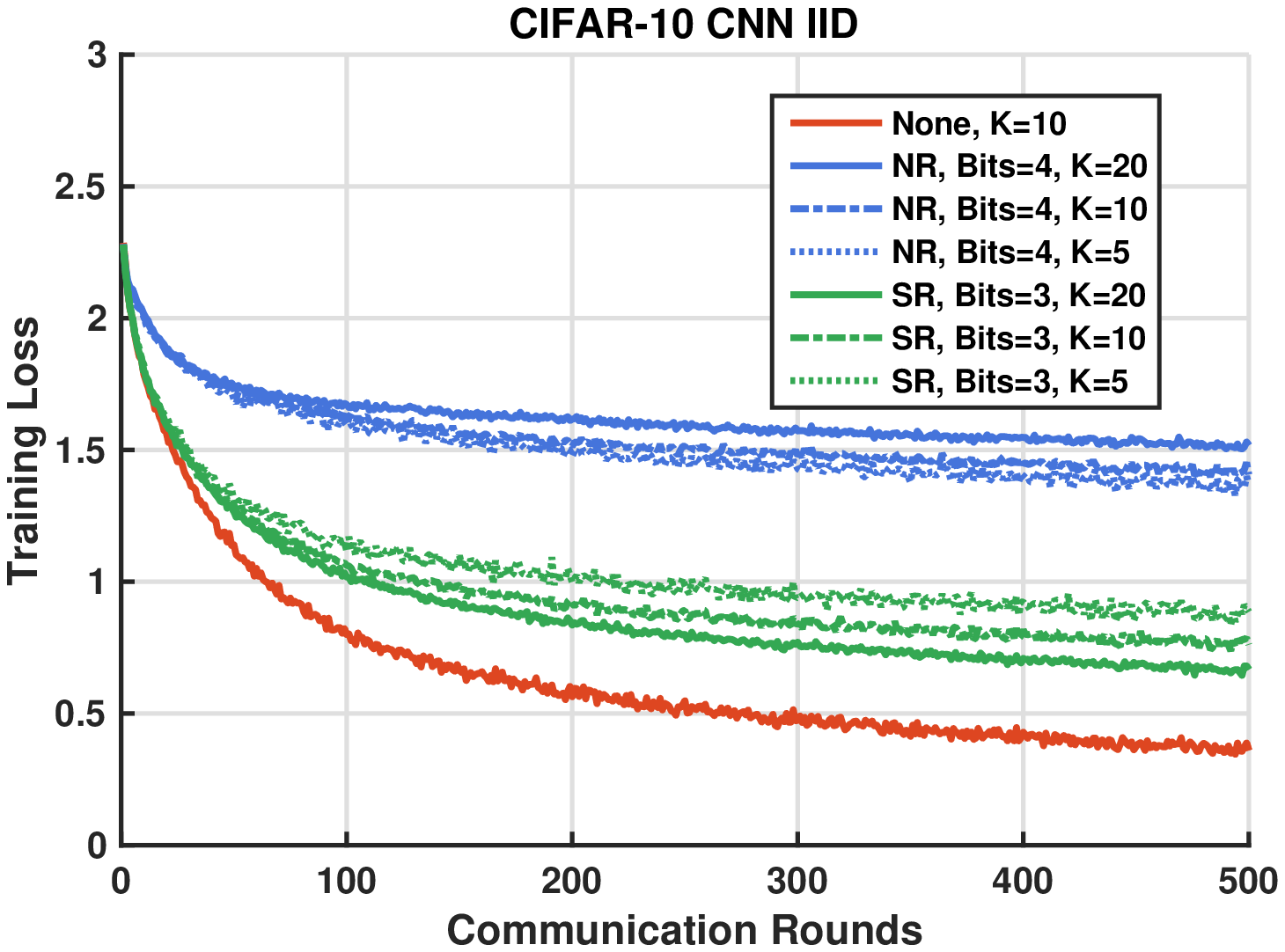}}
    \caption{Comparing the performance of Nearest Rounding (NR) and Stochastic Rounding (SR) on MNIST \rev{(top two subplots)} and CIFAR-10 \rev{(bottom two subplots)}.}
    \label{fig:upload2}
\end{figure*}

\mypara{Nearest rounding versus stochastic rounding.}
Our next experiment compares stochastic rounding (SR) and nearest rounding (NR). Although NR is widely used in communication systems, we see from Fig.~\ref{fig:upload2} that SR is significantly better in both the final model accuracy and the convergence speed even when fewer bits are used. In addition, we observe an interesting phenomenon that the impact of $K$ is different for NR and SR. \rev{For NR, having more clients participate in the model training may actually {\em degrade} the performance\footnote{\rev{We hypothesize that this is because NR, which is not an unbiased quantizer, might lead to error accumulation with more clients participating in the aggregation, and this detrimental effect may outweigh the benefit of more clients. We plan to investigate this aspect in a future work.}}, while this observation does not hold for SR, which is consistent with our theoretical results in {Theorem \ref{thm:1}}, where a larger $K$ reduces the value of $D$ and leads to a reduced upper bound of convergence error.} This observation is even more prominent in the results of CIFAR-10 in Fig.~\ref{fig:upload2}.  Nevertheless, the message from the experiment is clear -- one should adopt SR over NR when possible. We note that this is also consistent with the DNN compression literature \cite{lin2016fixed,han2016iclr}.

\begin{figure*}
    \hsize=\textwidth
    \centering
    \subfigure{
        \includegraphics[width=0.48\textwidth]{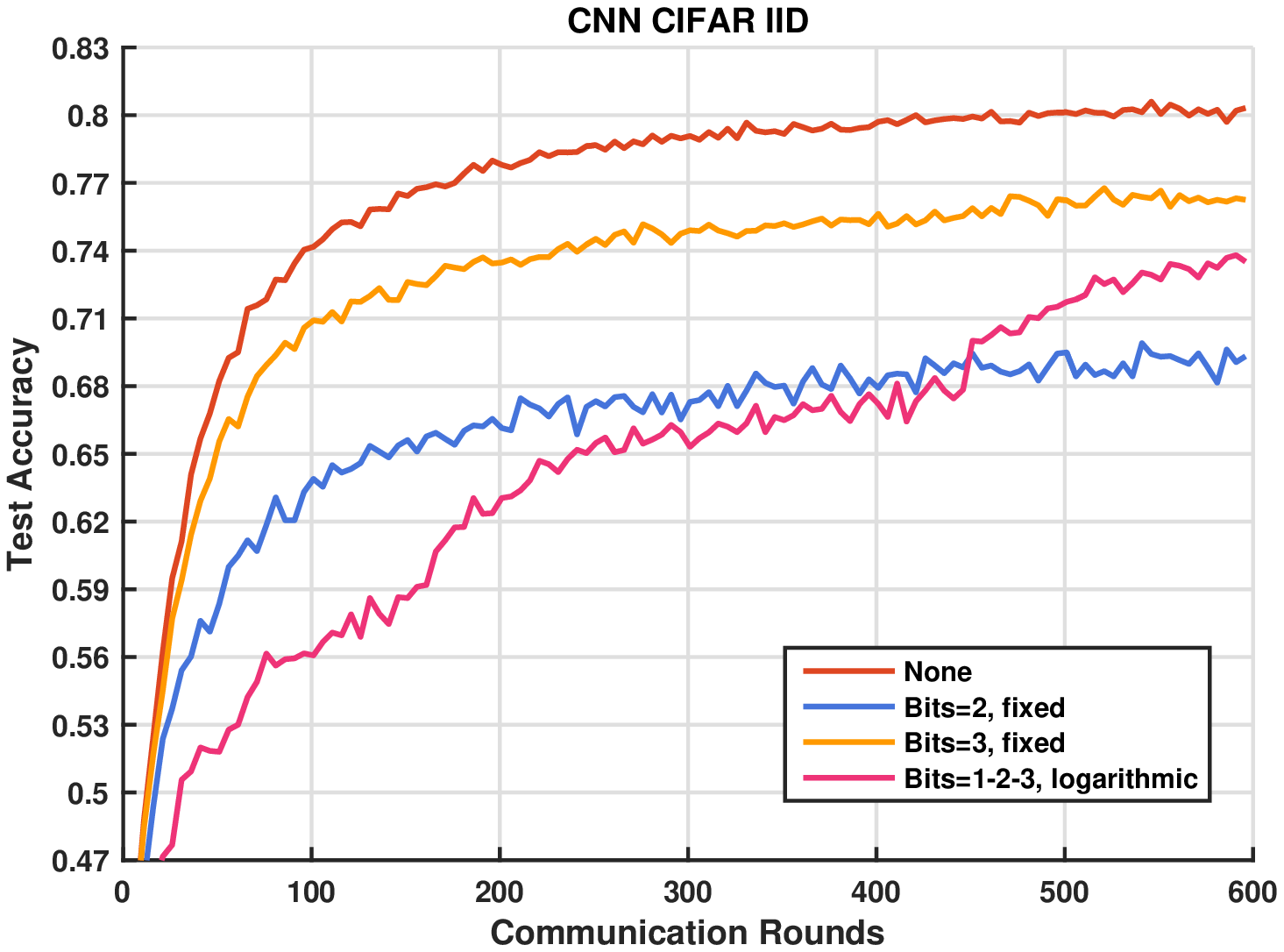}\label{sfig:inc_1}}
    \subfigure{
        \includegraphics[width=0.48\textwidth]{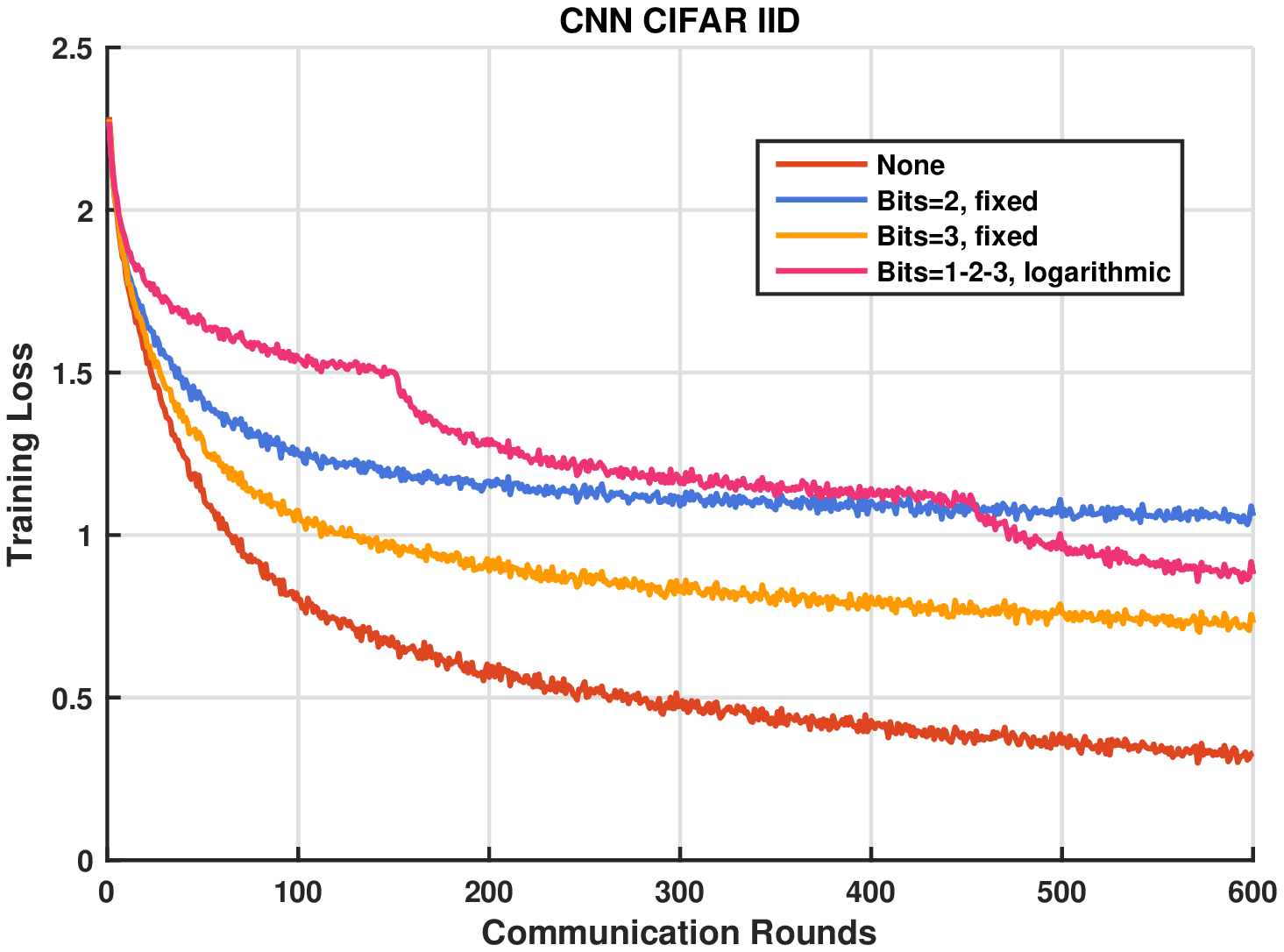}\label{sfig:inc_2}}
    \subfigure{
        \includegraphics[width=0.48\textwidth]{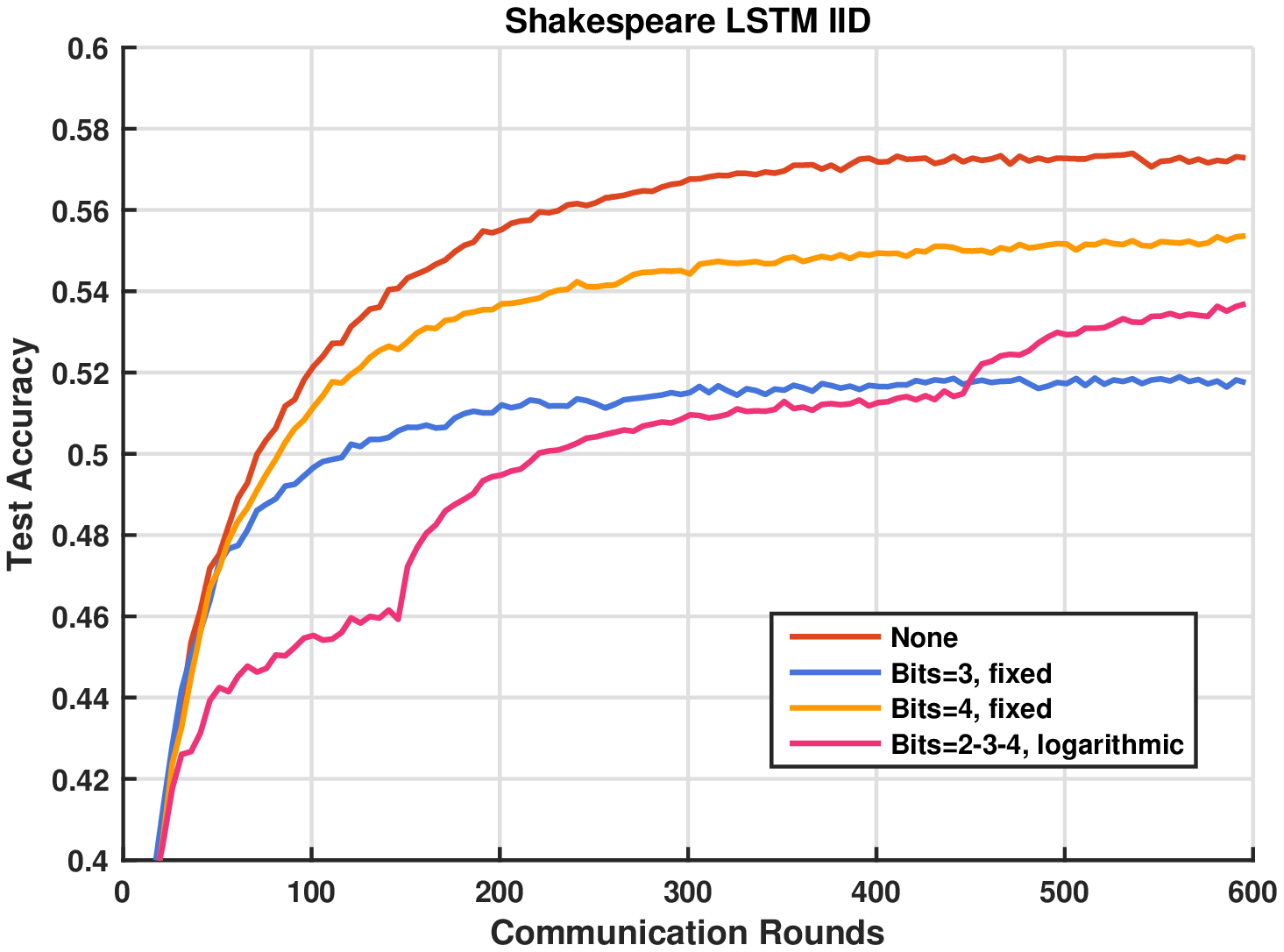}}
    \subfigure{
        \includegraphics[width=0.48\textwidth]{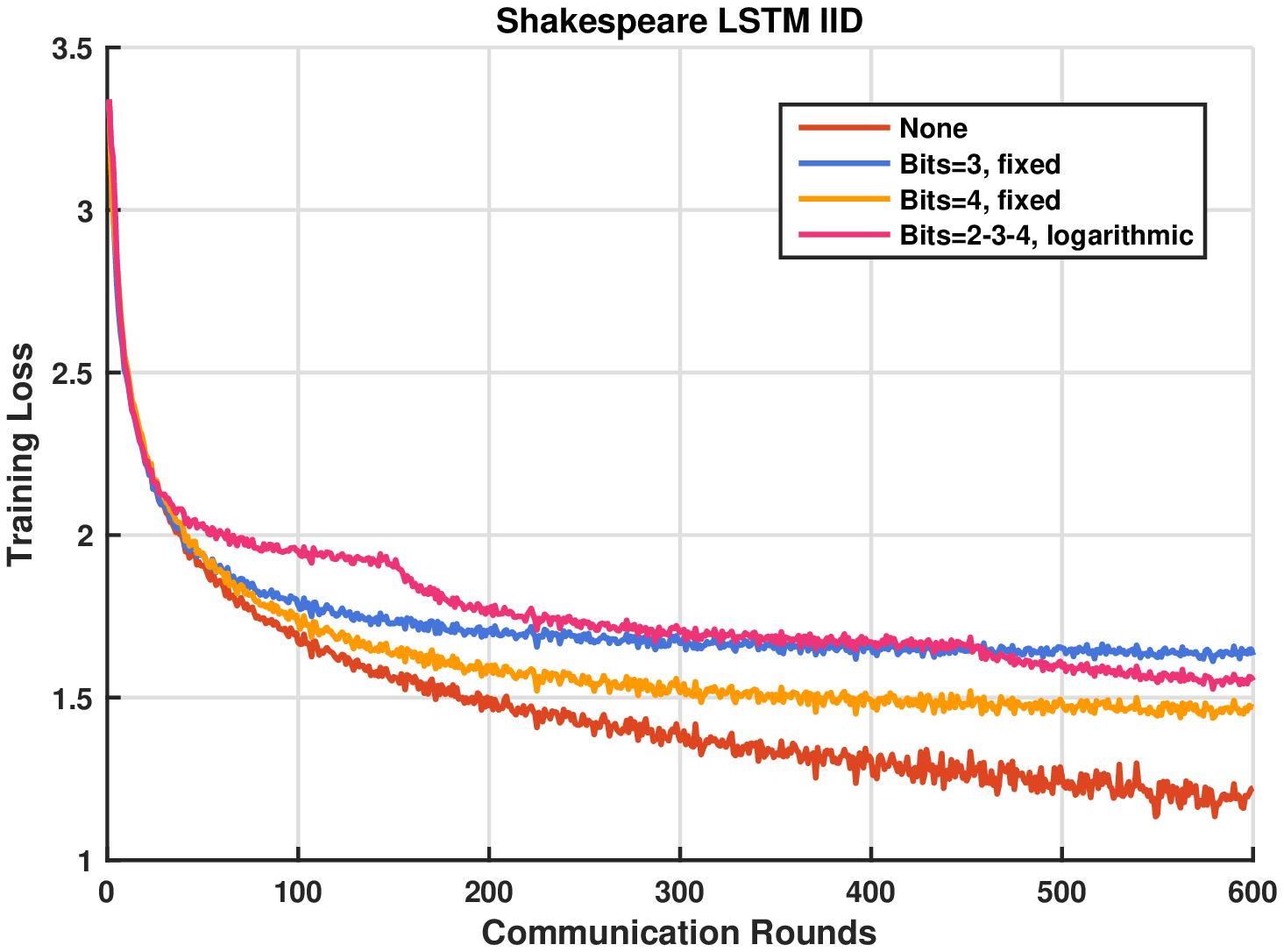}}
    \caption{\rev{Comparing the performance of fixed and increasing quantization level on CIFAR-10 (top two subplots) with 
    $f=2, p=75$ and Shakespeare (bottom two subplots) with $f=4, p=37.5$.}}
    \label{fig:ul_inc}
\end{figure*}

\rev{
\mypara{Benefits of increasing quantization level.}
The convergence analysis in Section \ref{ssec:conv_all} indicates that to achieve an $\mathcal{O}(\frac{1}{T})$ convergence rate with quantization, transmitting the weights without differential requires increasing the quantization level at a logarithmic rate. Our experimental results verify this conclusion. In Fig.\ref{fig:ul_inc}, the logarithmic approach increases the quantization bit-width according to $B=\lfloor\log_2\squab{f + (r-1)/p} \rfloor$, where $r=1,2,\cdots$ is the index of training round. By contrast, the fixed approach keep a constant bit-width throughout. In Fig.\ref{fig:ul_inc}, the average bit-width for each round of the logarithmic approach on CIFAR-10 dataset is 2, but we can see that it outperforms the result with fixed 2-bit in the final convergence accuracy. The average bit-width of the logarithmic approach on Shakespeare dataset is 3, which also have better performance compared to the fix-3bit quantization.
}

\begin{figure}[H]
    \centering
    \subfigure[weight]{
        \includegraphics[width=0.437\textwidth]{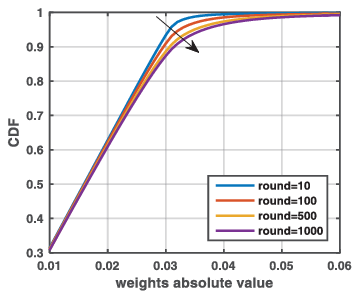}}
    \subfigure[differential weight]{
        \includegraphics[width=0.48\textwidth]{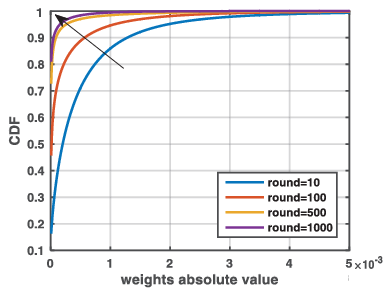}}
    \caption{Comparing the distribution and value range of the weight itself and differential weight (MNIST, i.i.d.). K is set to 20 and at specified rounds we make statistics of the weights of the 20 selected clients.}
    \label{fig:upload3}
\end{figure}

\mypara{Advantages of differential transmission.}
One of the key benefits in using DT is that the dynamic range of weight differential $\vect{d}_{t+1}^k$ is much smaller than the weight  $\vect{w}_{t+1}^k$ itself, and thus quantization will be more precise with the same bit-width $B$. We now empirically validate this point by plotting the empirical cumulative distribution function (CDF) of both representations in Fig. \ref{fig:upload3}. We can see that DT has a dynamic range that is an order-of-magnitude smaller than the weight itself, which suggests that the intuition is correct.  We also see that distribution of the weight differential gradually concentrates and the support also decreases as the training progresses towards the end. At round 10, 35\% of the weights is less than 9e-5 while at round 500, this proportion achieves 90\%. This is another useful observation, as it indicates that we may be able to decrease the quantization bit-width at the late stage of training.

\begin{figure}
    \centering
    \subfigure{
        \includegraphics[width=0.48\textwidth]{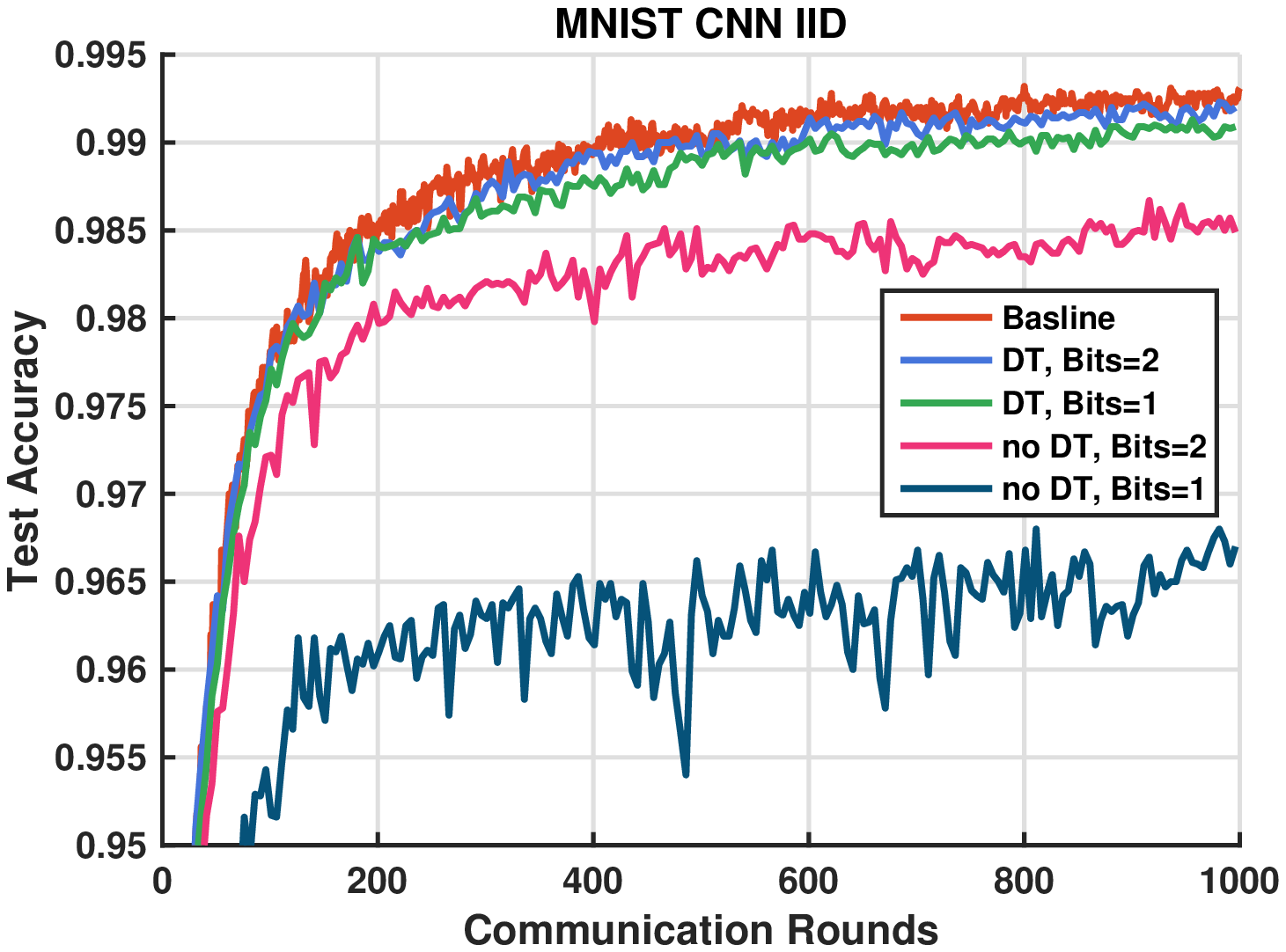}}
    \subfigure{
        \includegraphics[width=0.48\textwidth]{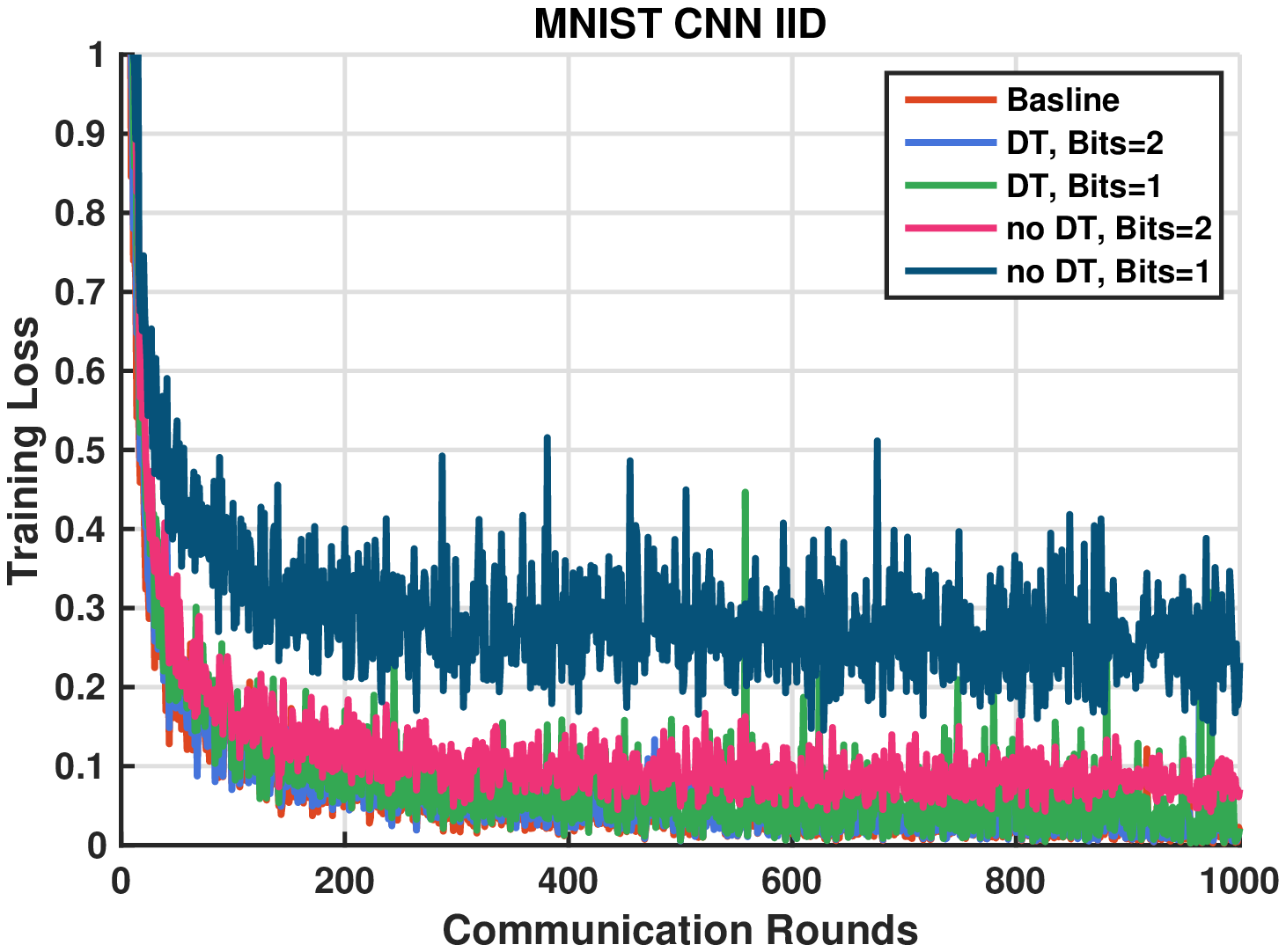}}
    \subfigure{
        \includegraphics[width=0.48\textwidth]{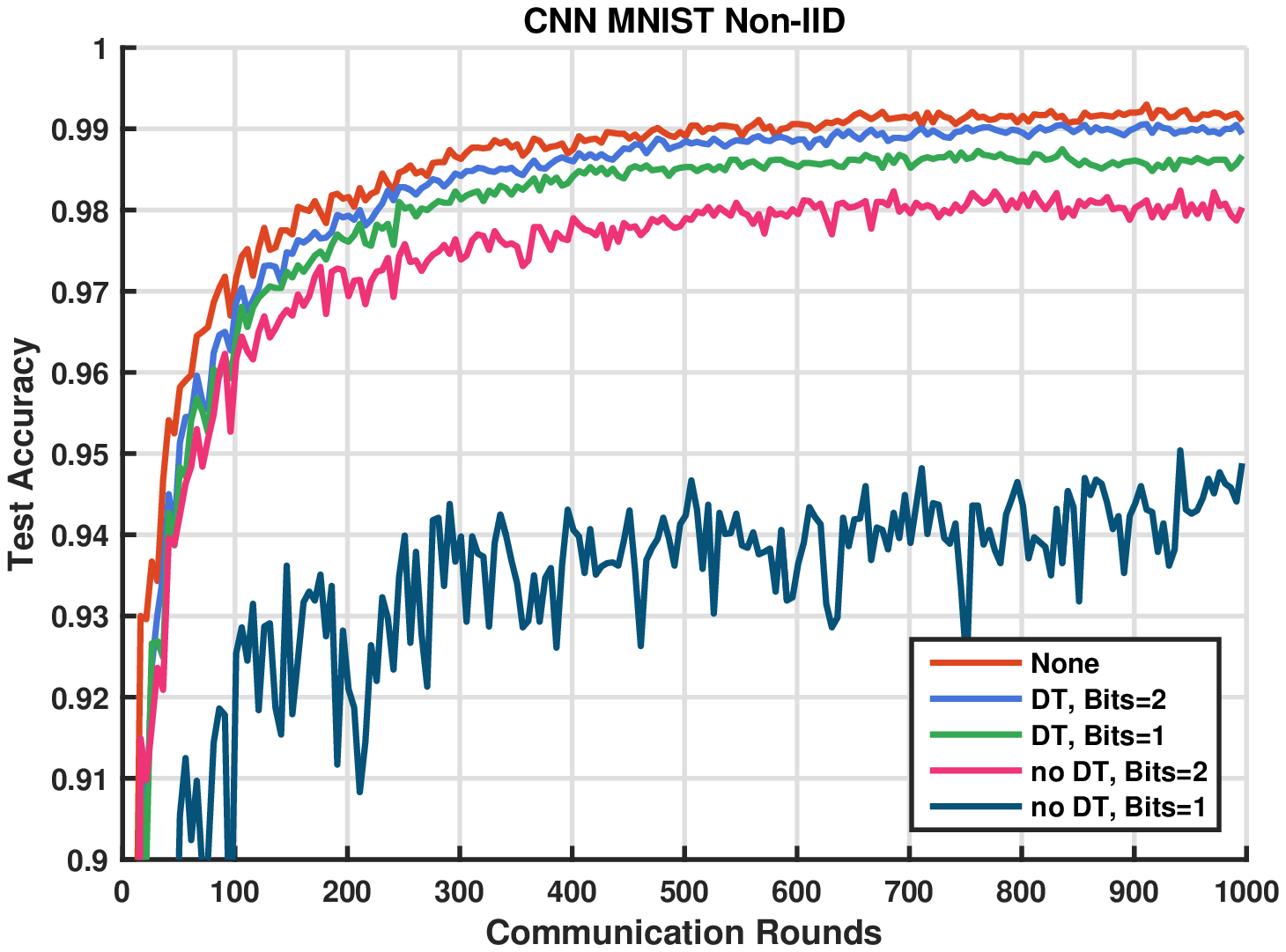}}
    \subfigure{
        \includegraphics[width=0.48\textwidth]{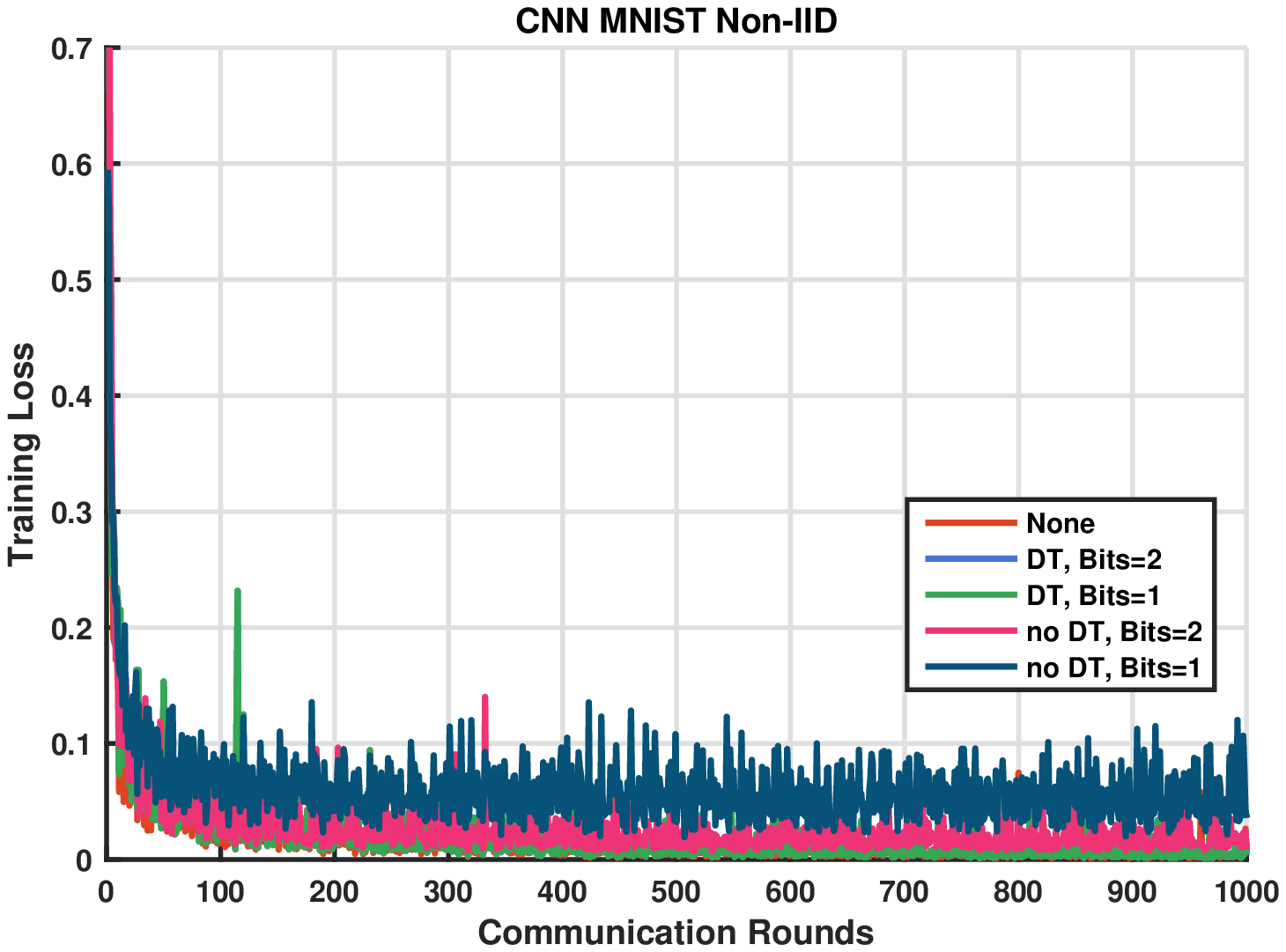}}
    \caption{Comparing the performance and transmission with and without DT on i.i.d. (top two subplots) and non-i.i.d. (bottom two subplots) MNIST dataset. Both are quantized with TQ and SR. For the 1-bit DT of the non-i.i.d. case, the learning rate is reduced to 0.03.}
    \label{fig:upload4}
\end{figure}

\mypara{Putting all techniques together.}
Finally, we report an experiment where TQ, SR and DT are combined, which represents the best quantization design for uplink communication in our paper. We are interested in evaluating how well this design performs, especially comparing against the floating-point baseline (no quantization). Fig.~\ref{fig:upload4} shows that, for both i.i.d. and non-i.i.d. cases, we are able to quantize the floating-point weight differential to 1-bit representations with almost negligible performance loss:
\begin{itemize}[leftmargin=12pt,topsep=0pt, itemsep=0pt,parsep=0pt]
\item \textbf{i.i.d.} 99.08\% accuracy (99.83\% of the baseline accuracy) for 1-bit (3.13\% of the baseline bandwidth); 99.18\% accuracy (99.93\% of the baseline accuracy) for 2-bit (6.25\% of the baseline bandwidth);
\rev{\item \textbf{non-i.i.d.} 98.59\% accuracy (99.41\% of the baseline accuracy) for 1-bit (3.13\% of the baseline bandwidth); and 98.99\% accuracy (99.81\% of the baseline accuracy) for 2-bit (6.25\% of the baseline bandwidth).}
\end{itemize}
These results suggest that the proposed design achieves the best communication efficiency in this FL task, to the best of the authors' knowledge. 

\begin{figure*}[htbp]
    \hsize=\textwidth
    \centering
    \subfigure{
        \includegraphics[width=0.48\textwidth]{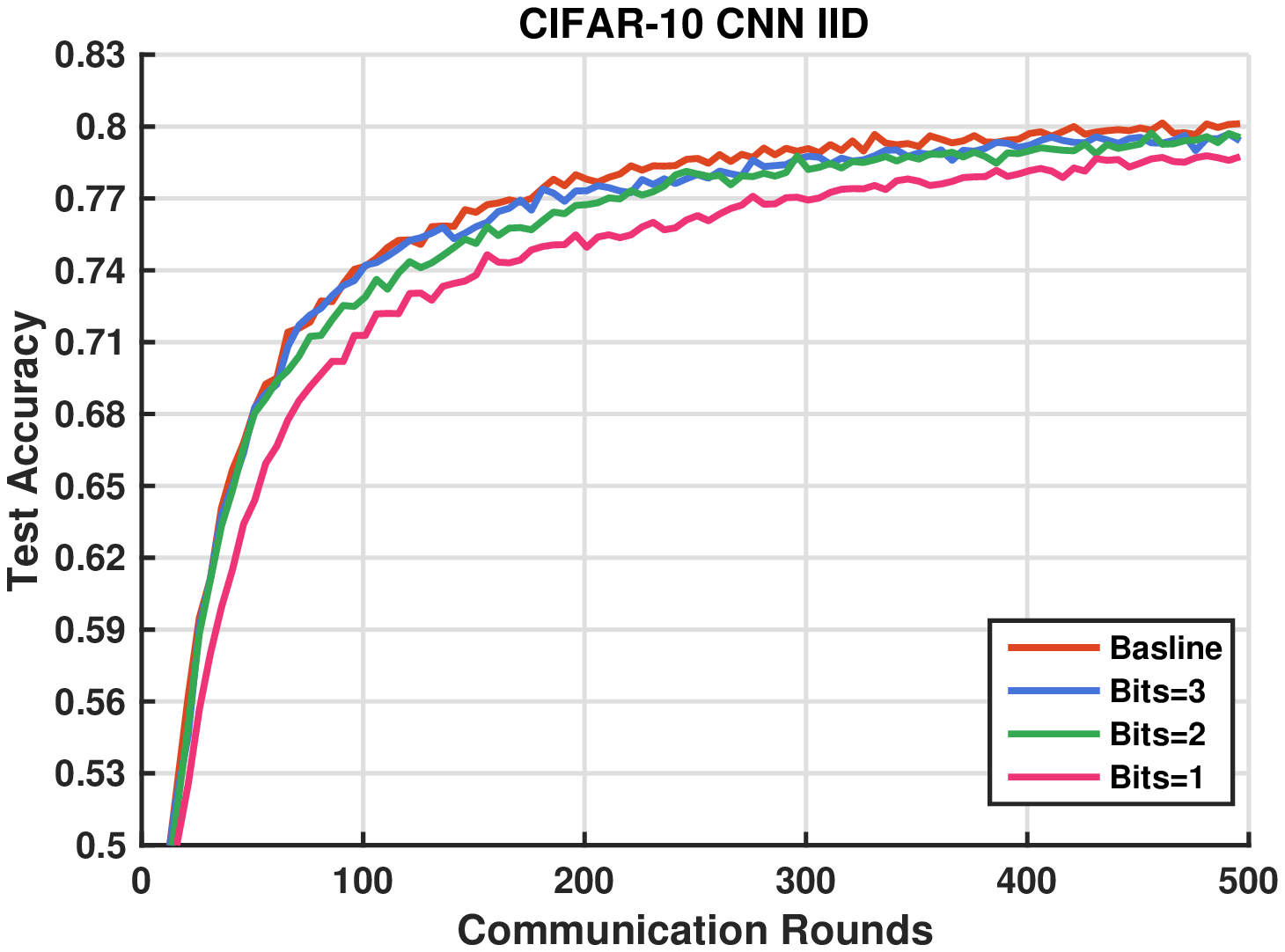}}
    \subfigure{
        \includegraphics[width=0.48\textwidth]{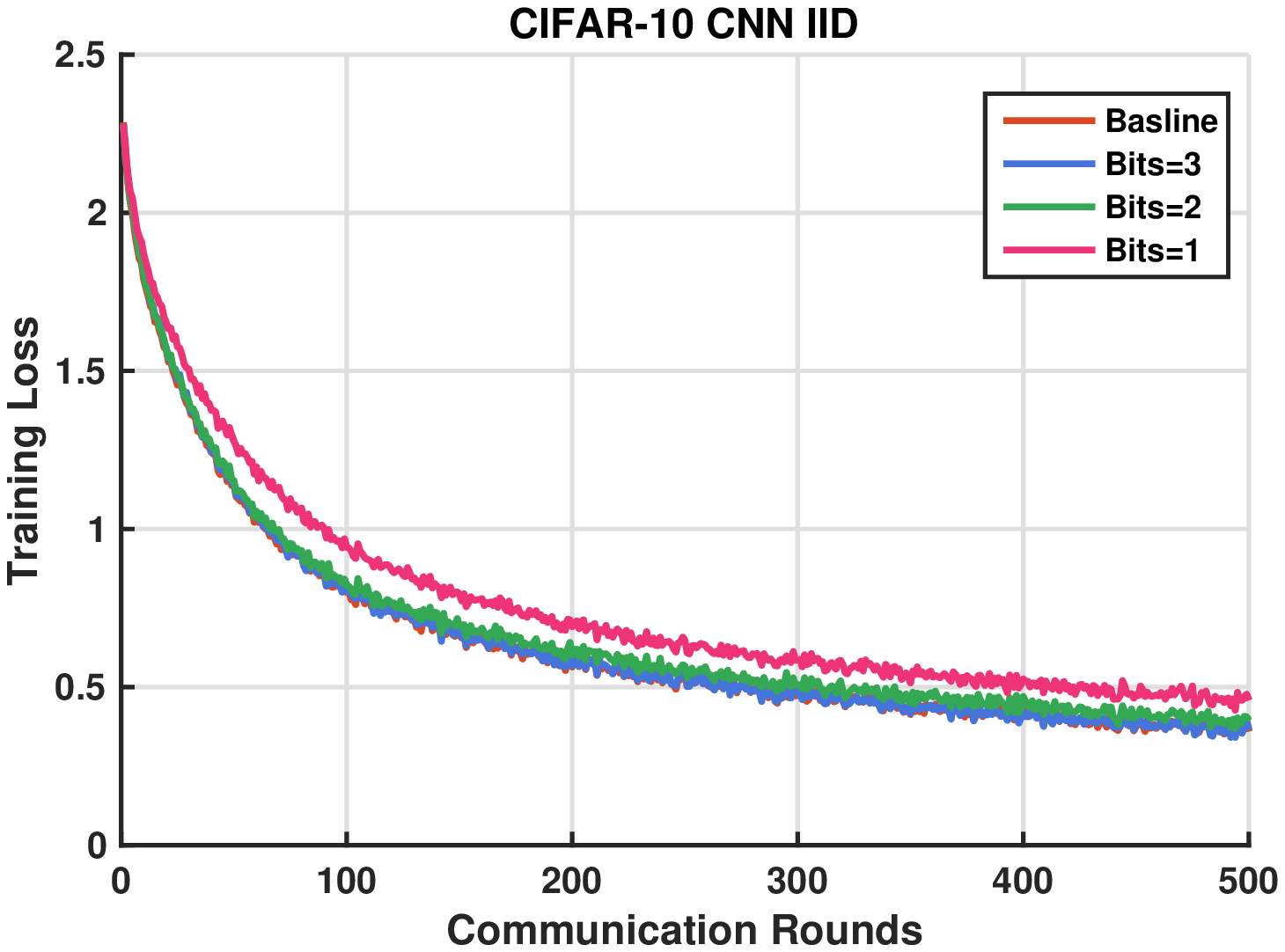}}
    \subfigure{
        \includegraphics[width=0.48\textwidth]{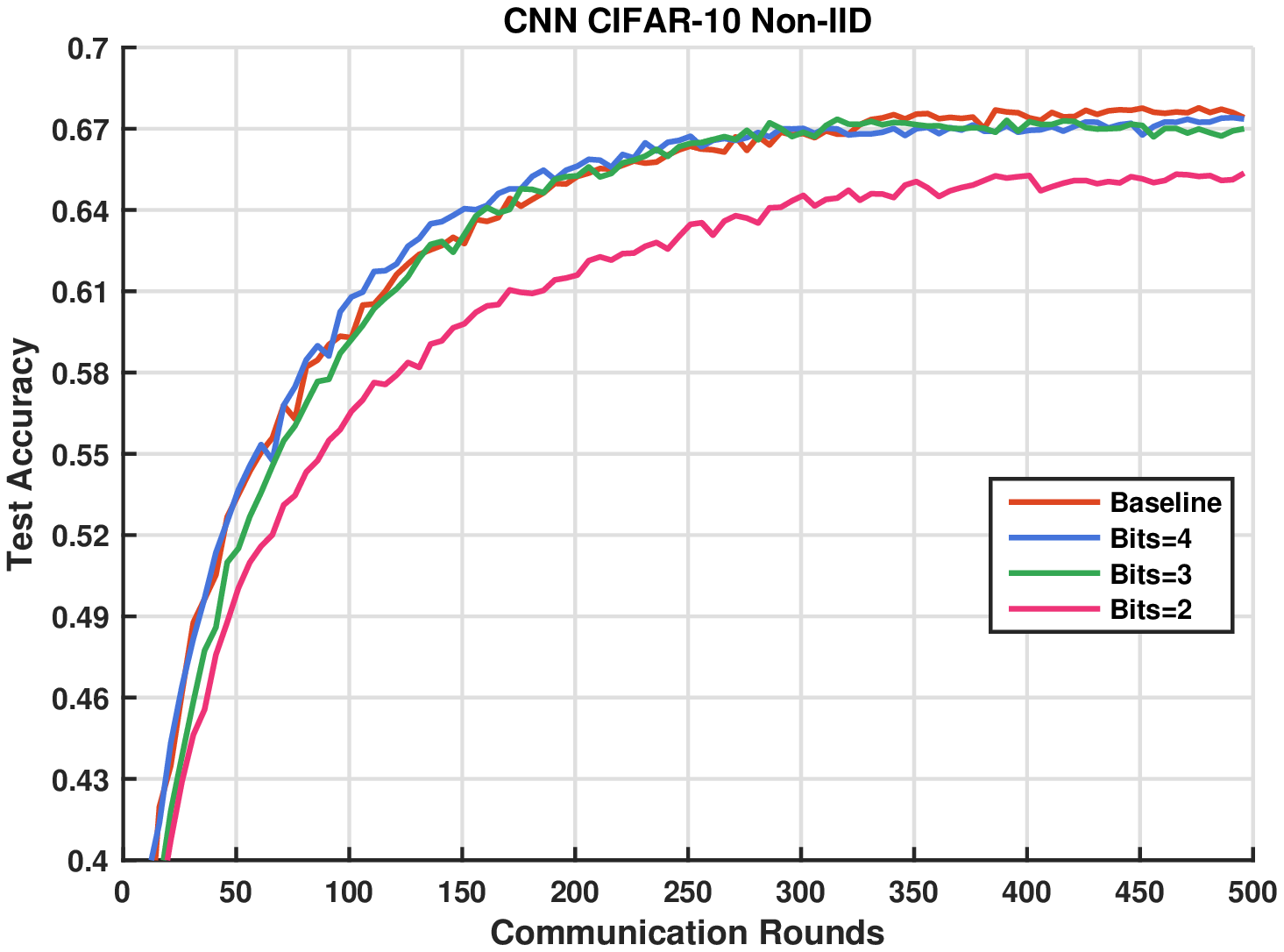}}
    \subfigure{
        \includegraphics[width=0.48\textwidth]{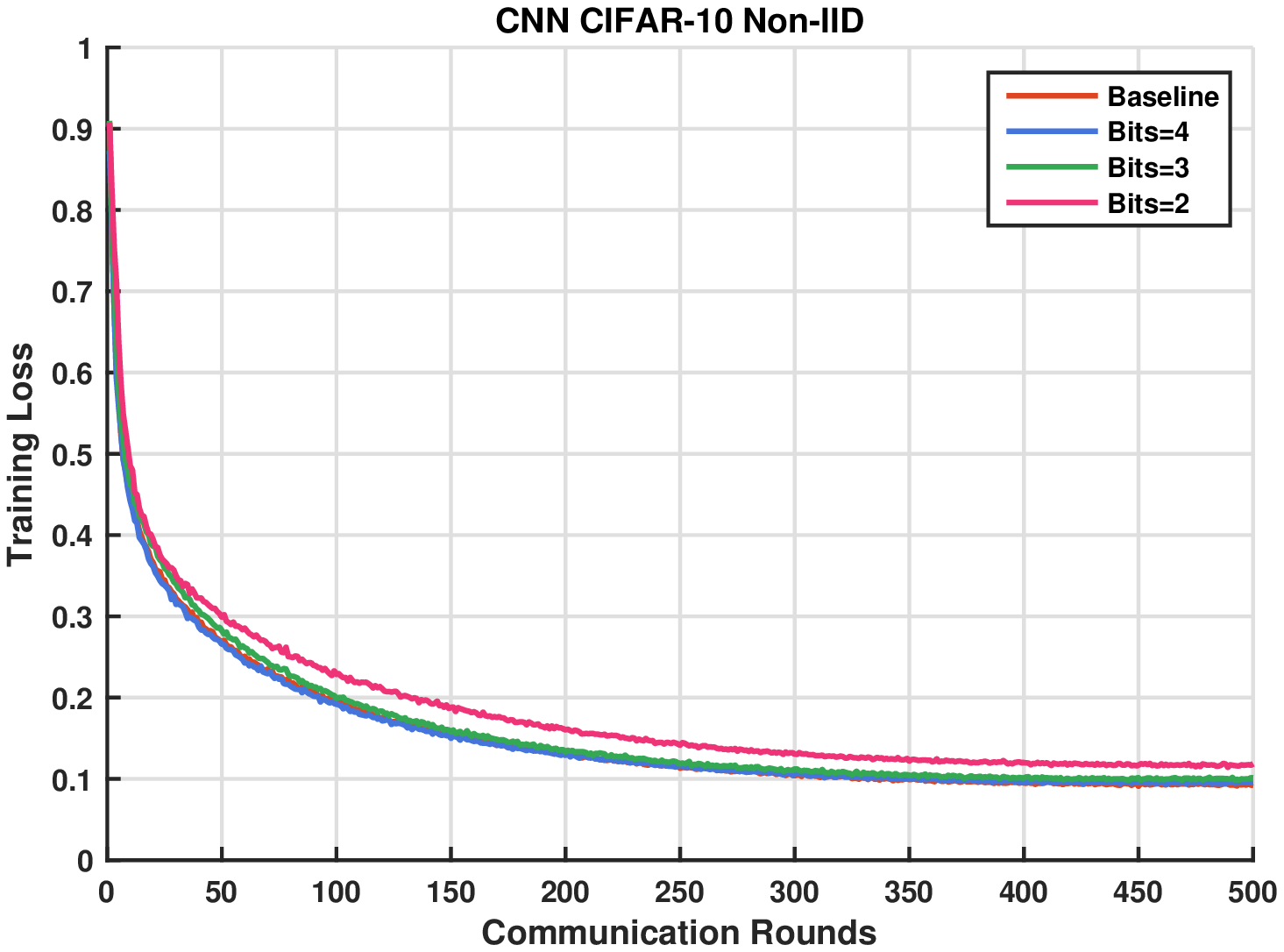}}
    \caption{The performance of uplink quantization on i.i.d. (top two subplots) and non-i.i.d. (bottom two subplots) CIFAR-10 dataset.}
    \label{fig:upload5}
\end{figure*}

\begin{figure*}[htbp]
    \hsize=\textwidth
    \centering
    \subfigure{
        \includegraphics[width=0.48\textwidth]{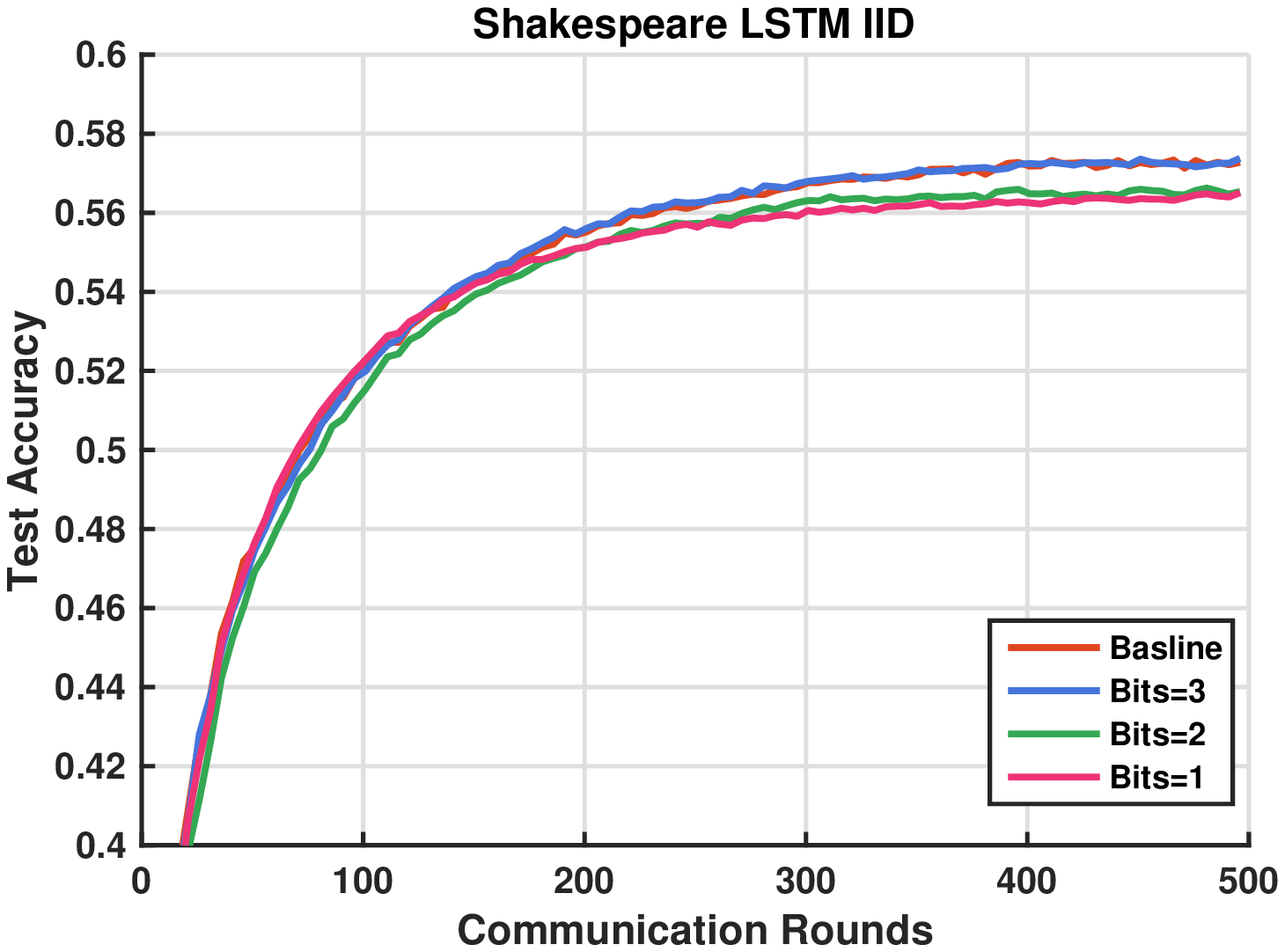}}
    \subfigure{
        \includegraphics[width=0.48\textwidth]{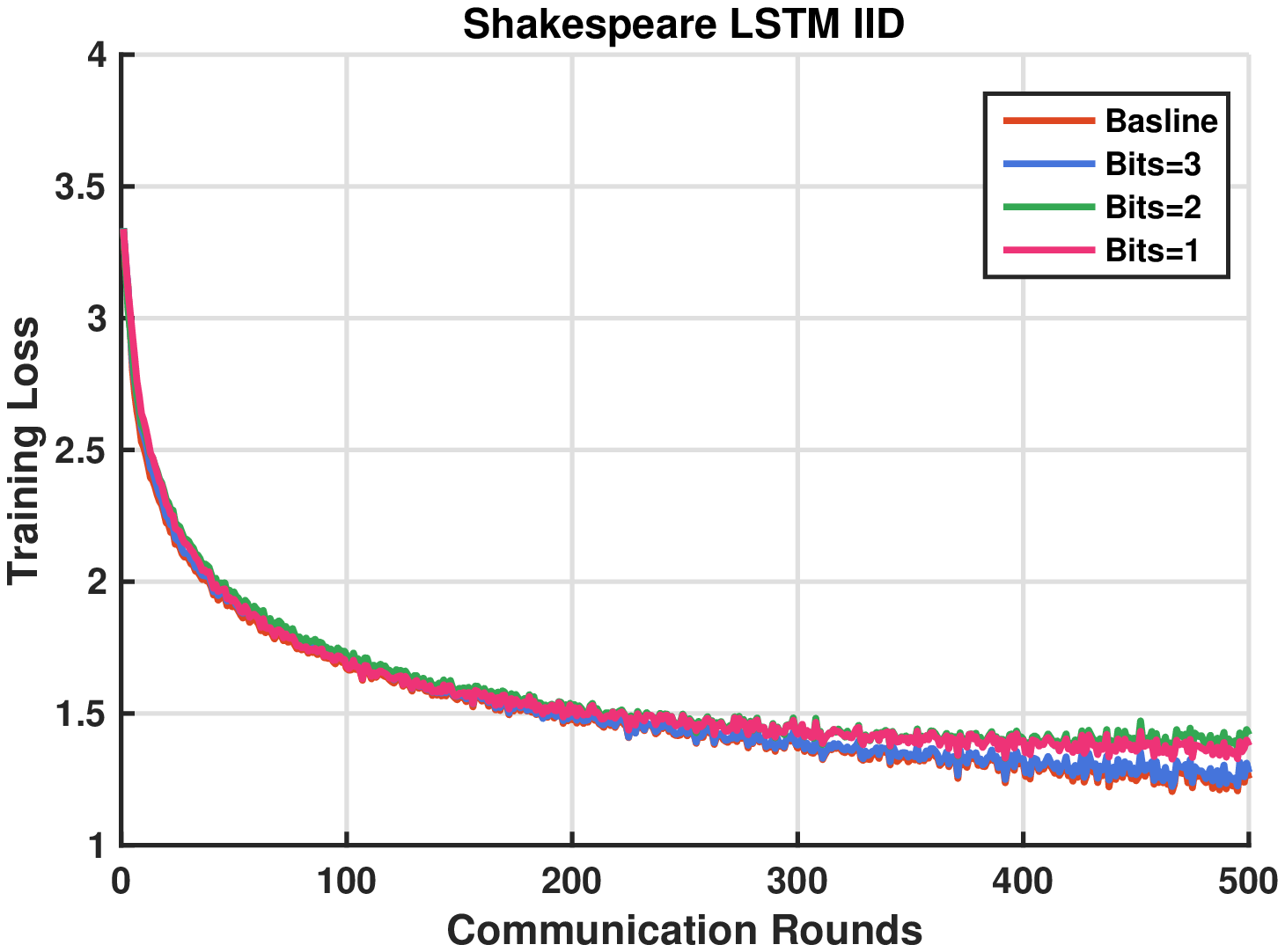}}
    \subfigure{
        \includegraphics[width=0.48\textwidth]{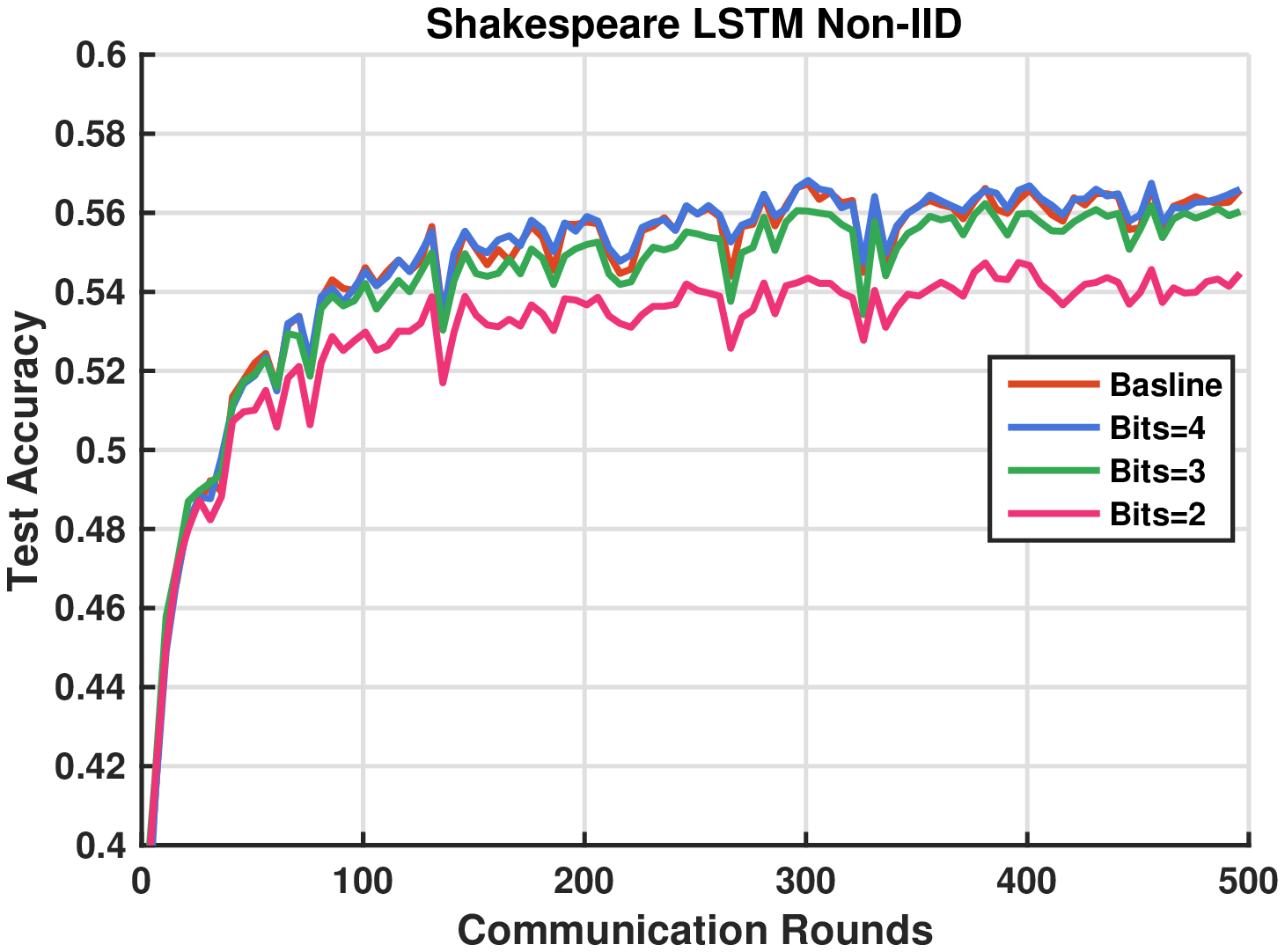}}
    \subfigure{
        \includegraphics[width=0.48\textwidth]{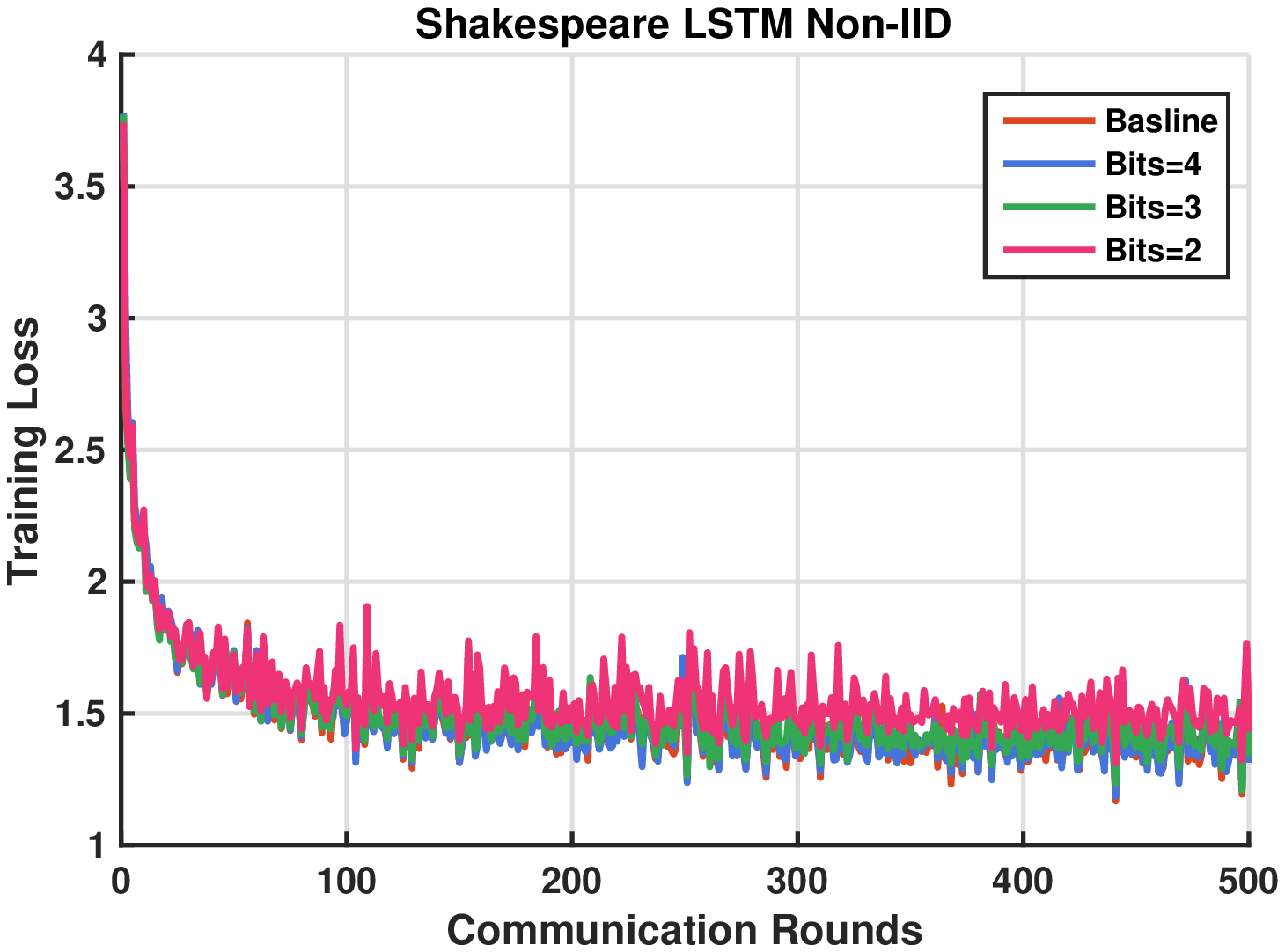}}
    \caption{The performance of uplink quantization on i.i.d. (top two subplots) and non-i.i.d. (bottom two subplots) Shakespeare dataset.}
    \label{fig:upload6}
\end{figure*}

\begin{figure}[htbp]
    \centering
    \subfigure{
        \includegraphics[width=0.48\columnwidth]{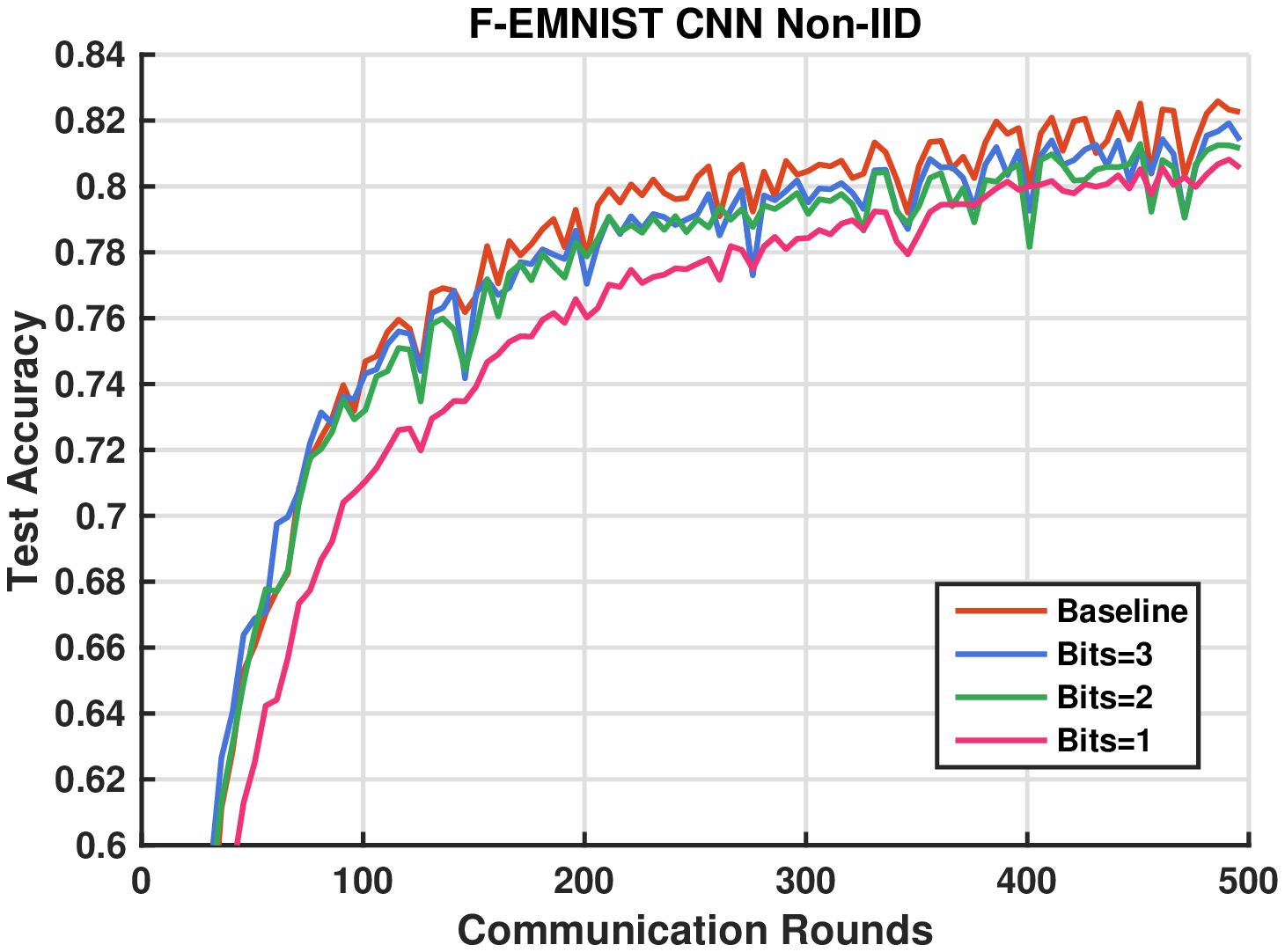}}
    \subfigure{
        \includegraphics[width=0.48\columnwidth]{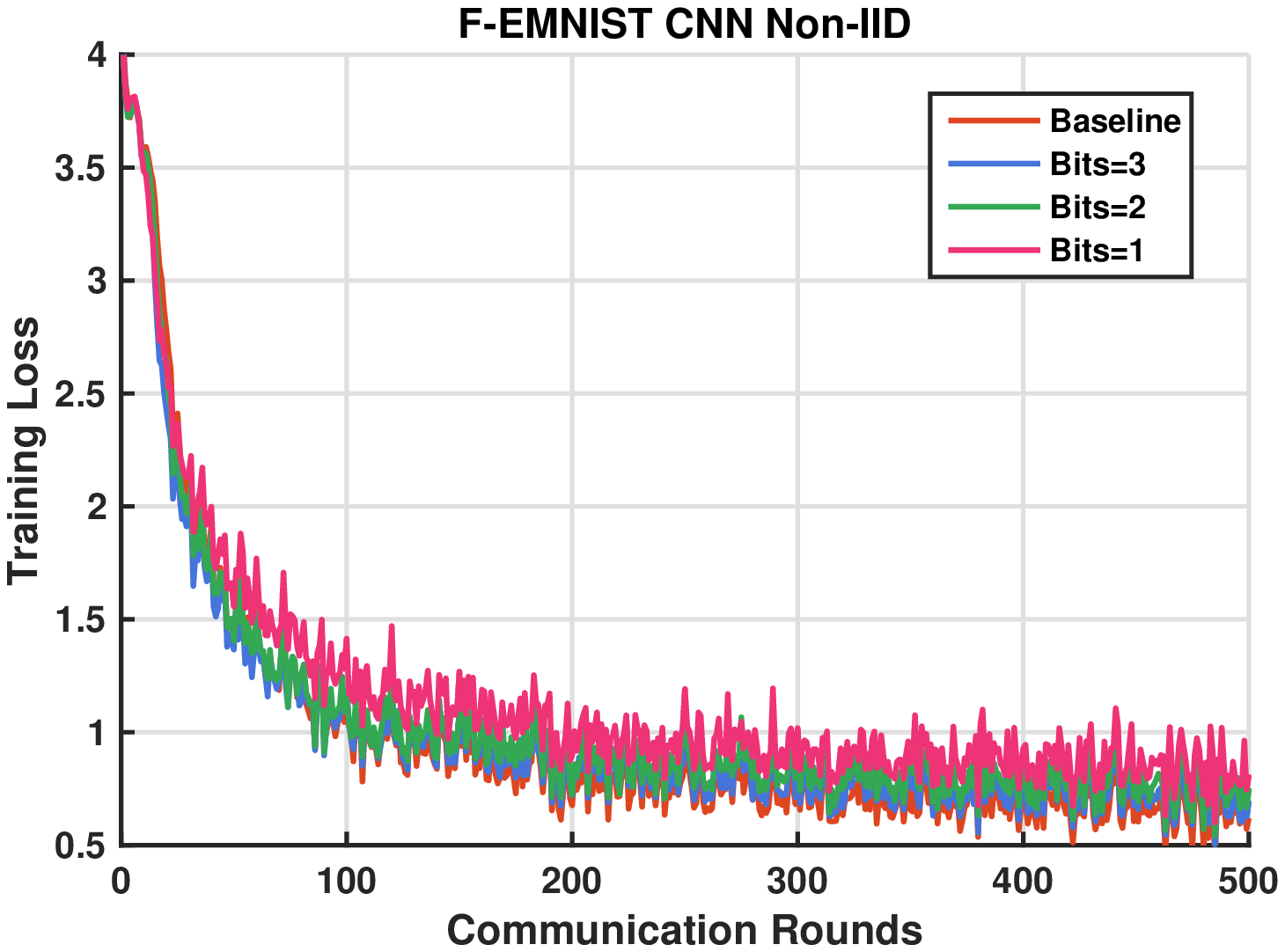}}
    \caption{The performance of uplink quantization on the naturally non-i.i.d. F-EMNIST dataset.}
    \label{fig:upload7}
\end{figure}

\mypara{Other datasets.} To further evaluate the performance of our uplink design, we also run experiments on \revsec{CIFAR-10, Shakespeare and F-EMNIST}, in which the training tasks are harder than classification on MNIST. \revsec{We report the results with the best quantization method (combining TQ, SR and DT) of these three datasets in Fig.~\ref{fig:upload5}, Fig.~\ref{fig:upload6} and Fig.~\ref{fig:upload7}, respectively. The results suggest that, with a well-designed uplink quantization, using 3 bits or fewer allows federated learning to achieve sufficiently good performance, for both i.i.d. and non-i.i.d. dataset.}

\subsection{Results for Downlink Communication}
\label{sec:res_dl}

\begin{figure}
    \centering
    \subfigure{
        \includegraphics[width=0.48\columnwidth]{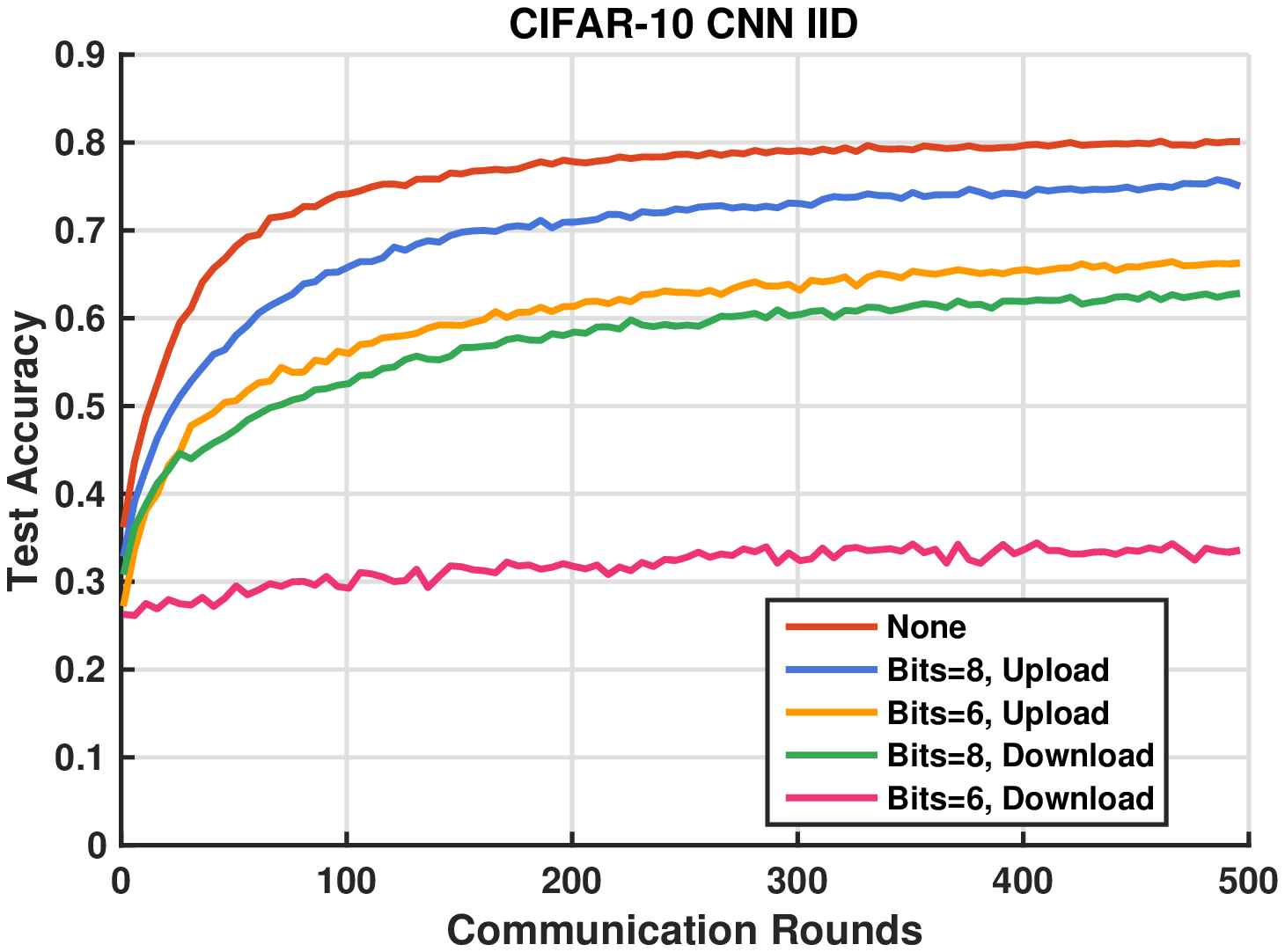}}
    \subfigure{
        \includegraphics[width=0.48\columnwidth]{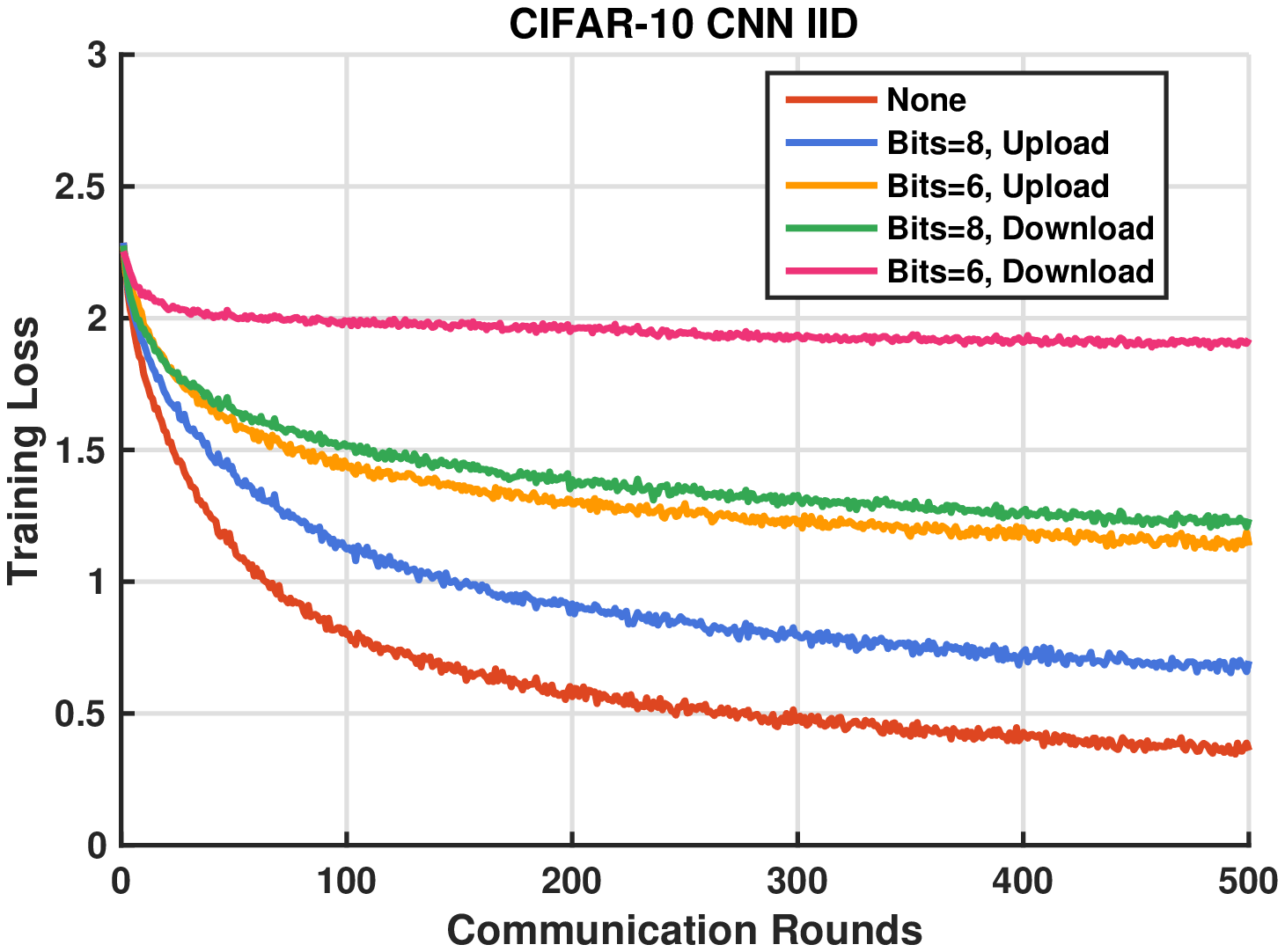}}
    \caption{Comparison of the impact of quantization (NQ and NR) on uplink and downlink communications.}
    \label{fig:download1}
\end{figure}

\mypara{Quantization has a bigger impact on downlink communication.} 
We evaluate the impact of low-precision quantization on downlink communication in this subsection. Our experimental results in Fig. \ref{fig:download1} suggest that a poorly designed quantization scheme (e.g., NQ with NR) for downlink can significantly degrade the performance of the overall FL -- for the same quantization level $B$ and the same quantization method, quantization in downlink has worse performance than quantization in uplink. This can be intuitively understood since the downloaded model is used by many clients, and hence the inaccuracy can manifest, resulting in a broader impact than upload inaccuracy \cite{lim2020federated}.

\begin{figure}
    \centering
    \subfigure{
        \includegraphics[width=0.48\columnwidth]{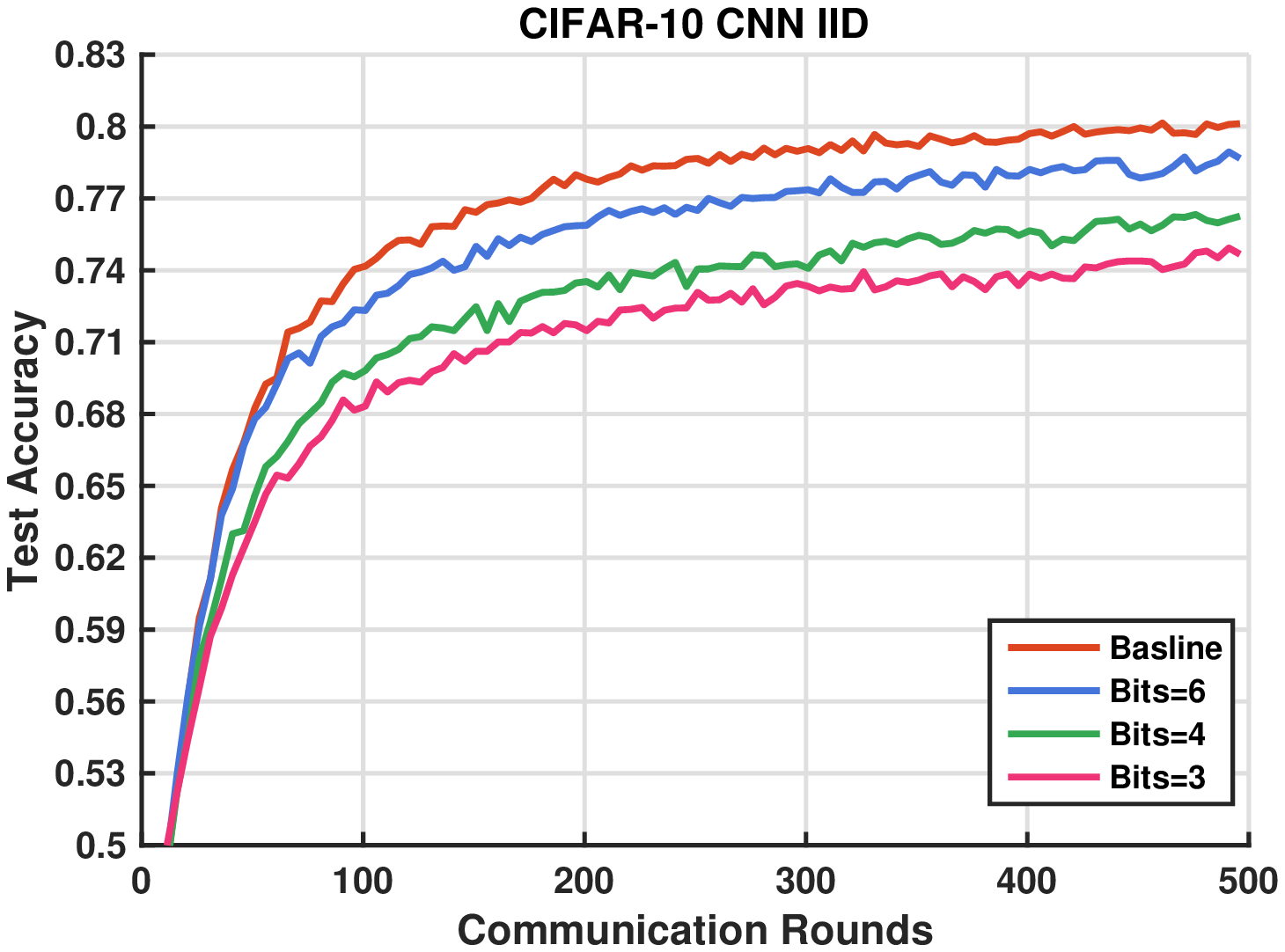}}
    \subfigure{
        \includegraphics[width=0.48\columnwidth]{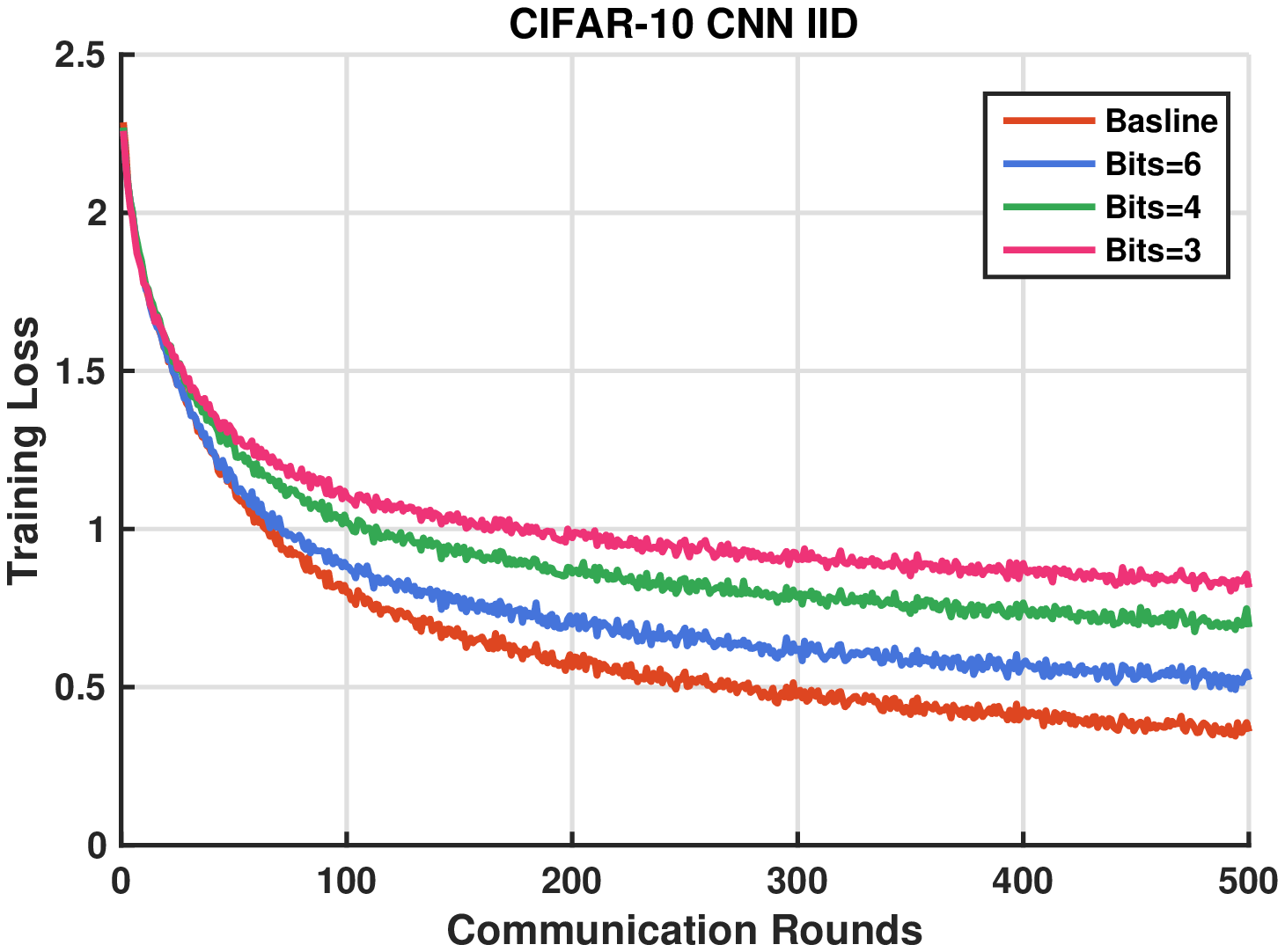}}
    \caption{Performance of quantization with TQ and SR in downlink communication.}
    \label{fig:download2}
\end{figure}

\mypara{Performance of quantization in downlink.}
We have mentioned in Section \ref{sec:comm_dl} that the quantization scheme designed for uplink communication can be adapted in downlink except DT. We now report the results of using quantization with TQ and SR on the CIFAR-10 dataset in Fig.~\ref{fig:download2}. The results show that with a well-design scheme, the downlink communication can also be made efficient and effective. Unlike the results in Fig.~\ref{fig:download1}, the accuracy of a 6-bit quantization can achieve 78.33\% accuracy (98\% of the baseline accuracy). However, without the support of DT, we see that there is a noticeable performance reduction when the bit-width falls below 3.

We now evaluate layered quantization (LQ) and see if it can improve the performance. We use the method described in Section \ref{sec:layered_dl} to carefully set an appropriate quantization gain for each layer. \revsec{The results reported in Fig.~\ref{fig:download3} suggest that LQ is effective for both i.i.d. and non-i.i.d. cases. For CIFAR-10 (F-EMNIST), it improves the performance of the 3-bit communication from 74.48\% to 77.15\% (75.23\% to 78.29\%) and the 4-bit communication from 76.04\% to 78.42\% (78.48\% to 80.46\%), respectively.}

\revsec{
\begin{figure*}[ht]
    \hsize=\textwidth
    \centering
    \subfigure{
        \includegraphics[width=0.48\textwidth]{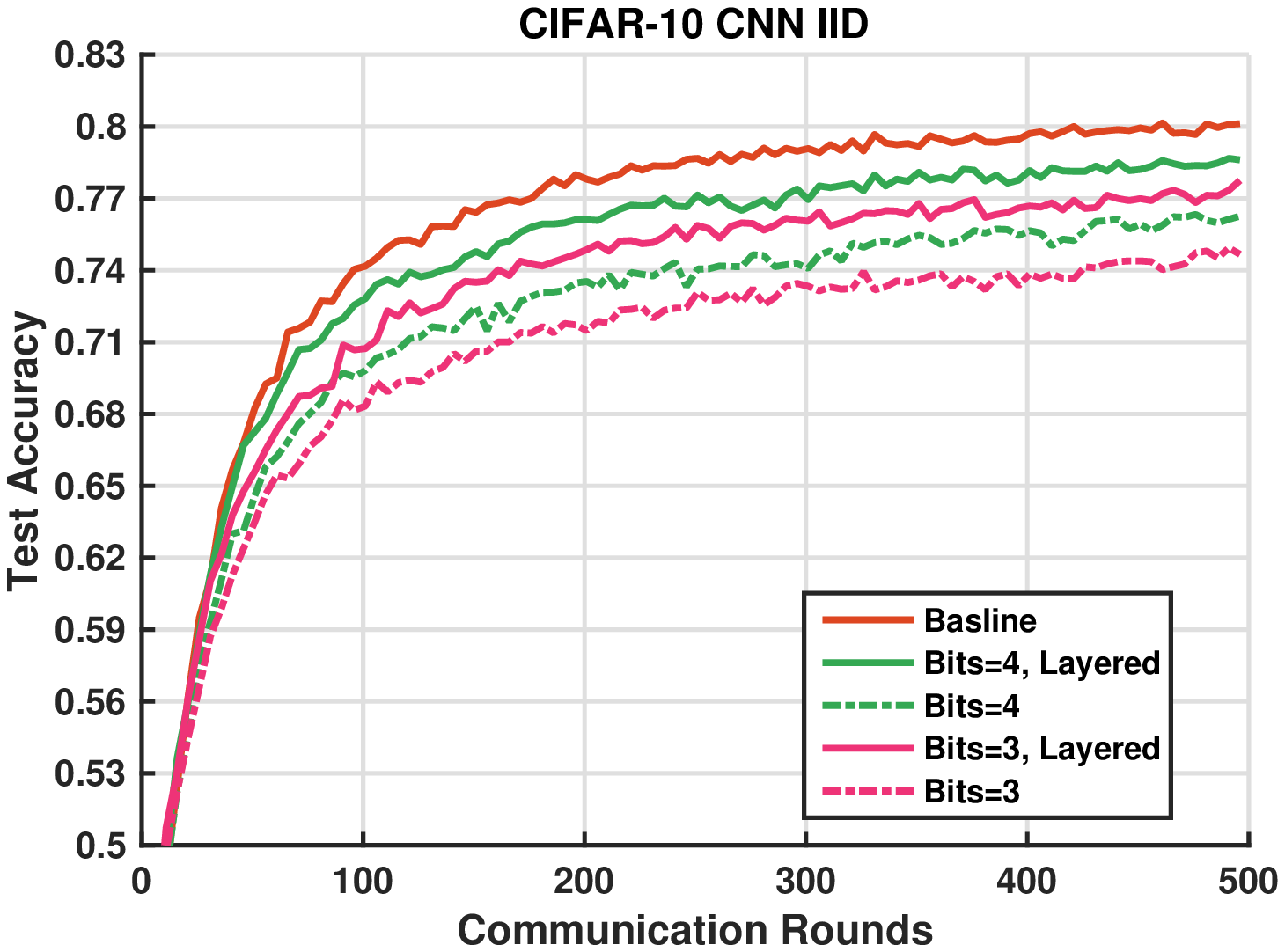}}
    \subfigure{
        \includegraphics[width=0.48\textwidth]{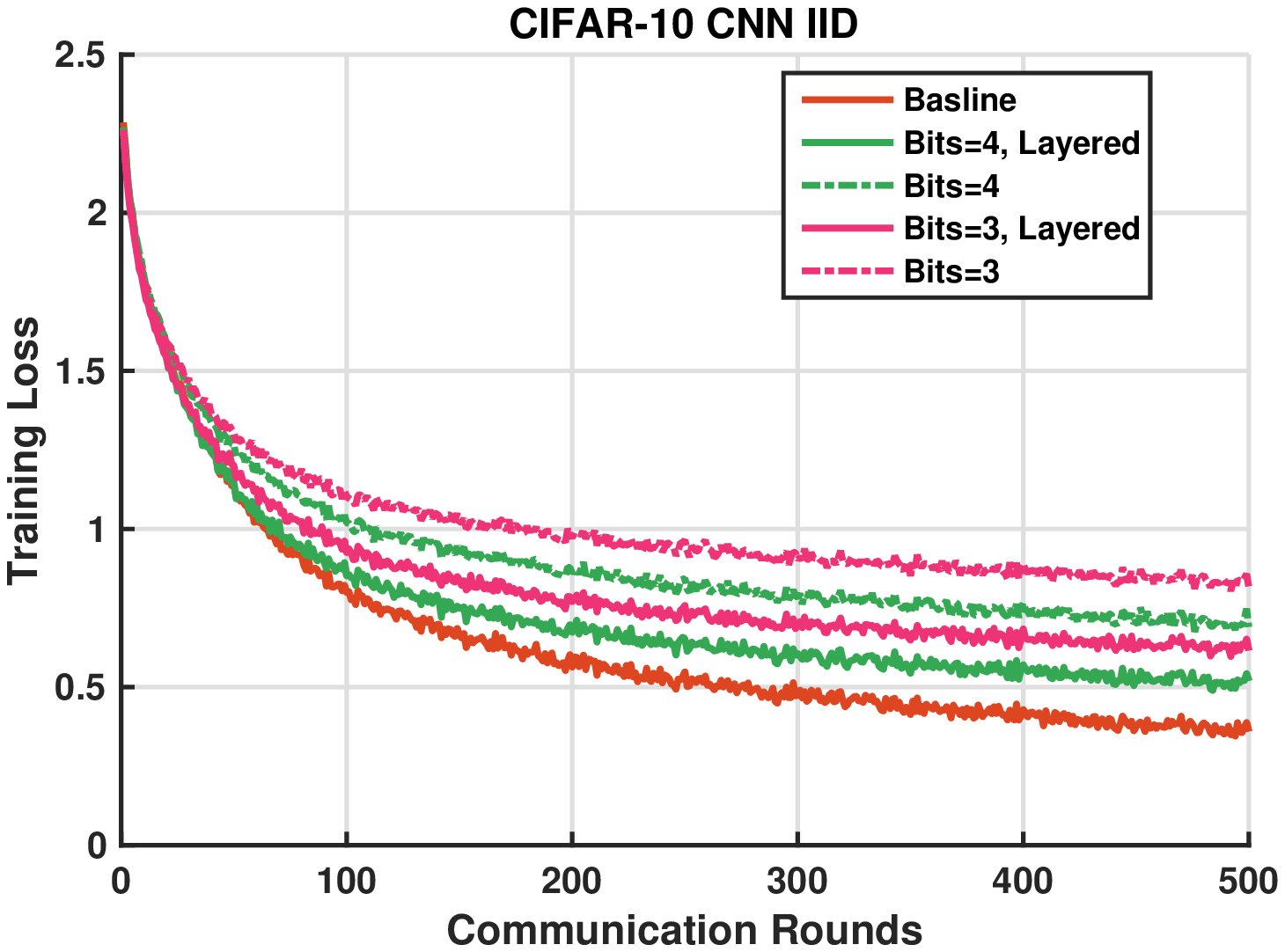}}
    \subfigure{
        \includegraphics[width=0.48\textwidth]{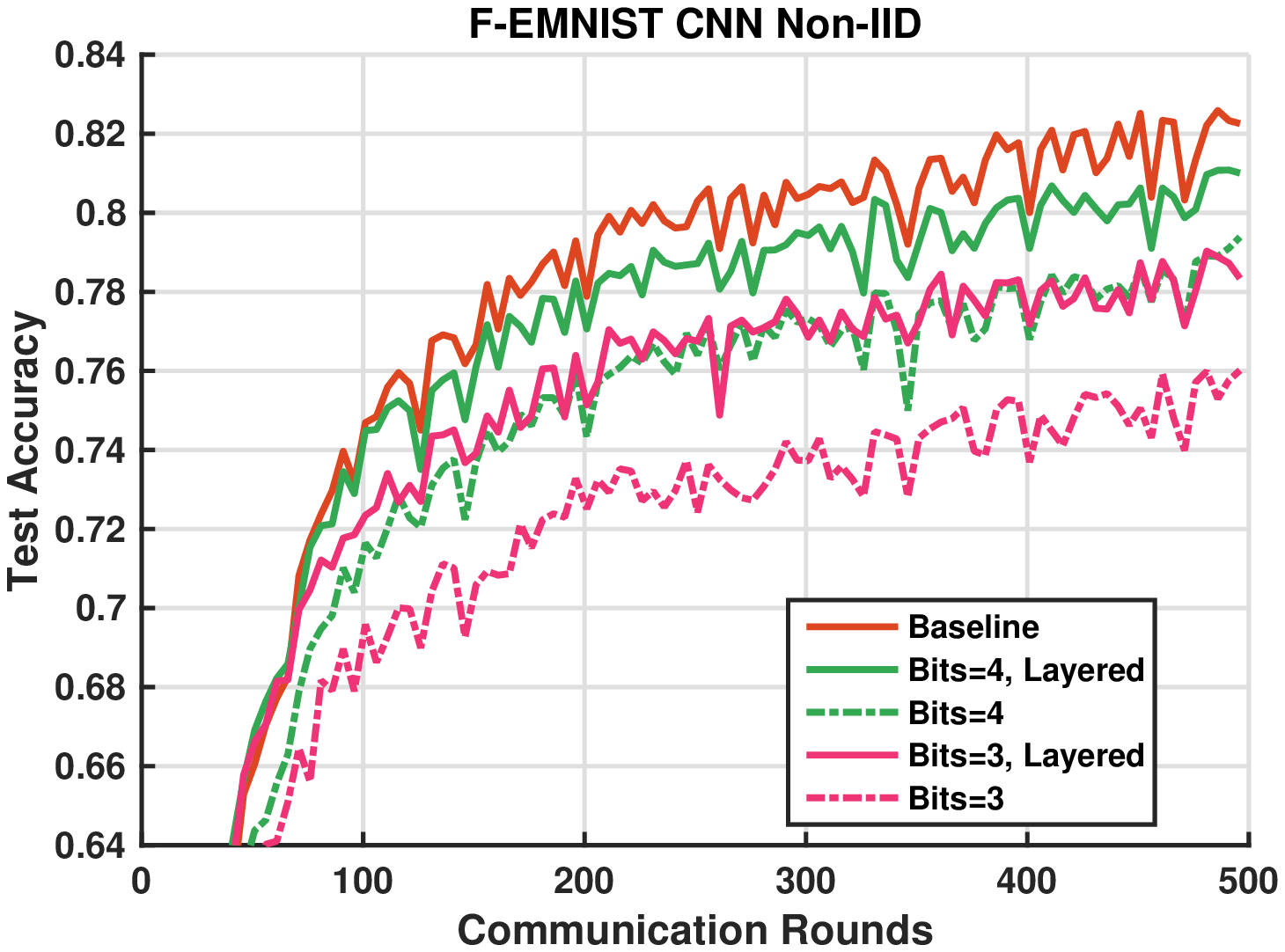}}
    \subfigure{
        \includegraphics[width=0.48\textwidth]{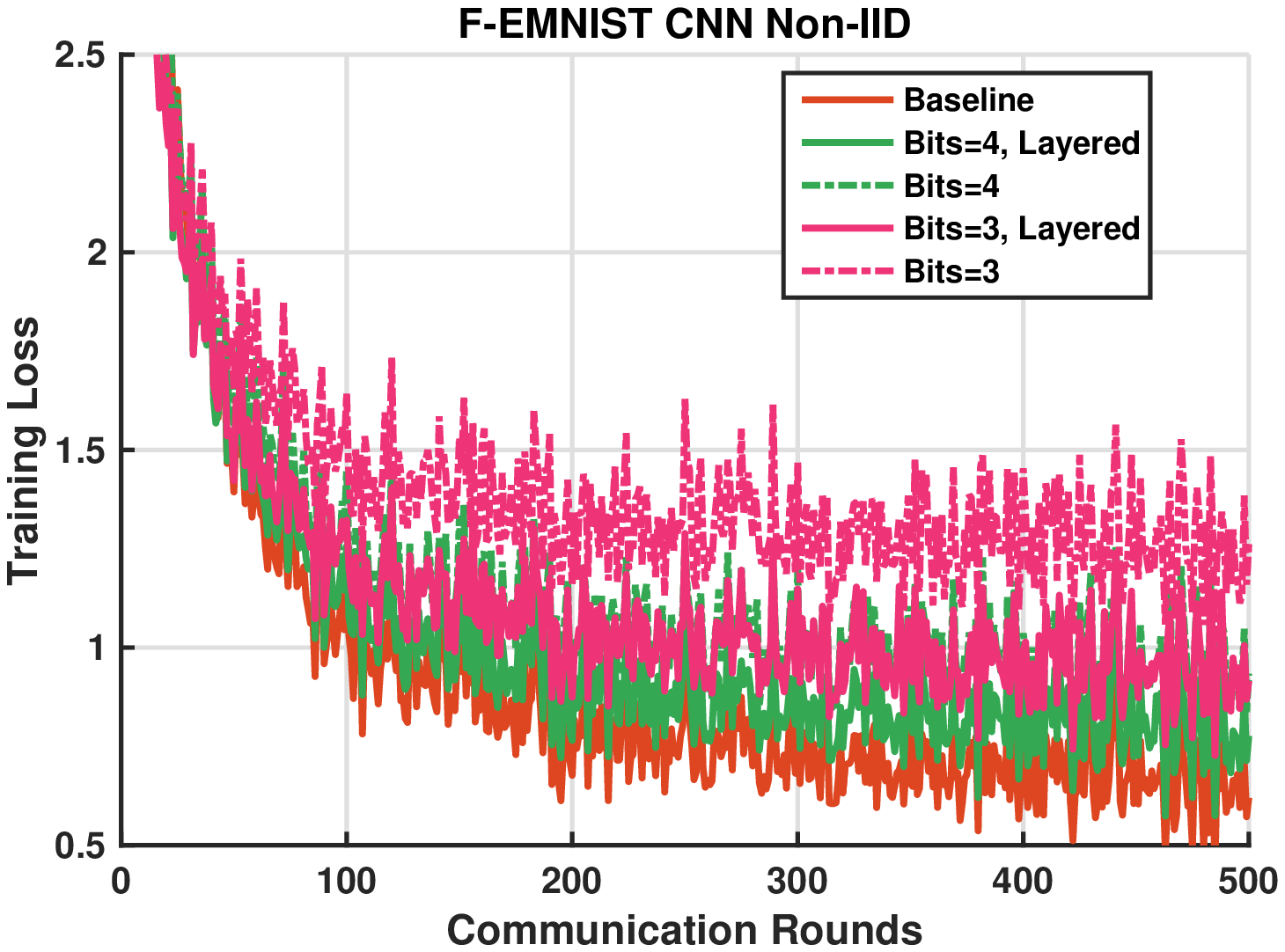}}
    \caption{Comparing the performance of quantization with and without layered quantization on CIFAR-10 (top two subplots) and F-EMNIST (bottom two subplots) datasets.}
    \label{fig:download3}
\end{figure*}
}

\begin{figure*}[ht]
    \hsize=\textwidth
    \centering
    \subfigure{
        \includegraphics[width=0.48\textwidth]{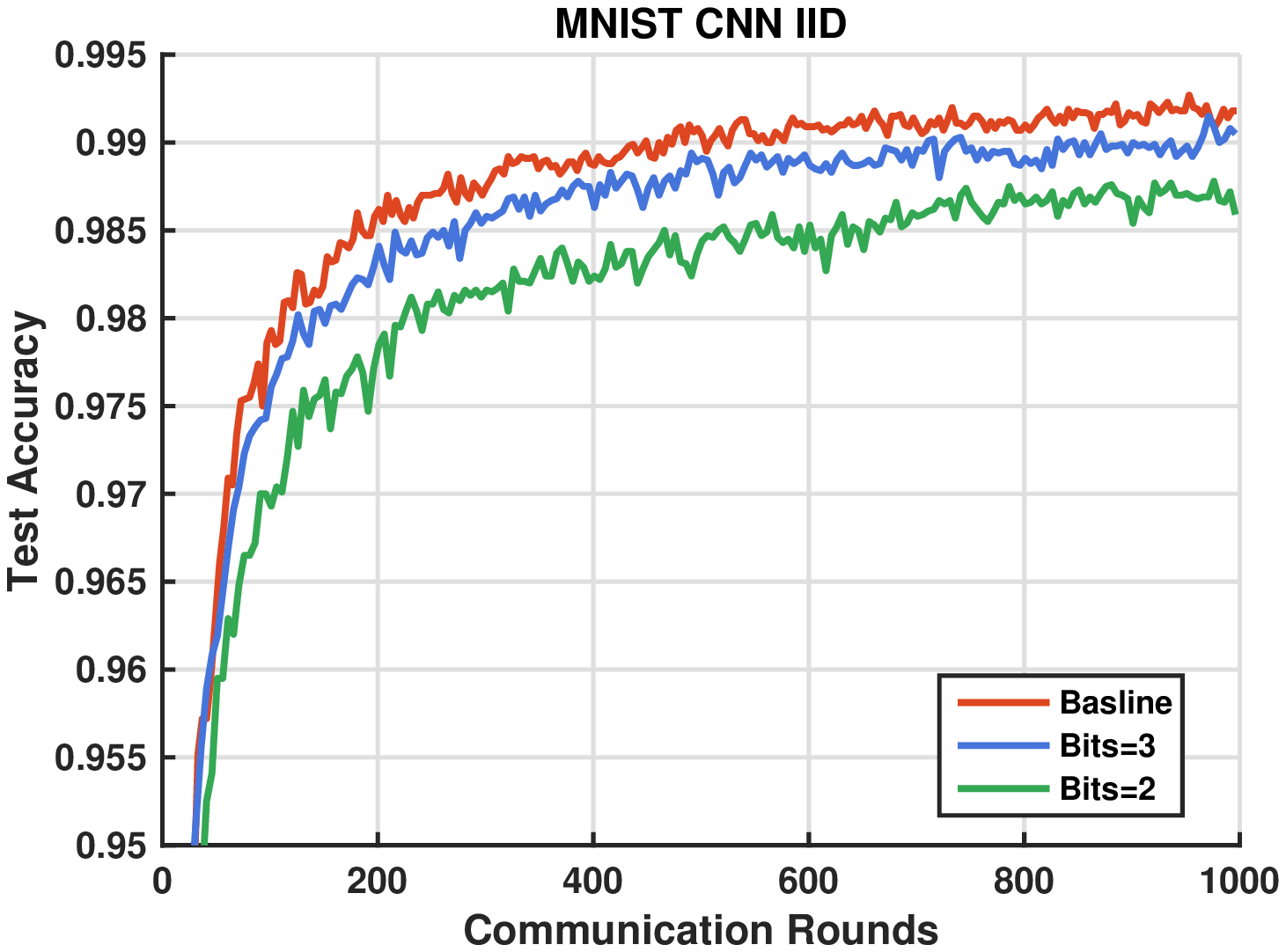}}
    \subfigure{
        \includegraphics[width=0.48\textwidth]{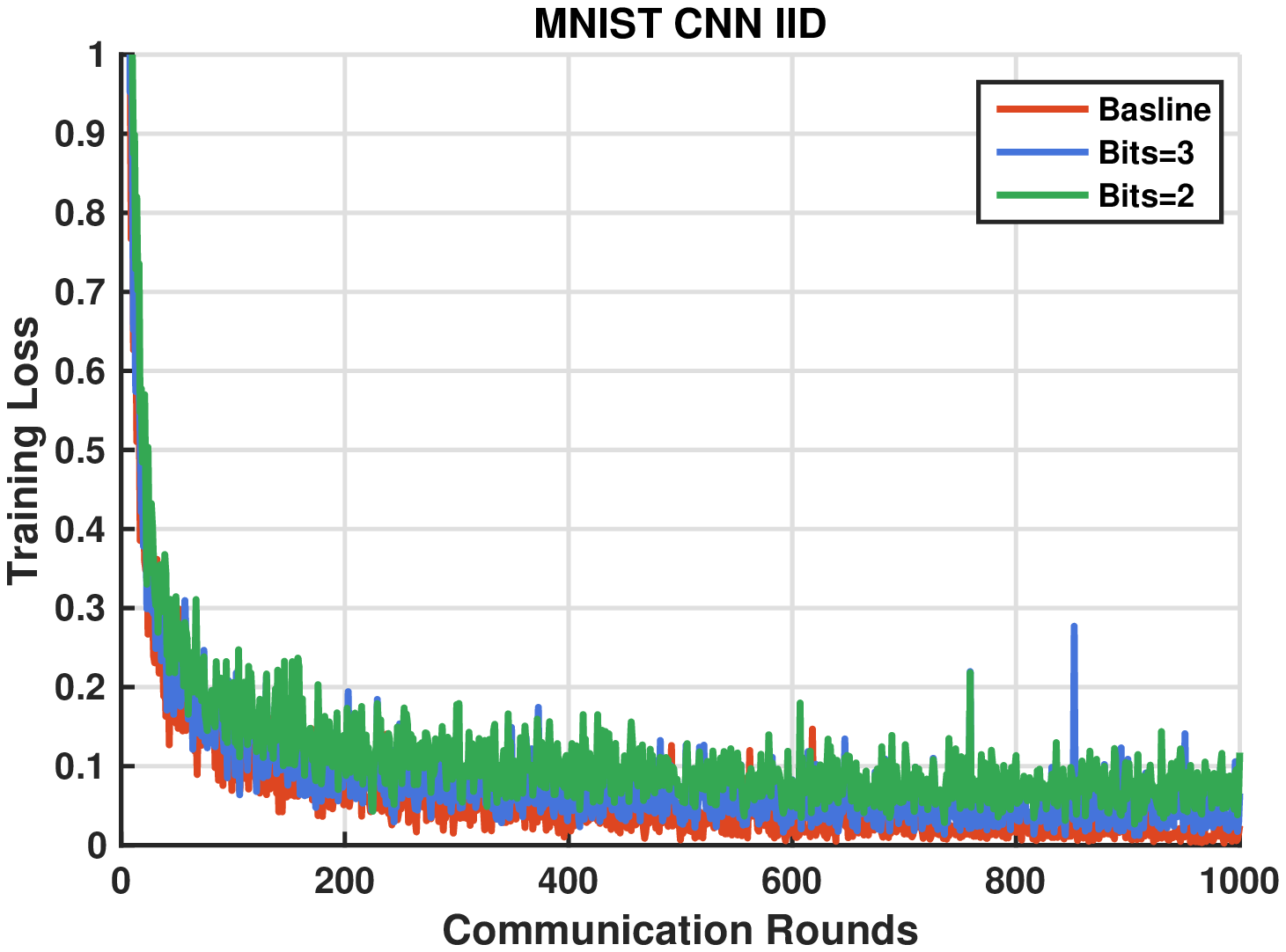}}
    \subfigure{
        \includegraphics[width=0.48\textwidth]{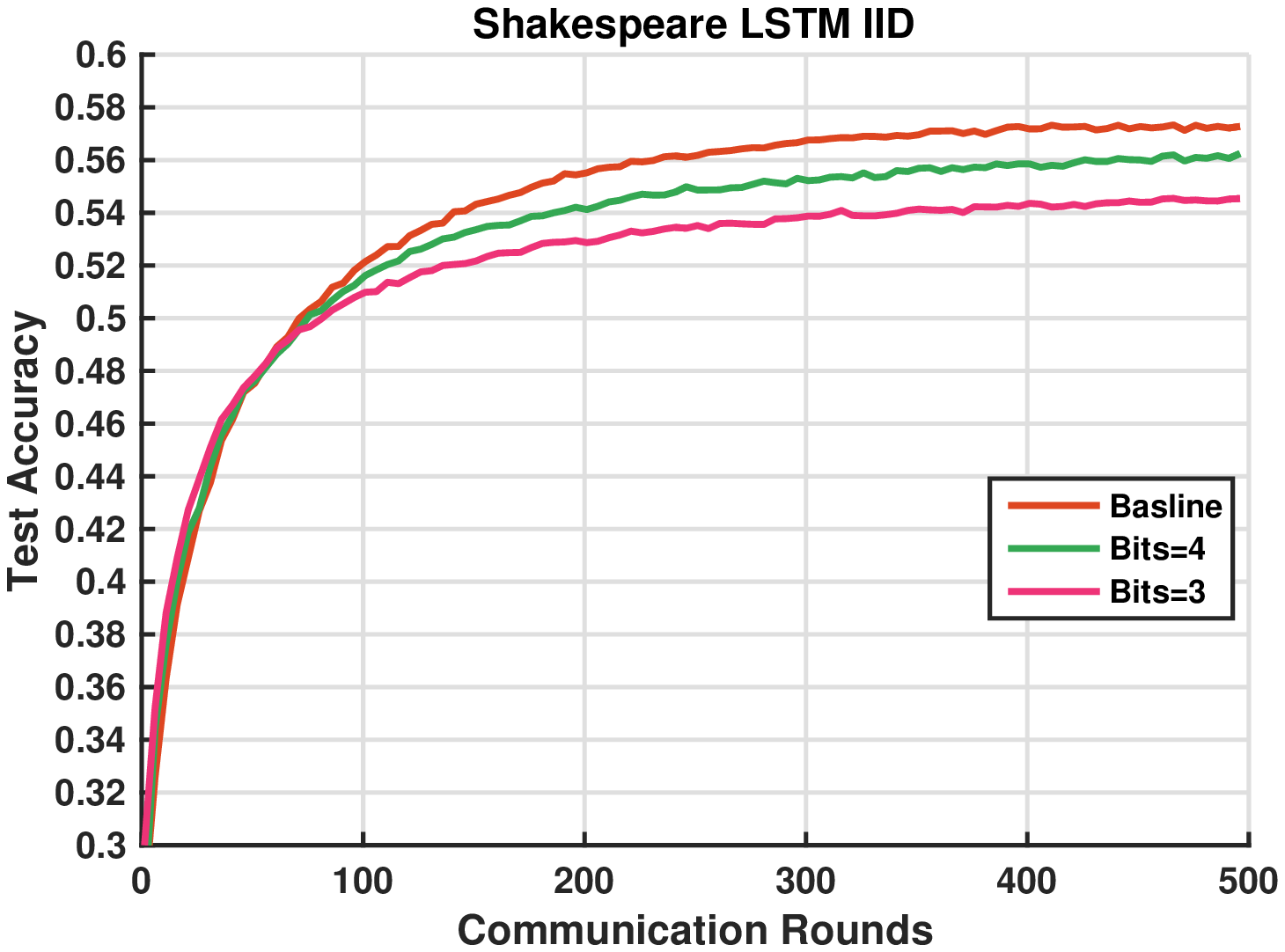}}
    \subfigure{
        \includegraphics[width=0.48\textwidth]{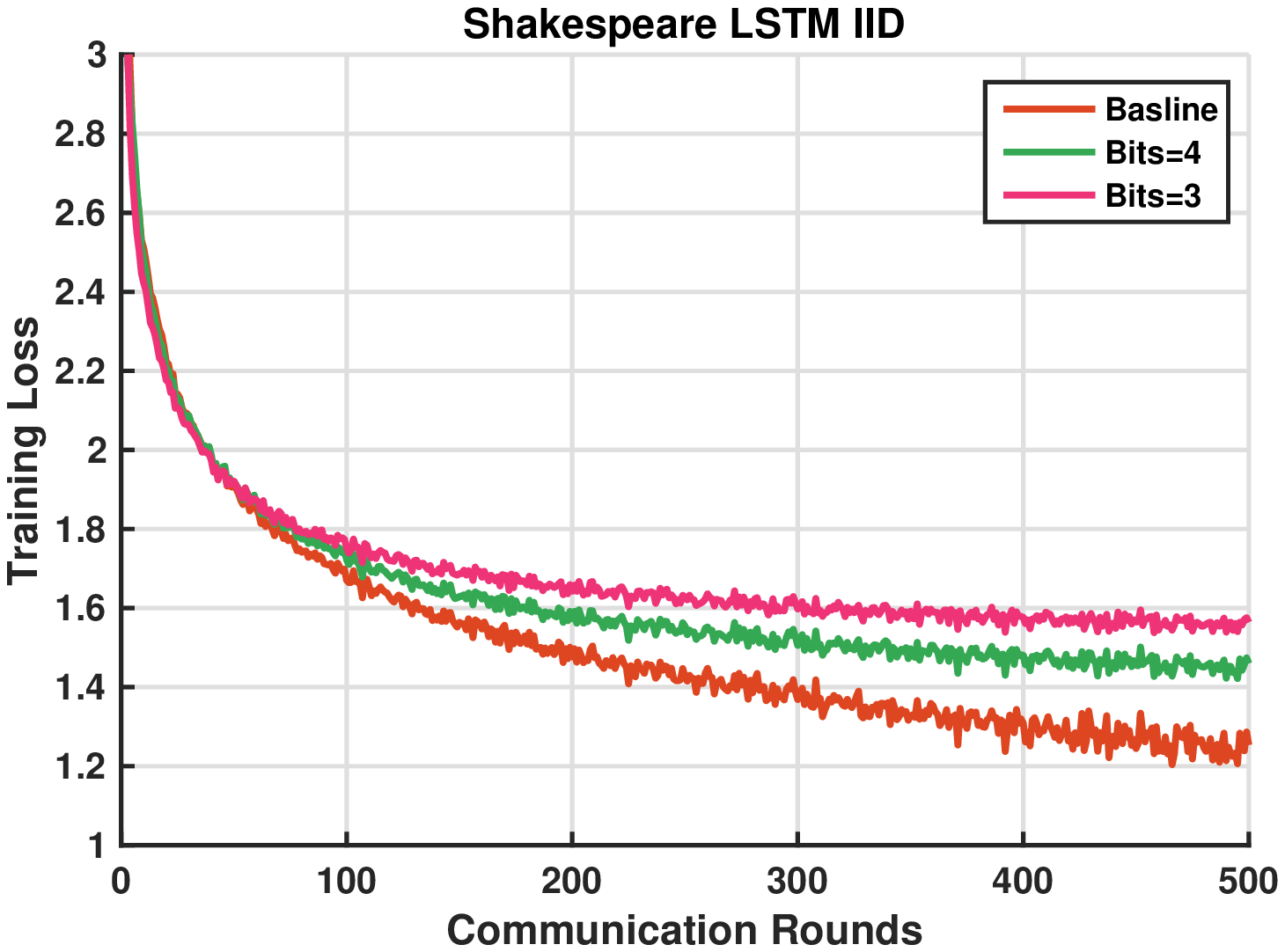}}
    \caption{The performance of the quantization scheme designed for downlink on MNIST (top two subplots) and Shakespeare (bottom two subplots) datasets.}
    \label{fig:download4}
\end{figure*}

\mypara{Results on other datasets.}
We now have identified the combination of TQ, SR and layered quantization as the best design options for downlink communication, and we now validate this combination on other datasets. The results from Fig.~\ref{fig:download4} further confirm that, at least for the three datasets we have evaluated, the proposed design can reduce the quantization bit-width to 4 (12.5\% of the baseline bandwidth) while achieving an accuracy degradation within 2\% of the baseline accuracy.

\subsection{Results of Quantization on Both Uplink and Downlink}
\label{sec:res_both}

Lastly, we carry out experiment with simultaneous quantization on both uplink and downlink communications.  The experimental results on different datasets are reported in Table \ref{table:effi}. We run 1000 rounds for MNIST and average the final 100 rounds as the final (convergence) accuracy (the fourth column). \revsec{As for CIFAR-10, Shakespeare and F-EMNIST}, we run 500 rounds and average the final 50 rounds. The last column shows the percentage of the baseline (using 32-bit float) can be achieved by the learning with quantized communications in both uplink and downlink. For all experiments, layered quantization with TQ and SR is used for downlink while DT with TQ and SR is used for uplink.

We evaluate how much communication payload can be reduced while maintaining a small accuracy loss (defined as less than 2\%). The results in Table \ref{table:effi} show that well designed quantization schemes are important to improve the communication efficiency. Take MNIST (i.i.d.) as an example, 2-bit for both downlink or uplink are sufficiently good, which can reduce (from the baseline) 93.75\% in communications for both uplink and downlink. Even for the more complex cases such as CIFAR-10 (non-i.i.d.), 6-bit for downlink and 4-bit for uplink have very good performance, reducing 81.25\% and 87.5\% of the communication bandwidth for each client on downlink and uplink respectively. Overall, we conclude that the proposed designs are effective in addressing the communication bottleneck of federated learning. 

\begin{table}[!t]
\caption{Performance of simultaneous quantization on both uplink and downlink.}
\label{table:effi}
\centering
\begin{tabular}{ccccc}
\hline
Dataset     & Downlink & Uplink & Accuracy (Baseline) & Percentage*        \\ \hline
\multicolumn{5}{c}{i.i.d.}                                                    \\ \hline
MNIST       & 2-bit    & 2-bit  & 98.46\% (99.11\%)          & 99.34\%     \\
CIFAR-10    & 5-bit    & 2-bit  & 78.43\% (79.93\%)          & 98.12\%     \\
Shakespeare & 5-bit    & 2-bit  & 56.17\% (57.25\%)          & 98.11\%     \\ \hline
\multicolumn{5}{c}{Non-i.i.d.}                                                \\ \hline
MNIST       & 2-bit    & 2-bit  & 97.41\% (99.10\%)          & 98.29\%     \\
CIFAR-10    & 6-bit    & 4-bit  & 61.67\% (62.61\%)          & 98.50\%     \\
Shakespeare & 5-bit    & 3-bit  & 55.16\% (56.16\%)          & 98.22\%     \\
\revsec{F-EMNIST}      & \revsec{5-bit}    & \revsec{3-bit}  & \revsec{80.24\% (81.82\%)}  & \revsec{98.07\%}     \\ \hline
\end{tabular}
\begin{center}
\small{*: The last column represents the percentage of FL accuracy against the baseline accuracy.}
\end{center}
\end{table}

\rev{
\subsection{Impact of hyperparameters}
\label{sec:res_param}

\begin{figure*}[h]
    \centering
    \subfigure[\scriptsize{CIFAR-10: varying local batch size}]{
        \includegraphics[width=0.47\textwidth]{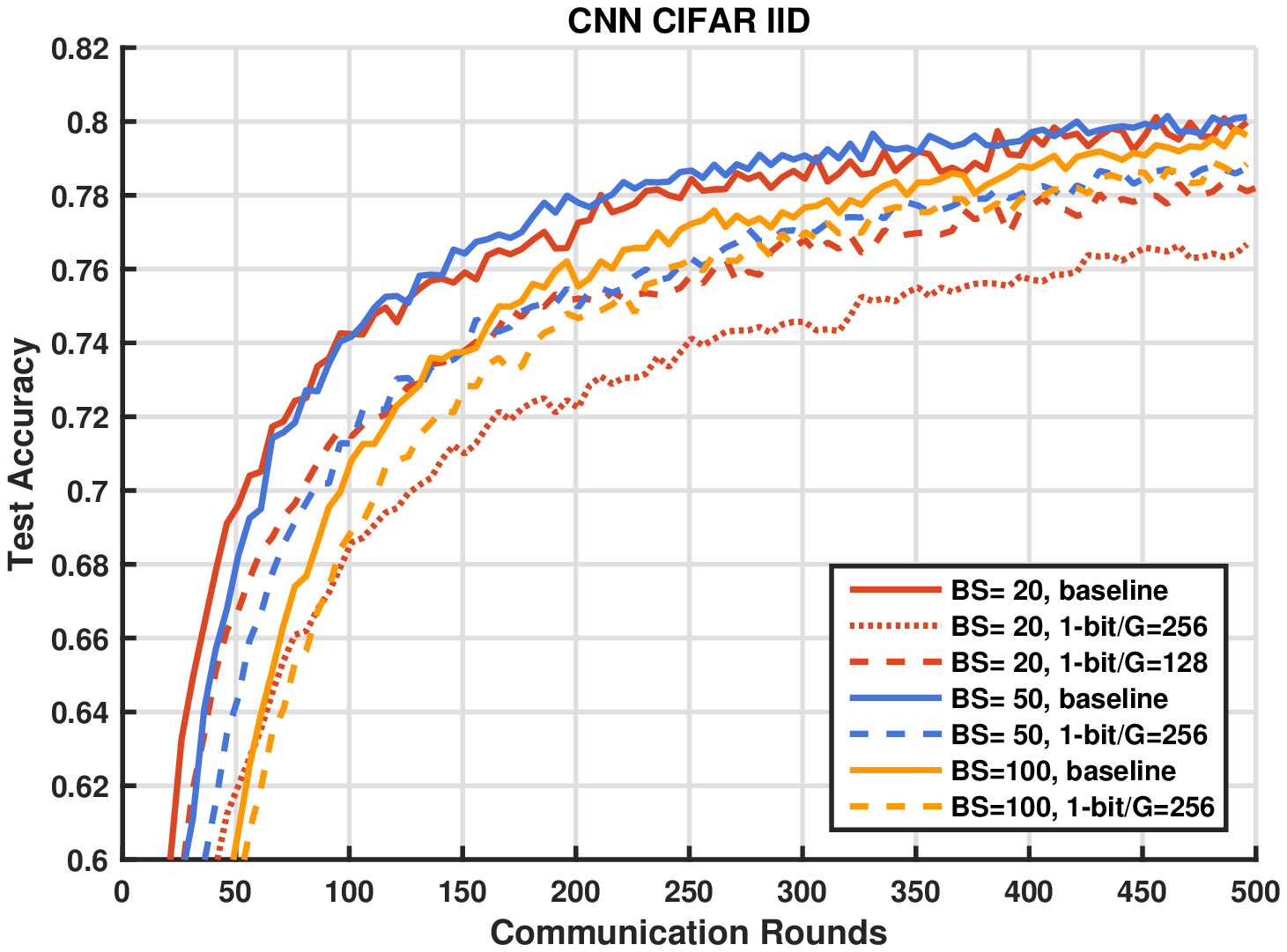}\label{sfig:cifar_E}}
    \subfigure[\scriptsize{CIFAR-10: varying local epoch}]{
        \includegraphics[width=0.47\textwidth]{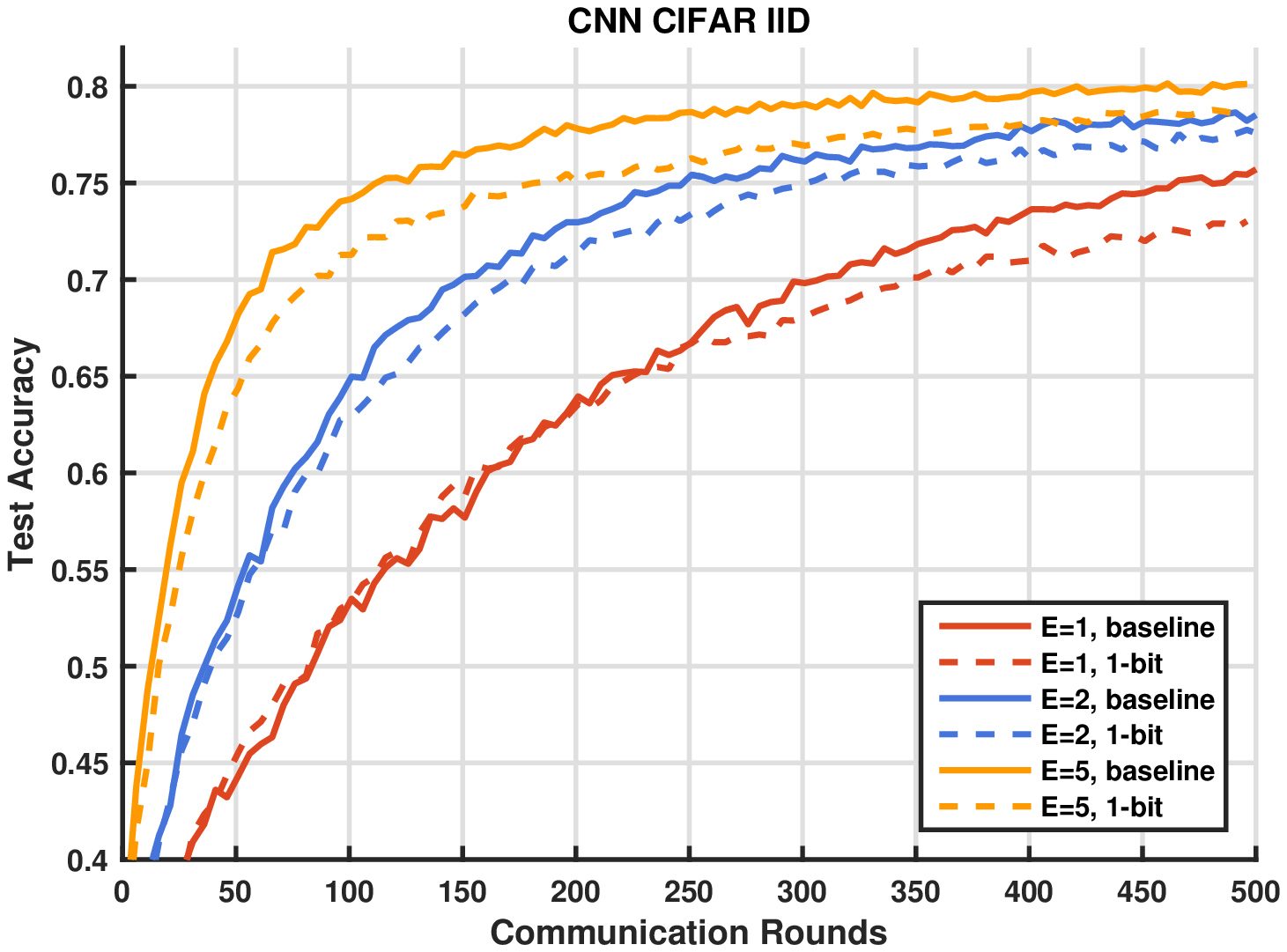}\label{sfig:cifar_BS}}
    \subfigure[\scriptsize{CIFAR-10: varying degrees of non-i.i.d.}]{
        \includegraphics[width=0.47\textwidth]{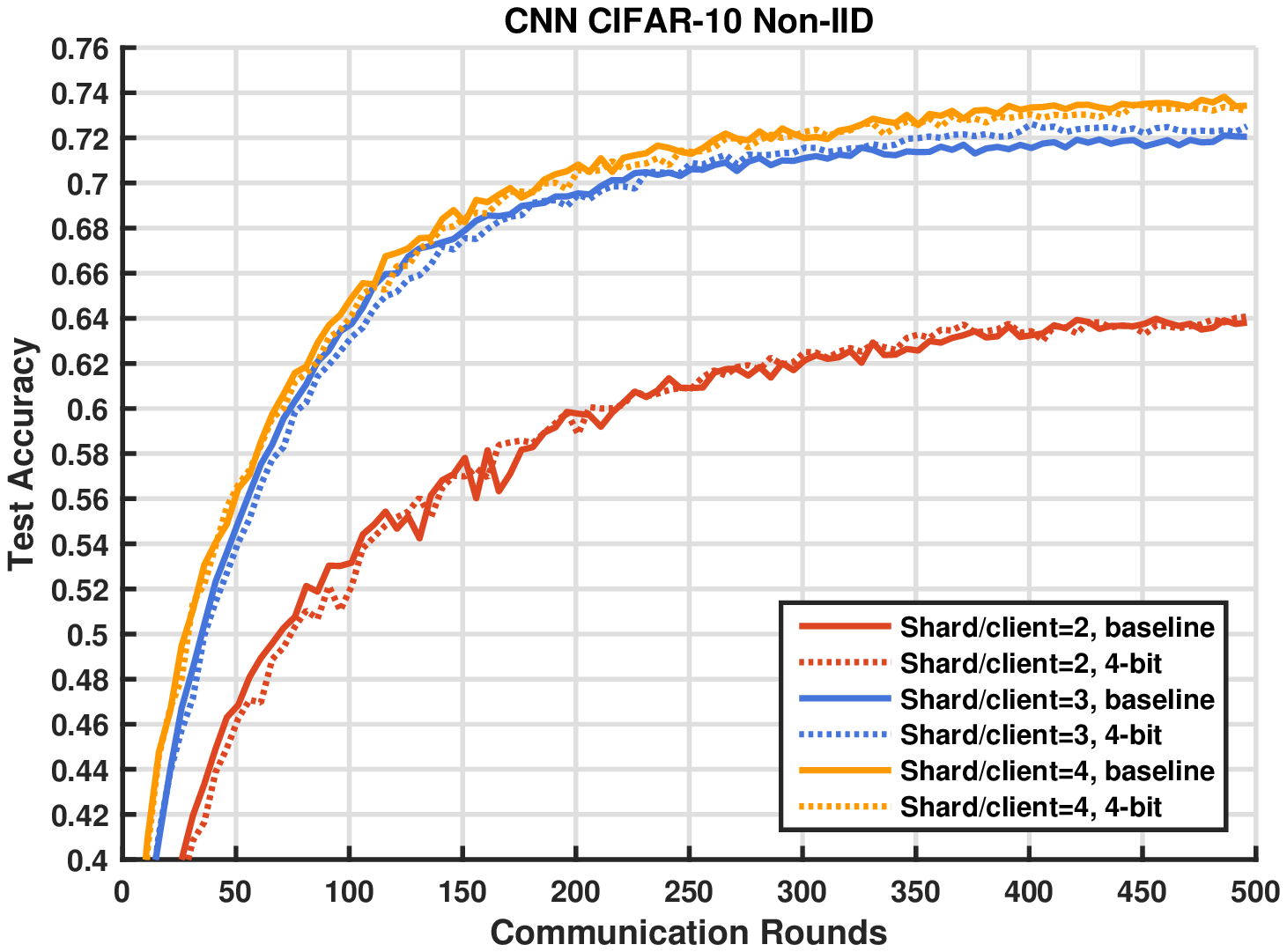}\label{sfig:cifar_noniidness}}
    \subfigure[\scriptsize{MNIST: varying client scale}]{
        \includegraphics[width=0.47\textwidth]{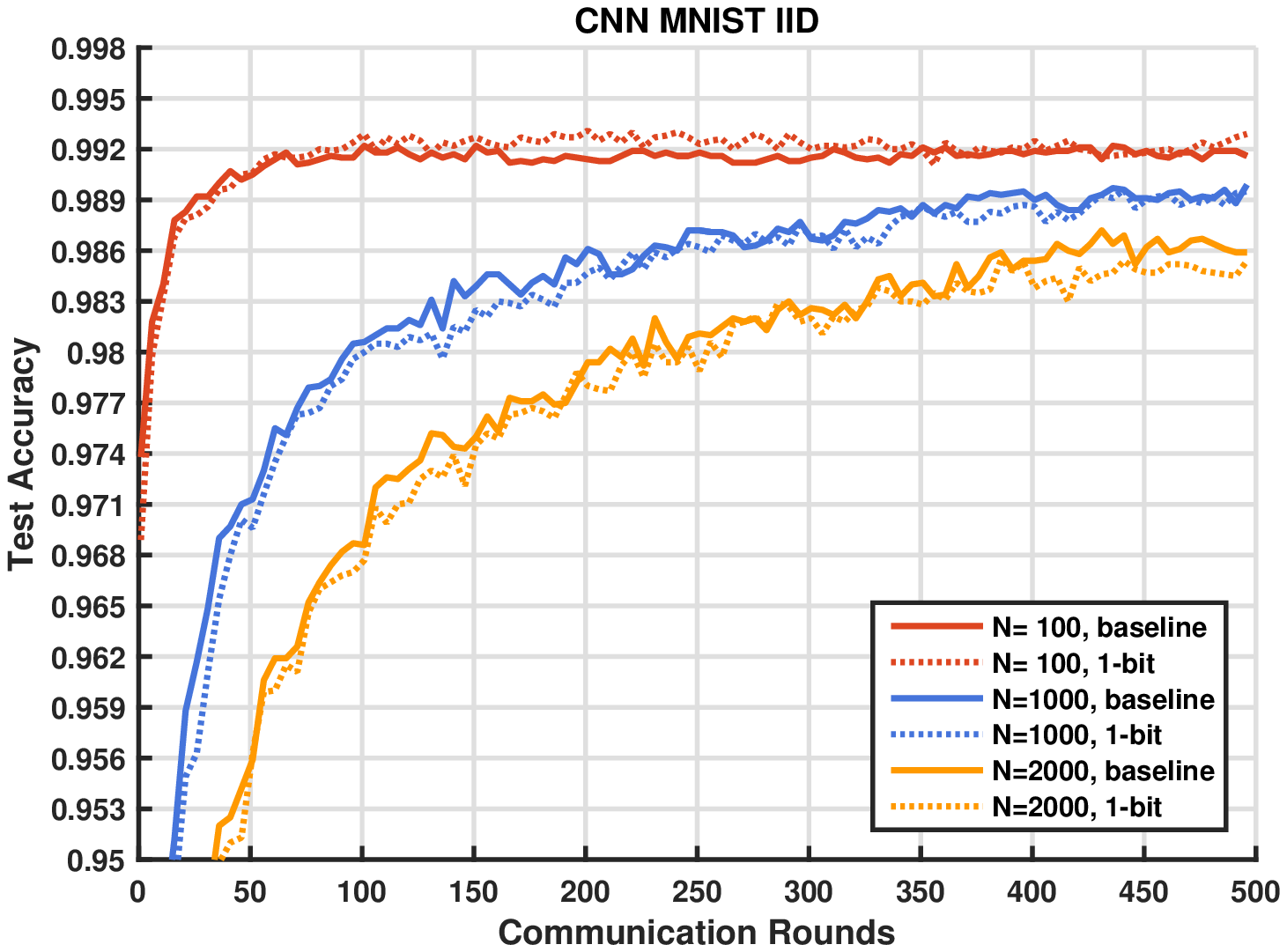}\label{sfig:mnist_client}}
    \caption{Impact of different hyperparameters with quantized uplink transmission in FL. Differential transmission is used for all the experiments. The baseline curves mean the results without any quantization.}
    \label{fig:hyper}
\end{figure*}

There are several hyperparameters that impact the training of FL, some of which have been discussed in \cite{mcmahan2017fl}. Fig.~\ref{fig:hyper} shows the relationship between the quantization and these hyperparameters. Local batch size ($BS$) is suggested to be small in most cases, and we can see in Fig.~\ref{sfig:cifar_E} that the baseline performance of $BS=20$ and 50 are better. However, we notice that the accuracy loss between the baseline and 1-bit quantization is the smallest for $BS=100$. For the setting of $BS=20$, smaller quantization gain $G$ should be used; otherwise the performance is severely degraded. The possible reason is that a smaller $BS$ brings more local iterations on the clients at each round and then increases the dynamic range of weight differentials, which indicates that there is a tradeoff between increasing computation per client and better quantization performance. As for the local epochs, although a larger $E$ might also increase the dynamic range, this becomes less important, compared to the benefit in convergence speed, as shown in Fig.~\ref{sfig:cifar_BS}. Therefore, even with quantization, a relative large $E$ can be adopted, especially for the i.i.d. cases. Fig.~\ref{sfig:cifar_noniidness} and Fig.~\ref{sfig:mnist_client} imply that the proposed quantization scheme is not very sensitive to the degree of non-i.i.d. or client scale, since the theoretical results have shown that the impact of quantization is decoupled with $\Gamma$ or $N$. 
}

\section{Conclusions}
\label{sec:conclusions}

We have studied the design and analysis of physical layer quantization and transmission methods for wireless federated learning. If nothing else, this paper showed that the communication design must tailor to the characteristics of FL. In particular, the choice of \textit{what} to transmit and \textit{how} to transmit them has a profound impact on the performance of federated learning in a wireless system, and we established this conclusion both theoretically, via convergence analysis of various quantization and transmission options in the well-known \textsc{FedAvg}, and experimentally, via comprehensive evaluation on real-world datasets. An important theoretical convergence result was established, which states that in order to achieve an $\mathcal{O}({1}/{T})$ convergence rate with quantization, transmitting the weight requires increasing the quantization level at a \textit{logarithmic} rate, while transmitting the weight differential can keep a constant quantization level. As a crown jewel of the experimental study, we were able to achieve a significant milestone: 1-bit quantization ($3.1\%$ of the floating-point baseline bandwidth) achieves $99.8\%$ of the floating-point baseline accuracy at almost the same convergence rate on MNIST, representing the best known bandwidth-accuracy tradeoff to the best of the authors' knowledge. 

\rev{In addition to enabling efficient communication design for FL, we have noticed that quantization can also be combined with communication resource allocation. For example, the theoretical result of Theorem~\ref{thm:1} naturally leads to a resource (bit) allocation problem where one is given a total budget of uplink bandwidth and asked to allocate the bits over communication rounds to optimize the learning performance. Another interesting future research direction is the combination of quantization and client selection. For example, for a given total uplink bandwidth budget, how to balance the increased number of clients and reduced quantization precision.}


\appendices

\section{Proof of Theorem~\ref{thm:1}}
\label{app:proof_thm1}

\subsection{Notations}
\label{sec:notation}

In our analysis, there are three sources of randomness: stochastic gradients, random sampling of clients,  and  stochastic rounding. To distinguish them, we respectively use the notation $\expt_{SG}[\cdot]$, $\expt_{\mathcal{S}_t}[\cdot]$ and $\expt_{SR}[\cdot]$  and use $\expt[\cdot]$ for expectation over all three of them. With a slight abuse of notation, we change the timeline to be with respect to the SGD iteration time instead of the communication round. Let $\vect{w}_t^k$ be the model weights on the $k$th client at the $t$th iteration and $\vect{w}_{t}$ be the global model at the $t$th iteration. In \textsc{FedAvg}, clients perform $E$ local iterations before global aggregation. Hence,  $\vect{w}_{t}$ is only accessible for specific $t \in \mathcal{I}_E$, where $\mathcal{I}_E=\{nE ~|~n=1,2,\dots\}$, i.e. the time for communication. 

For client $k$, it trains the model locally with
\begin{equation}
\label{eqn:local_update}
    \vect{v}_{t+1}^k = \vect{w}_{t}^k - \eta_t \nabla F_k (\vect{w}_{t}^k, \xi_t^k).
\end{equation}
If $t+1 \notin \mathcal{I}_E$, the next-step result is $\vect{w}_{t+1}^k = \vect{v}_{t+1}^k$ since no global aggregation takes place. If $t+1 \in \mathcal{I}_E$, all client $k \in \mathcal{S}_{t+1}$ upload their quantized weights $Q(\vect{v}_{t+1}^k)$. The global model is updated with
$    \vect{w}_{t+1} = \frac{1}{K} \sum_{k \in S_{t+1}} Q(\vect{v}_{t+1}^k)$.
Since we do not model downlink quantization, selected clients update their local weights as $\vect{w}_{t+1}^k = \vect{w}_{t+1}$ and start the next local training period. We define the following three variables to summarize the aforementioned steps:
\begin{align*}
    \vect{v}_{t+1}^k & \triangleq \vect{w}_t^k - \eta_t \nabla F_k(\vect{w}_t^k, \xi_t^k); \\
    \vect{u}_{t+1}^k & \triangleq \begin{cases}
        \vect{v}_{t+1}^k & \text{if~} t+1 \notin \mathcal{I}_E, \\
        \frac{1}{K} \sum_{i \in S_{t+1}} \vect{v}_{t+1}^i & \text{if~} t+1 \in \mathcal{I}_E; 
    \end{cases} \\
    \vect{w}_{t+1}^k & \triangleq \begin{cases}
        \vect{v}_{t+1}^k & \text{if~} t+1 \notin \mathcal{I}_E, \\
        \frac{1}{K} \sum_{i \in S_{t+1}} Q(\vect{v}_{t+1}^i) & \text{if~} t+1 \in \mathcal{I}_E. 
    \end{cases}
\end{align*}
We define three \textbf{virtual sequences} $\avgvect{v}_t=\frac{1}{N}\sum_{k=1}^N \vect{v}_t^k$, $\avgvect{w}_{t} = \frac{1}{N}\sum_{k=1}^N \vect{w}_t^k$ and $\avgvect{u}_{t} = \frac{1}{N}\sum_{k=1}^N \vect{u}_t^k$ to facilitate the analysis. For convenience, we  define $\avgvect{g}_{t} = \frac{1}{N}\sum_{k=1}^N \nabla F_k(\vect{w}_t^k)$ and $\vect{g}_{t} = \frac{1}{N}\sum_{k=1}^N \nabla F_k(\vect{w}_t^k, \xi_t^k)$. Therefore, $\avgvect{v}_t = \avgvect{w}_t - \eta_t \vect{g}_t$ and $\expt_{SG}\squab{\vect{g}_t} = \avgvect{g}_t$. Notice that we take average over all $N$ instead of $K$ clients, which is because for $t+1 \in \mathcal{I}_E$, we have
\begin{equation}\label{eqn:avg_ie}
    \avgvect{w}_{t+1} = \frac{1}{N}\sum_{k=1}^N \vect{w}_t^k = \frac{1}{K} \sum_{k \in S_{t+1}} Q(\vect{v}_{t+1}^k)
\end{equation}
and the global model is meaningful only at $t+1 \in \mathcal{I}_E$.

\subsection{Lemmas}
\label{sec:key_lemma1}
We present some necessary lemmas that are useful in the proof of Theorem~\ref{thm:1}.  Lemmas \ref{lemma:sgd}, \ref{lemma:var_bound} and \ref{lemma:div_bound} have been established in \cite{li2019convergence} for floating-point weights. Because (1) Lemma 1 is derived based on the smoothness and convexity of $F_k(\cdot)$; (2) Lemma 2 is derived based on the bounded variance of SGD; and (3) Lemma 3 is derived based on the bounded gradient of $F_k(\cdot)$, these lemmas still hold under Assumption \ref{as:F} for quantized \textsc{FedAvg}.

\begin{lemma}[Result of one step SGD]
\label{lemma:sgd}
    Let Assumption \ref{as:F}-1) and 2) hold. If $\eta_t \leq \frac{1}{4L}$, we have
        \begin{equation*}
            \expt_{SG} \norm{\avgvect{v}_{t+1} - \vect{w}^*}^2 \leq (1-\eta_t \mu) \expt_{SG} \norm{\avgvect{w}_t - \vect{w}^*}^2 
             + \eta_t^2 \expt_{SG} \norm{\vect{g}_t - \avgvect{g}_t}^2 + 6 L \eta_t^2 \Gamma 
             + 2\expt_{SG} \squab{\frac{1}{N}\sum_{k=1}^N \norm{\avgvect{w}_t - \vect{w}_t^k}^2}.
        \end{equation*}
\end{lemma}

\begin{lemma}[Bounding the variance]
\label{lemma:var_bound}
    Let Assumption \ref{as:F}-3) hold, If follows that
    \begin{equation*}
      \expt_{SG} \norm{\vect{g}_t - \avgvect{g}_t}^2 \leq \rev{\sum_{k=1}^{N}\frac{\sigma_k^2}{N^2}}.
    \end{equation*}
\end{lemma}

\begin{lemma}[Bonding the divergence of ${\vect{w}_t^k}$]
\label{lemma:div_bound}
    Let Assumption \ref{as:F}-4) hold, $\eta_t$ is non-increasing and $\eta_t \leq 2 \eta_{t+E}$ for all $t \geq 0$. It follows that
    \begin{equation*}
    \expt_{SG} \squab{\frac{1}{N} \sum_{k=1}^N \norm{\avgvect{w}_t - \vect{w}_t^k}^2} \leq 4\eta_t^2 (E-1)^2 H^2.
    \end{equation*}
\end{lemma}

Lemmas~\ref{lemma:sample} to \ref{lemma:quan} are specific for uplink quantization of \textsc{FedAvg}, whose proofs are deferred to Appendix~\ref{sec:proof1_lemma}.

\begin{lemma}[Unbiased and variance bounded sampling]
\label{lemma:sample}
    Let Assumption \ref{as:F}-4) hold. For $t+1 \in \mathcal{I}_E$, assume that $\eta_t \leq 2\eta_{t+E}$ for all $t \geq 0$. We have
    \begin{eqnarray}    
        \expt_{\mathcal{S}_t} \squab{\avgvect{u}_{t+1}} &=& \avgvect{v}_{t+1}, \label{eqn:sam_unbiased} \\
        \expt \norm{\avgvect{v}_{t+1} - \avgvect{u}_{t+1}}^2 &\leq & \frac{N-K}{N-1} \frac{4}{K} \eta_t^2 E^2 H^2. \nonumber
    \end{eqnarray}
\end{lemma}

\begin{lemma}[Properties of stochastic rounding]
\label{lemma:sr}
    For a vector $\vect{w} \in \mathbb{R}^d$ satisfying $\norm{\vect{w}}_{\infty} \leq M$, let $Q(\vect{w})$ be the quantization of $\vect{w}$ with stochastic rounding, quantization level $B$ and quantization gain $G=\frac{2^{B_t-1}}{M}$. Then we have: 
    \begin{eqnarray*}
    \expt_{SR} \squab{Q(\vect{w})} &=& \vect{w},\\
    \expt_{SR} \squab{\norm{Q(\vect{w}) - \vect{w}}^2} &\leq& d \left ( \frac{M}{2^B-1} \right )^2.
    \end{eqnarray*}
\end{lemma}

\begin{lemma}[Unbiased and variance bounded quantization]
\label{lemma:quan}
    Let Assumption \ref{as:w_bounded} hold. With stochastic rounding and the quantization level set to $B_{t+1}$, we have
    \begin{eqnarray}
        \expt_{SR}\squab{\avgvect{w}_{t+1}} &=& \avgvect{u}_{t+1}, \label{eqn:unbiased} \\
        \expt_{SR}\norm{\avgvect{w}_{t+1} - \avgvect{u}_{t+1}}^2 &\leq& q_{t+1}^2 \cdot \frac{dM^2}{K} \nonumber
    \end{eqnarray}
    for $t+1 \in \mathcal{I}_{E}$, where $q_{t+1} = 1/(2^{B_{t+1}}-1)$.
\end{lemma}

\subsection{Proof of Theorem 1}
\label{sec:compl_proof1}
    If $t+1 \notin \mathcal{I}_E$, $\avgvect{w}_{t+1} = \avgvect{v}_{t+1}$, then using Lemmas \ref{lemma:sgd} to \ref{lemma:div_bound}, we first take expectation over the randomness of stochastic gradient and get
        \begin{align*}
             &  \expt_{SG}  \norm{\avgvect{w}_{t+1} - \vect{w}^*}^2= \expt_{SG} \norm{\avgvect{v}_{t+1} - \vect{w}^*}^2  \\
            & \leq (1-\eta_t \mu) \expt_{SG} \norm{\avgvect{w}_t - \vect{w}^*}^2 + \eta_t^2 \expt_{SG} \norm{\vect{g}_t - \avgvect{g}_t}^2 
            + 6 L \eta_t^2 \Gamma + 2\expt_{SG} \squab{\frac{1}{N}\sum_{k=1}^N \norm{\avgvect{w}_t - \vect{w}_t^k}^2}  \\
            & \leq (1 \!- \!\eta_t \mu) \expt_{SG} \norm{\avgvect{w}_t -  \vect{w}^*}^2 
             + \eta_t^2 \left [{\sum_{k=1}^{N}\frac{\sigma_k^2}{N^2}} + 6L \Gamma  + 8(E-1)^2 H^2 \right ].
        \end{align*}
    We then take expectation over the randomness of $\mathcal{S}_t$ and stochastic rounding to have
    \begin{equation} \label{eqn:noInIE}
        \expt \norm{\avgvect{w}_{t+1} - \vect{w}^*}^2 \leq  (1-\eta_t \mu) \expt \norm{\avgvect{w}_t - \vect{w}^*}^2 
        + \eta_t^2 \squab{\rev{\sum_{k=1}^{N}\frac{\sigma_k^2}{N^2}} + 6L \Gamma + 8(E-1)^2 H^2}.
    \end{equation}
    If $t+1 \in \mathcal{I}_{E}$, note that
    \begin{equation} 
    \label{eqn:depart1}
        \begin{split}
            & \norm{\avgvect{w}_{t+1} - \vect{w}^*}^2 = \norm{\avgvect{w}_{t+1} - \avgvect{u}_{t+1} + \avgvect{u}_{t+1}- \vect{w}^*}^2 \\
            & = \underbrace{\norm{\avgvect{w}_{t+1} - \avgvect{u}_{t+1}}^2}_{A_1} + \underbrace{\norm{\avgvect{u}_{t+1} - \vect{w}^*}^2}_{A_2} 
            + \underbrace{2\dotp{\avgvect{w}_{t+1} - \avgvect{u}_{t+1}}{\avgvect{u}_{t+1}- \vect{w}^*}}_{A_3}.
        \end{split}
    \end{equation}
    When the expectation is taken over the randomness of stochastic rounding, the last term $A_3$ vanishes since we have $\expt_{SR} \squab{\avgvect{w}_{t+1} - \avgvect{u}_{t+1}} = \vect{0}$ (from Eqn.~\eqref{eqn:unbiased}). $A_1$ can be bounded using Lemma \ref{lemma:quan}. As for $A_2$, we have
    \begin{equation*}
        \begin{split}
            & \norm{\avgvect{u}_{t+1} - \vect{w}^*}^2 = \norm{\avgvect{u}_{t+1} - \avgvect{v}_{t+1} + \avgvect{v}_{t+1}- \vect{w}^*}^2 \\
            & = \underbrace{\norm{\avgvect{u}_{t+1} - \avgvect{v}_{t+1}}^2}_{B_1} + \underbrace{\norm{\avgvect{v}_{t+1} - \vect{w}^*}^2}_{B_2} 
            + \underbrace{2\dotp{\avgvect{u}_{t+1} - \avgvect{v}_{t+1}}{\avgvect{v}_{t+1}- \vect{w}^*}}_{B_3}.
        \end{split}
    \end{equation*}
    When expectation is taken over the randomness of $\mathcal{S}_{t}$, the last term $B_3$ vanishes because $\expt_{\mathcal{S}_{t}} \squab{\avgvect{u}_{t+1} - \avgvect{v}_{t+1}} = \vect{0}$ (from Eqn.~\eqref{eqn:sam_unbiased}). $B_1$ can be bounded using Lemma \ref{lemma:sample}, and $B_2$ can be bounded using Lemmas \ref{lemma:sgd} to \ref{lemma:div_bound}. In summary, by taking expectation over all the three randomnesses on Eqn.~\eqref{eqn:depart1}, we finally have
        \begin{equation} 
        \label{eqn:inIE}
        \begin{split}
           & \expt \norm{\avgvect{w}_{t+1} - \vect{w}^*}^2 \leq (1-\eta_t \mu) \expt \norm{\avgvect{w}_t - \vect{w}^*}^2 + q_{t+1}^2  \frac{dM^2}{K} \\
            & + \eta_t^2 \squab{\rev{\sum_{k=1}^{N}\frac{\sigma_k^2}{N^2}} + 6L \Gamma + 8(E-1)^2 H^2 + \frac{N-K}{N-1} \frac{4}{K} E^2 H^2}. 
        \end{split}
        \end{equation}
    If we increase the quantization level $B_{t+1}$ following $$B_{t+1} = \log_2{\left ( 1/\eta_t +1 \right )},$$ then  $$q_{t+1} = 1/(2^{B_{t+1}} -1 ) = \eta_t.$$ Let $\Delta_t = \expt \norm{\avgvect{w}_{t}- \vect{w}^*}^2$. From Eqn.~\eqref{eqn:noInIE} and Eqn.~\eqref{eqn:inIE}, it is clear that no matter whether $t+1 \in \mathcal{I}_E$ or $t+1 \notin \mathcal{I}_E$, we always have
    \begin{equation*}
      \Delta_{t+1} \leq (1-\eta_t\mu) \Delta_{t} + \eta_t^2 D
    \end{equation*}
    where
    \begin{equation*}
      D = \rev{\sum_{k=1}^{N}\frac{\sigma_k^2}{N^2}} + 6L \Gamma + 8(E-1)^2 H^2 + \frac{N-K}{N-1} \frac{4}{K} E^2 H^2 + \frac{dM^2}{K}. 
    \end{equation*}
    We decay the learning rate with $\eta_t=\frac{\beta}{t+\gamma}$ for some $\beta \geq \frac{1}{\mu}$ and $\gamma \geq 0$ such that $\eta_1 \leq \min\{\frac{1}{\mu}, \frac{1}{4L}\}=\frac{1}{4L}$ and $\eta_t \leq 2 \eta_{t+E}$. Now we prove that $\Delta_{t} \leq \frac{v}{\gamma +t}$ where $$v = \max \{ \frac{\beta^2 D}{\beta \mu -1}, (\gamma+1) \Delta_{0} \}$$ by induction. First, the definition of $v$ ensures that it holds for $t=0$. Assume the conclusion holds for some $t>0$, it follows that
    \begin{equation*}
    \begin{split}
        \Delta_{t+1} & \leq (1 - \eta \mu) \Delta_t + \eta_t^2 D  \\
        & = \left ( 1-\frac{\beta\mu}{t+\gamma} \right ) \frac{v}{t+\gamma} + \frac{\beta^2D}{(t+\gamma)^2} \\
        & = \frac{t+\gamma-1}{(t+\gamma)^2}v + \squab{\frac{\beta^2D}{(t+\gamma)^2} - \frac{\mu \beta -1}{(t+\gamma)^2}v} \\ 
        & \leq \frac{v}{t+\gamma+1}.
    \end{split}
    \end{equation*}
    Then by the strong convexity of $F(\cdot)$,
    \begin{equation*}
    \expt \squab{F(\avgvect{w}_t)} - F^* \leq \frac{L}{2} \Delta_t \leq \frac{L}{2}\frac{v}{\gamma+t}. 
    \end{equation*}
    Specially, if we choose $\beta = \frac{2}{\mu}$, $\gamma = \max\{ 8\frac{L}{\mu}-1, E\}$ and denote $\kappa = \frac{L}{\mu}$, then $\eta_t = \frac{2}{\mu}\frac{1}{\gamma+t}$. Using $\max\{a, b\} \leq a+b$, we have
    \begin{equation*}
    \begin{split}
         v  & \leq \frac{\beta^2 D}{\beta \mu -1} + (\gamma+1) \Delta_{0} \\
        & = 4\frac{D}{\mu^2} + (\gamma+1)\Delta_0 \\
         & \leq 4\frac{D}{\mu^2} + \left(8\frac{L}{\mu} - 1 + E + 1\right )\Delta_0 \\
         & = 4\frac{D}{\mu^2} + \left(8\frac{L}{\mu} \!+ \!E \right )\norm{\vect{w}_0 - \vect{w}^*}^2.
        \end{split}
    \end{equation*}
    Therefore,
    \begin{equation*}
        \begin{split}
            \expt \squab{F(\avgvect{w}_t)} - F^*  
            & \leq \frac{L}{2(\gamma+t)} \squab{4\frac{D}{\mu^2} + (8\frac{L}{\mu} + E)\norm{\vect{w}_0 - \vect{w}^*}^2} \\
            & = \frac{2\kappa}{\gamma+t} \squab{\frac{D}{\mu} + \left ( 2L + \frac{E\mu}{4} \right )\norm{\vect{w}_0 - \vect{w}^*}^2}.
        \end{split}
    \end{equation*}

\subsection{Deferred proofs of lemmas}
\label{sec:proof1_lemma}

\mypara{Proof of Lemma \ref{lemma:sample}.}
    Let $\mathcal{S}_{t+1}$ denote the set of chosen indexes. Note that the number of possible $\mathcal{S}_{t+1}$ is $C_N^K$ and we denote the $l$th possible result as $\mathcal{S}_{t+1}^l = \{i_1^l, \dots, i_K^l\}$, where $l=1,\dots,C_N^K$. Therefore,
    \begin{equation*}
    \begin{split}
       \sum_{j=1}^{C_N^K} \sum_{k=1}^K \vect{v}_{t+1}^{i_k^l} = \frac{K \cdot C_N^K}{N} \sum_{i=1}^N {\vect{v}_{t+1}^k} 
       = C_{N-1}^{K-1}\sum_{i=1}^N {\vect{v}_{t+1}^k}.
    \end{split}
    \end{equation*}
    Since when $t+1 \in \mathcal{I}_E$, $\vect{u}_{t+1}^k = \frac{1}{K} \sum_{k \in S{t+1}} \vect{v}_{t+1}^k$ for all $k$, we have
    \begin{equation*}
   \avgvect{u}_{t+1} = \sum_{k=1}^N \vect{u}_{t+1}^k = \frac{1}{K} \sum_{k \in S_{t+1}} \vect{v}_{t+1}^k.
    \end{equation*}
    Then
    \begin{equation*}
        \expt_{\mathcal{S}_{t}}\squab{\avgvect{u}_{t+1}}  = \sum_{l=1}^{C_N^K} \mathbb{P}\left (\mathcal{S}_{t+1} = \mathcal{S}_{t+1}^l \right )  \frac{1}{K} \sum_{k \in S_{t+1}^l} \vect{v}_{t+1}^k 
        = \frac{1}{C_N^K} \frac{1}{K} \sum_{j=1}^{C_N^K} \sum_{k=1}^K \vect{v}_{t+1}^{i_k^l} = \frac{C_{N-1}^{K-1}}{C_N^K} \frac{1}{K}\sum_{k=1}^N {\vect{v}_{t+1}^k} 
        = \avgvect{v}_{t+1}.
    \end{equation*}
    As for the variance, we have \cite{li2019convergence}
    \begin{equation}
    \label{eqn:uv_st}
        \begin{split}
           &\expt_{\mathcal{S}_{t}}  \norm{\avgvect{u}_{t+1}-\avgvect{v}_{t+1}}^2
           = \expt_{\mathcal{S}_{t}} \norm{\frac{1}{K} \sum_{i \in S_{t+1}} \vect{v}_{t+1}^{i}-\avgvect{v}_{t+1}} 
           =  \frac{1}{K^2} \expt_{\mathcal{S}_{t}} \norm{\sum_{i=1}^{N} \mathbb{I}\left\{i \in S_{t}\right\}\left(\vect{v}_{t+1}^{i}-\avgvect{v}_{t+1}\right)}^2 \\
           & =  \frac{1}{K^{2}}\left[\sum_{i \in[N]} \mathbb{P}\left(i \in S_{t+1}\right) \norm{\vect{v}_{t+1}^{i}-\avgvect{v}_{t+1}}^2 
           + \sum_{i \neq j} \mathbb{P}\left(i, j \in S_{t+1}\right) \dotp{\vect{v}_{t+1}^{i}-\avgvect{v}_{t+1}}{\vect{v}_{t+1}^{j}-\avgvect{v}_{t+1}} \right ] \\
            &= \frac{1}{K N} \sum_{i=1}^{N} \norm{\vect{v}_{t+1}^{i}-\avgvect{v}_{t+1}}^2 
            + \sum_{i \neq j} \frac{K-1}{K N(N-1)} \dotp{\vect{v}_{t+1}^{i}-\avgvect{v}_{t+1}}{\vect{v}_{t+1}^{j}-\avgvect{v}_{t+1}} \\
            & =  \frac{1-\frac{K}{N}}{K(N-1)} \sum_{i=1}^{N}\norm{\vect{v}_{t+1}^{i}-\avgvect{v}_{t+1}}^2
        \end{split}
    \end{equation}
    where we use the following results: $$\mathbb{P}\left(i \in S_{t+1}\right) = \frac{K}{N}$$ and $$\mathbb{P}\left(i,j \in S_{t+1}\right) = \frac{K(K-1)}{N(N-1)}$$ for all $i \neq j$, and $$\sum_{i \in [N]} \norm{\vect{v}_{t+1}^i -\avgvect{v}_{t+1}}^2 + \sum_{i \neq j} \dotp{\vect{v}_{t+1}^{i}-\avgvect{v}_{t+1}}{\vect{v}_{t+1}^{j}-\avgvect{v}_{t+1}} = 0.$$    
    Since $t+1 \in \mathcal{I}_E$, we know that $t_0=t-E+1 \in \mathcal{I}_E$ is the communication time, implying that $\{\vect{u}_{t_0}^k \}_{k=1}^N$ are identical. Then
    \begin{equation*}
        \begin{split}
            & \sum_{i=1}^N \norm{\vect{v}_{t+1}^i - \avgvect{v}_{t+1}}^2 \\
            & = \sum_{i=1}^N \norm{(\vect{v}_{t+1}^i - \avgvect{u}_{t_0}) - (\avgvect{v}_{t+1} - \avgvect{u}_{t_0})}^2 \\
            & = \sum_{i=1}^N \norm{\vect{v}_{t+1}^i - \avgvect{u}_{t_0}}^2 - 2 \dotp{\sum_{i=1}^N \vect{v}_{t+1}^i - \avgvect{u}_{t_0}}{\avgvect{v}_{t+1}-\avgvect{u}_{t_0}} 
            + \sum_{i=1}^N \norm{\avgvect{v}_{t+1} - \avgvect{u}_{t_0}}^2 \\
            & = \sum_{i=1}^N \norm{\vect{v}_{t+1}^i - \avgvect{u}_{t_0}}^2 - \sum_{i=1}^N \norm{\avgvect{v}_{t+1} - \avgvect{u}_{t_0}}^2 \\
            & \leq \sum_{i=1}^N \norm{\vect{v}_{t+1}^i - \avgvect{u}_{t_0}}^2
        \end{split}
    \end{equation*}
    Taking expectation over the randomness of stochastic gradient on Eqn.~\eqref{eqn:uv_st}, we have
    \begin{equation*}
        \begin{split}
            \expt & \squab{\frac{1}{K(N-1)}\left(1-\frac{K}{N}\right) \sum_{k=1}^N \norm{\vect{v}_{t+1}^i - \avgvect{v}_{t+1}}^2} \\
            & \leq \frac{N-K}{K(N-1)} \frac{1}{N} \sum_{k=1}^N \expt \norm{\vect{v}_{t+1}^i - \avgvect{u}_{t_0}}^2 \\
            & \leq \frac{N-K}{K(N-1)}\frac{1}{N} \sum_{k=1}^N E \sum_{i=t_0}^t \expt \norm{\eta_i \nabla F_k{(\vect{u}_i^k, \xi_i^k)}}^2  \\
            & \leq \frac{N-K}{K(N-1)} E^2 \eta_{t_0}^2 H^2  \\
            & \leq \frac{N-K}{N-1}\frac{4}{K} E^2 \eta_t^2 H^2
        \end{split}
    \end{equation*}
    where the last line is because $\eta_t$ is non-increasing and $\eta_{t_0} \leq 2\eta_t$.

\mypara{Proof of Lemma \ref{lemma:sr}.}
    Let $w$ be an arbitrary element of $\vect{w}$. Then $|w| \leq M$. With $B$-bit quantization we can divide $[-M,+M]$ into $\zeta$ smaller intervals $I_1=[s_1,s_2],I_2=[s_2,s_3],...,I_{\zeta}=[s_\zeta,s_{\zeta+1}]$, with $\zeta=2^{B-1}$. Suppose $w$ is located at the $i$th interval, i.e
    \begin{equation*}
     s_i \leq w \leq s_{i+1}.
    \end{equation*}
    Using stochastic rounding, we get the quantized result as
    \begin{equation*}
        Q(w) = \begin{cases}
             s_i, & \text{w.p.~}\frac{s_{i+1}-w}{s_{i+1}-s_i}, \\
             s_{i+1}, & \text{w.p.~}\frac{w-s_{i}}{s_{i+1}-s_i}.
        \end{cases}
    \end{equation*}
    Then
    \begin{equation*}
        \expt_{SR} \squab{Q(w)} = s_i  \frac{s_{i+1}-w}{s_{i+1}-s_i} + s_{i+1}  \frac{w-s_{i}}{s_{i+1}-s_i}
        = \frac{w(s_{i+1}-s_{i})}{s_{i+1}-s_i}=w,
    \end{equation*}
    and
    \begin{equation*}
       \begin{split}
            &\expt_{SR} \squab{(Q(w) - w)^2} \\
            & =  (s_i-w)^2  \frac{s_{i+1}-w}{s_{i+1}-s_i} + (s_{i+1}-w)^2  \frac{w-s_{i}}{s_{i+1}-s_i} \\
             & = (w-s_i)(s_{i+1}-w) \\
             &\leq \left (\frac{s_{i+1}-s_i}{2} \right )^2 = \left (\frac{M}{2^{B}-1} \right)^2.
        \end{split}
    \end{equation*}
    Hence, for $\vect{w}=[w_1, w_2, \dots, w_d]$, we have
    \begin{equation*}
            \expt_{SR} \squab{Q(\vect{w})} 
             =  \left[ \expt_{SR}[Q(w_1)], \expt_{SR}[Q(w_2)], \dots, \expt_{SR}[Q(w_d)] \right]  =  \vect{w},
    \end{equation*}
    and
    \begin{equation*}
        \expt_{SR} \norm{Q(\vect{w}) - \vect{w}}^2 = \sum_{i=1}^d \expt_{SR} \squab{(Q(w_{i}) - w_{i})^2} 
        \leq d \left ( \frac{M}{2^{B}-1} \right )^2.
    \end{equation*}

\mypara{Proof of Lemma \ref{lemma:quan}.}
    According to Eqn.~\eqref{eqn:avg_ie} and Lemma \ref{lemma:sr}, for $t+1 \in \mathcal{I}_E$, we have
    \begin{equation*}
        \begin{split}
            & \expt_{SR} \squab{\avgvect{w}_{t+1}}  = \expt_{SR} \squab{\frac{1}{K}\sum_{k \in \mathcal{S}_{t+1}} Q(\vect{v}_{t+1}^k)} \\
            & = \frac{1}{K} \sum_{k \in \mathcal{S}_{t+1}} \expt_{SR} \squab{Q(\vect{v}_{t+1}^k)}\\
            & = \frac{1}{K} \sum_{k \in \mathcal{S}_{t+1}} \vect{v}_{t+1}^k= \avgvect{u}_{t+1}.
        \end{split}
    \end{equation*}
    As the quantization level is set to $B_{t+1}$, with Lemma \ref{lemma:sr}, we know that for all $k\in[K]$,
    \begin{equation}\label{eqn:var_qt}
        \expt_{SR} \norm{Q(\vect{v}_{t+1}^k) - \vect{v}_{t+1}^k}^2 \leq q_{t+1}^2  dM^2   
    \end{equation}
    where $q_{t+1} = 1/(2^{B_{t+1}}-1)$. Then
        \begin{equation*}
        \begin{split}
            \expt_{SR} \norm{\avgvect{w}_{t+1} - \avgvect{u}_{t+1}}^2 
            &= \expt_{SR} \norm{\frac{1}{K}\!\sum_{k \in \mathcal{S}_{t+1}}\! Q(\vect{v}_{t+1}^k) - \frac{1}{K}\!\sum_{k \in \mathcal{S}_{t+1}}\! \vect{v}_{t+1}^k}^2 \\
             & = \frac{1}{K^2}\expt_{SR} \norm{\sum_{k \in \mathcal{S}_{t+1}} (Q(\vect{v}_{t+1}^k) - \vect{v}_{t+1}^k)}^2.
        \end{split}
    \end{equation*}
    Let $\vect{e}_{t+1}^k = Q(\vect{v}_{t+1}^k) - \vect{v}_{t+1}^k$, then
    \begin{equation*}
        \expt_{SR} \norm{\avgvect{w}_{t+1} - \avgvect{u}_{t+1}}^2  = \frac{1}{K^2} \sum_{k \in \mathcal{S}_{t+1}} \expt_{SR} \norm{\vect{e}_{t+1}^k}^2 
        + \frac{1}{K^2} \expt_{SR} \squab{\sum_{i,j \in \mathcal{S}_{t+1}, i\neq j} \dotp{\vect{e}_{t+1}^i}{{\vect{e}_{t+1}^j}}}.
    \end{equation*}
    We know $\expt_{SR} \squab{\vect{e}_{t+1}^k} = \vect{0}$ from Lemma \ref{lemma:sr}, and $\vect{e}_{t+1}^i$ and $\vect{e}_{t+1}^j$ are independent if $i \neq j$. Therefore,
    \begin{equation}\label{eqn:dotp_zero}
        \expt_{SR} \left [ \sum_{i\neq j} \dotp{\vect{e}_{t+1}^i}{{\vect{e}_{t+1}^j}} \right ]  = \sum_{i \neq j} \expt_{SR} \left [ \dotp{\vect{e}_{t+1}^i}{{\vect{e}_{t+1}^j}} \right ] 
        = \sum_{i \neq j} \dotp{\expt_{SR}[\vect{e}_{t+1}^i]}{\expt_{SR}[\vect{e}_{t+1}^j]}=0.
    \end{equation}
    With Eqn.~\eqref{eqn:var_qt}, we have
    \begin{equation*}
        \begin{split}
            \expt&_{SR}  \norm{\avgvect{w}_{t+1} - \avgvect{u}_{t+1}}^2 = \frac{1}{K^2} \expt_{SR} \sum_{k \in \mathcal{S}_{t+1}} \norm{\vect{e}_{t+1}^k}^2 
            \\
            & = \frac{1}{K^2} \sum_{k \in \mathcal{S}_{t+1}} \expt_{SR} \norm{Q(\vect{v}_{t+1}^k) - \vect{v}_{t+1}^k}^2 
            \leq q_{t+1}^2 \frac{dM^2}{K}.
        \end{split}
    \end{equation*}

\section{Proof of Theorem~\ref{thm:2}}
\label{app:proof_thm2}
\subsection{Notations} 
All of the notations in Appendix~\ref{app:proof_thm1} can be extended for DT unless $\vect{w}_{t+1}^k$ is slightly different. For quantized differential transmission, if $t+1 \in \mathcal{I}_E$, each client in $\mathcal{S}_{t+1}$ uploads the quantized differential weights $Q(\vect{d}_{t+1}^k)$ where $\vect{d}_{t+1}^k = \vect{v}_{t+1}^k - \vect{w}_{t+1-E}$ and $\vect{w}_{t+1-E}$ means the most recent global model it downloaded from the server. And the global aggregation is $\vect{w}_{t+1} = \vect{w}_{t+1-E} + \frac{1}{K} \sum_{k \in \mathcal{S}_{t+1}} Q(\vect{d}_{t+1}^k)$. Hence, we can redefine $\vect{w}_{t+1}^k$ as
\begin{align*}
    \vect{w}_{t+1}^k & = \begin{cases}
        \vect{v}_{t+1}^k & if~ t+1 \notin \mathcal{I}_E, \\
        \vect{w}_{t+1-E} + \frac{1}{K} \sum_{k \in S_{t+1}} Q(\vect{d}_{t+1}^k) & if~ t+1 \in \mathcal{I}_E. 
    \end{cases}
\end{align*}

\subsection{Lemma}
\begin{lemma}[Unbiased and variance bounded quantization]
\label{lemma:dt_quan}
    With stochastic rounding and quantization level $B$ and assuming the quantization gain for $\vect{d}_{t+1}^k$ is $G=2^{B-1}/\norm{\vect{d}_{t+1}^k}_{\infty}$, for all $t+1 \in \mathcal{I}_E, k\in \mathcal{S}_{t+1}$, we have
    \begin{equation*}
      \expt_{SR}\squab{\avgvect{w}_{t+1}} = \avgvect{u}_{t+1},
    \end{equation*}  
    and
    \begin{equation*}
      \expt \norm{\avgvect{w}_{t+1} - \avgvect{u}_{t+1}}^2 \leq \frac{4d}{K \left( 2^B-1 \right )^2} \eta_t^2 E^2 H^2.
    \end{equation*}
\end{lemma}
\mypara{Proof of Lemma \ref{lemma:dt_quan}.} 
    Considering the special case of Lemma \ref{lemma:sr}, say $M=\norm{\vect{w}}_{\infty}$ and the corresponding $G = 2^{B-1}/\norm{\vect{w}}_{\infty}$, we have
    \begin{equation}\label{eqn:sr_spec}
            \expt_{SR} \squab{\norm{Q(\vect{w}) - \vect{w}}^2} \leq d \left ( \frac{M}{2^B-1} \right )^2 
            = d \frac{\norm{\vect{w}}_{\infty}^2}{(2^B-1)^2} \leq d \frac{\norm{\vect{w}}^2}{(2^B-1)^2}
    \end{equation}
    Then, for $t+1 \in \mathcal{I}_E$,
    \begin{eqnarray*}
       && \avgvect{w}_{t+1} = \frac{1}{N}\sum_{k=1}^N \vect{w}_{t+1}^k = \vect{w}_{t+1-E} + \frac{1}{K} \sum_{k \in S_{t+1}} Q(\vect{d}_{t+1}^k)\\
       && \avgvect{u}_{t+1} = \frac{1}{N}\sum_{k=1}^N \vect{u}_{t+1}^k = \frac{1}{K} \sum_{k \in S_{t+1}} \vect{v}_{t+1}^k
    \end{eqnarray*}
    Therefore, we get
    \begin{equation*}
        \begin{split}
            \expt_{SR} \squab{ \avgvect{w}_{t+1} } & = \vect{w}_{t+1-E} + \frac{1}{K} \sum_{k \in S_{t+1}} \expt_{SR} \squab{Q(\vect{d}_{t+1}^k)} \\
            & = \vect{w}_{t+1-E} + \frac{1}{K} \sum_{k \in S_{t+1}} \vect{d}_{t+1}^k \\
            & = \vect{w}_{t+1-E} + \frac{1}{K} \sum_{k \in S_{t+1}} (\vect{v}_{t+1}^k - \vect{w}_{t+1-E}) \\
            & = \frac{1}{K} \sum_{k \in S_{t+1}} \vect{v}_{t+1}^k = \avgvect{u}_{t+1}
        \end{split}
    \end{equation*}
    As for the variance, we have
    \begin{equation*}
        \begin{split}
            \expt&_{SR}\norm{\avgvect{w}_{t+1} - \avgvect{u}_{t+1}}^2 
            = \expt_{SR} \norm{\vect{w}_{t+1-E} + \frac{1}{K} \sum_{k \in S_{t+1}} Q(\vect{d}_{t+1}^k) - \frac{1}{K} \sum_{k \in S_{t+1}} \vect{v}_{t+1}^k}^2 \\
            & = \frac{1}{K^2} \expt_{SR} \norm{\sum_{k \in S_{t+1}} Q(\vect{d}_{t+1}^k) - \sum_{k \in S_{t+1}} (\vect{v}_{t+1}^k - \vect{w}_{t+1-E} )}^2 \\
            & = \frac{1}{K^2} \expt_{SR} \norm{\sum_{k \in S_{t+1}} \left (Q(\vect{d}_{t+1}^k)-\vect{d}_{t+1}^k \right )}^2 
            = \frac{1}{K^2} \sum_{k \in S_{t+1}} \expt_{SR} \norm{Q(\vect{d}_{t+1}^k) - \vect{d}_{t+1}^k}^2 
        \end{split}
    \end{equation*}
    where the last equality is due to $\expt_{SR}[Q(\vect{d}_{t+1}^k)-\vect{d}_{t+1}^k]=\vect{0}$ (see the proof of Eqn.~\eqref{eqn:dotp_zero}). Since we set $G=1/\norm{\vect{d}_{t+1}^k}$ for all $\vect{d}_{t+1}^k$, with Eqn.~\eqref{eqn:sr_spec}, we get
    \begin{equation*}
        \begin{split}
            \expt&_{SR}\norm{\avgvect{w}_{t+1} - \avgvect{u}_{t+1}}^2 = \frac{1}{K^2} \sum_{k \in S_{t+1}} \expt_{SR} \norm{Q(\vect{d}_{t+1}^k) - \vect{d}_{t+1}^k}^2 \\
            & \leq \frac{1}{K^2} \sum_{k \in S_{t+1}} \frac{d}{\left( 2^B-1 \right )^2} \norm{\vect{d}_{t+1}^k}^2 \\
            & = \frac{d}{K^2 \left( 2^B-1 \right )^2} \sum_{k \in S_{t+1}} \norm{\sum_{\tau=t+1-E}^{t} \eta_{\tau}\nabla F_k(\vect{w}_{\tau}^k, \xi_{\tau}^k)}^2 \\
            & \leq \frac{dE}{K^2 \left( 2^B-1 \right )^2} \sum_{k \in S_{t+1}} \sum_{\tau=t+1-E}^{t} \eta_{\tau}^2 \norm{\nabla F_k(\vect{w}_{\tau}^k, \xi_{\tau}^k)}^2
        \end{split}
    \end{equation*}
   By further taking expectation over the randomness of stochastic gradient, we get
    \begin{equation*}
        \begin{split}
            \expt &\norm{\avgvect{w}_{t+1} - \avgvect{u}_{t+1}}^2 \\
            & \leq \frac{dE}{K^2 \left( 2^B-1 \right )^2} \sum_{k \in S_{t+1}} \sum_{\tau=t+1-E}^{t} \eta_{\tau}^2 \expt_{SG} \norm{\nabla F_k(\vect{w}_{\tau}^k, \xi_{\tau}^k)}^2 \\
            & \leq \frac{dE}{K^2 \left( 2^B-1 \right )^2} \sum_{k \in S_{t+1}} \sum_{\tau=t+1-E}^{t} \eta_{t+1-E}^2 H^2 \\
            & = \frac{dE^2}{K \left( 2^B-1 \right )^2} \eta_{t+1-E}^2 H^2 \leq \frac{4d}{K \left( 2^B-1 \right )^2} \eta_t^2 E^2 H^2
        \end{split}
    \end{equation*}
    where we use the fact that $\eta_t$ is non-increasing and $2\eta_{t+1-E} \leq 2 \eta_{t}$.

\subsection{Proof of Theorem 2}
We use Lemma \ref{lemma:dt_quan} to update Eqn.~\eqref{eqn:inIE} to
\begin{equation*}
\begin{split}
    & \expt \norm{\avgvect{w}_{t+1} - \vect{w}^*}^2  \leq (1-\eta_t \mu) \expt \norm{\avgvect{w}_t - \vect{w}^*}^2 \\
    & \!+\! \eta_t^2 \!\left [\frac{4d}{K \!\left( 2^B-1 \!\right )^2} E^2 H^2 \!+\! \frac{\sigma_k^2}{N} \!+ \!6L \Gamma \!+ \!8(E\!-\!1)^2 H^2 
    + \frac{N-K}{N-1} \frac{4}{K} E^2 H^2 \right ].
\end{split}
\end{equation*}
    Let $\Delta_t = \expt \norm{\avgvect{w}_t - \vect{w}^*}^2$, therefore for $t+1 \in \mathcal{I}_E$ or $t+1 \notin \mathcal{I}_E$, we have
    \begin{equation*}
    \Delta_{t+1} \leq (1-\eta_t\mu) \Delta_{t} + \eta_t^2 D,
    \end{equation*}
    with
    \begin{equation*}
      D = \rev{\sum_{k=1}^{N}\frac{\sigma_k^2}{N^2}} + 6L \Gamma + 8(E-1)^2 H^2 
      + \frac{N-K}{N-1} \frac{4}{K} E^2 H^2 + \frac{4d}{K \left( 2^B-1 \right )^2} E^2 H^2.
    \end{equation*}
We can then apply the same induction as in Appendix \ref{sec:compl_proof1} to get the final result.

\section{Proof of Theorem~\ref{thm:3}}
\label{app:proof_thm3}

\mypara{Notations.} Again we extend the notations in Appendix~\ref{app:proof_thm1} to  downlink quantization. The global model aggregation is $\vect{w}_{t+1} = \frac{1}{K} \sum_{k \in S_{t+1}} \vect{v}_{t+1}^k$ and its quantized version $Q(\vect{w}_{t+1})$ is broadcast to $K$ randomly selected clients for the next round. All notations are similarly defined. We further note that the analysis of convergence should be on $\norm{\avgvect{u}_{t+1}-\vect{w}^*}^2$ instead of $\norm{\avgvect{w}_{t+1}-\vect{w}^*}^2$, since the server has access to unquantized global model aggregation.

\mypara{Proof of Theorem~\ref{thm:3}.} Under Assumption~\ref{as:F}, Lemmas \ref{lemma:sgd} to \ref{lemma:div_bound} still hold. We need to consider four cases.

\noindent $1)$ \textbf{$t+1 \notin \mathcal{I}_E$ and $t \notin \mathcal{I}_E$.}
    By taking expectation over all the three randomness, we can get
    \begin{equation} \label{eqn:dl_noInIE}
        \expt \norm{\avgvect{v}_{t+1} - \vect{w}^*}^2 \leq (1-\eta_t \mu) \expt \norm{\avgvect{w}_t - \vect{w}^*}^2 
        + \eta_t^2 \squab{\frac{\sigma_k^2}{N} + 6L \Gamma + 8(E-1)^2 H^2}.
    \end{equation}
    since $t \notin \mathcal{I}_E$, we have $ \avgvect{w}_t = \avgvect{u}_t = \avgvect{v}_t = \frac{1}{N} \sum_{k=1}^N \vect{v}_{t}^k$.
    Hence, we can transform Eqn.~\eqref{eqn:dl_noInIE} into 
    \begin{equation*}
        \expt \norm{\avgvect{u}_{t+1} - \vect{w}^*}^2 \leq (1-\eta_t \mu) \expt \norm{\avgvect{u}_t - \vect{w}^*}^2 
         + \eta_t^2 \squab{\rev{\sum_{k=1}^{N}\frac{\sigma_k^2}{N^2}} + 6L \Gamma + 8(E-1)^2 H^2}.
    \end{equation*}
    
\noindent $2)$ \textbf{$t+1 \notin \mathcal{I}_E$ and $t \in \mathcal{I}_E$.}
    We still have $\avgvect{u}_{t+1} = \avgvect{v}_{t+1}$ and $\expt \norm{\avgvect{v}_{t+1} - \vect{w}^*}^2 = \expt \norm{\avgvect{u}_{t+1} - \vect{w}^*}^2$. But now $\avgvect{w}_t  = Q( \frac{1}{K} \sum_{k \in S_{t}} \vect{v}_{t}^k)$ and $\avgvect{u}_t  = \frac{1}{K} \sum_{k \in S_{t}} \vect{v}_{t}^k$.
    Under Assumption \ref{as:w_bounded}, we have that $\norm{\vect{v}_t^k}_{\infty} \leq M$, which suggests $\norm{\avgvect{u}_t}_{\infty} \leq M$. Using Lemma~\ref{lemma:sr}, we have
     \begin{equation}\label{eqn:dl_unbias}
         \expt_{SR} \squab{\avgvect{w}_t} = \expt_{SR} \squab{Q(\avgvect{u}_t)} = \avgvect{u}_t
     \end{equation}
     \begin{equation}\label{eqn:dl_var}
         \expt_{SR} \squab{\norm{\avgvect{w}_t - \avgvect{u}_t}^2} = \expt_{SR} \squab{\norm{Q(\avgvect{u}_t) - \avgvect{u}_t}^2} \leq d \cdot q_{t}^2 M^2
     \end{equation}
    where $q_t = 1/(2^{B_t} -1)$ and $B_{t}$ is the quantization level for the $t$th iteration. Therefore,
    \begin{equation}\begin{split}
       \norm{\avgvect{w}_{t} - \vect{w}^*}^2  & = \norm{\avgvect{w}_{t} - \avgvect{u}_{t} + \avgvect{u}_{t}- \vect{w}^*}^2  \\
       & = \underbrace{\norm{\avgvect{w}_{t} - \avgvect{u}_{t}}^2}_{A_1} + \underbrace{\norm{\avgvect{u}_{t}- \vect{w}^*}^2}_{A_2} 
       + \underbrace{2\dotp{\avgvect{w}_{t} - \avgvect{u}_{t}}{\avgvect{u}_{t}- \vect{w}^*}}_{A_3}.
    \end{split}\end{equation}
    When expectation is taken over the randomness of stochastic rounding, the last term $A_3$ vanishes because of Eqn.~\eqref{eqn:dl_unbias} and $A_3$ can be bounded using Eqn.~\eqref{eqn:dl_var}. We further have
    \begin{equation}\label{eqn:w_u}
        \expt \norm{\avgvect{w}_t - \vect{w}^*}^2 \leq \expt \norm{\avgvect{u}_t - \vect{w}^*}^2 + d \cdot q_{t}^2 M^2,
    \end{equation}
    which transforms Eqn.~\eqref{eqn:dl_noInIE} into
    \begin{equation*}
    \begin{split}
        & \expt \norm{\avgvect{u}_{t+1} - \vect{w}^*}^2  \leq (1-\eta_t \mu) \expt \norm{\avgvect{u}_t - \vect{w}^*}^2  \\ 
        &+ (1-\eta_t \mu)d q_{t}^2 M^2 
        + \eta_t^2 \squab{\rev{\sum_{k=1}^{N}\frac{\sigma_k^2}{N^2}} + 6L \Gamma + 8(E-1)^2 H^2}.
    \end{split}
    \end{equation*}

\noindent $3)$ \textbf{$t+1 \in \mathcal{I}_E$ and $t \notin \mathcal{I}_E$.}
    We still have $\avgvect{u}_{t+1} = \avgvect{w}_{t+1}$ and $\expt \norm{\avgvect{u}_{t+1} - \vect{w}^*}^2 = \expt \norm{\avgvect{w}_{t+1} - \vect{w}^*}^2$. But now $\avgvect{v}_{t+1}  = \frac{1}{N} \sum_{k=1}^N \vect{v}_{t+1}^k$ and $\avgvect{u}_{t+1}  = \frac{1}{K} \sum_{k \in S_{t+1}} \vect{v}_{t+1}^k$, and
    \begin{equation}
    \begin{split}
        & \norm{\avgvect{u}_{t+1} - \vect{w}^*}^2 = \norm{\avgvect{u}_{t+1} - \avgvect{v}_{t+1} + \avgvect{v}_{t+1}- \vect{w}^*}^2 \\
        & = \underbrace{\norm{\avgvect{u}_{t+1} - \avgvect{v}_{t+1}}^2}_{B_1} + \underbrace{\norm{\avgvect{v}_{t+1}- \vect{w}^*}^2}_{B_2} 
        + \underbrace{2\dotp{\avgvect{u}_{t+1} - \avgvect{v}_{t+1}}{\avgvect{v}_{t+1}- \vect{w}^*}}_{B_3}.
    \end{split}
    \end{equation}
    Lemma~\ref{lemma:sample} indicates $\expt_{\mathcal{S}_{t}} \squab{\avgvect{u}_{t+1} - \avgvect{v}_{t+1}} = \vect{0}$, so when expectation is taken over the randomness of $\mathcal{S}_{t}$, the last term $B_3$ vanishes. $B_1$ can be bounded by Eqn.~\eqref{eqn:sam_unbiased}. We finally have    
    \crt{
    \begin{equation} \label{eqn:u_v}
        \expt \norm{\avgvect{u}_{t+1} - \vect{w}^*}^2 \leq \expt \norm{\avgvect{v}_{t+1} - \vect{w}^*}^2 + \frac{N-K}{N-1} \frac{4}{K} \eta_t^2 E^2 H^2,
    \end{equation}
    }
    \crt{With Eqn.~\eqref{eqn:dl_noInIE}, and $\expt \norm{\avgvect{w}_{t} - \vect{w}^*}^2=\expt \norm{\avgvect{u}_{t} - \vect{w}^*}^2$ since $t \notin \mathcal{I}_E$, we can further have}
    \begin{equation*} 
    \begin{split}
        & \expt  \norm{\avgvect{u}_{t+1} - \vect{w}^*}^2 \leq (1-\eta_t \mu) \expt \norm{\avgvect{u}_t - \vect{w}^*}^2 \\
        & + \eta_t^2 \squab{\rev{\sum_{k=1}^{N}\frac{\sigma_k^2}{N^2}} + 6L \Gamma + 8(E-1)^2 H^2 + \frac{N-K}{N-1} \frac{4}{K} E^2 H^2}.
    \end{split}
    \end{equation*}
    
\noindent $4)$ \textbf{$t+1 \in \mathcal{I}_E$ and $t \in \mathcal{I}_E$.} 
This case is only possible for $E=1$. In this case, $\avgvect{v}_{t+1} \neq \avgvect{u}_{t+1}$ and $\avgvect{u}_{t+1} \neq \avgvect{w}_{t+1}$. We use both Eqn.~\eqref{eqn:w_u} and Eqn.~\eqref{eqn:u_v} to transform Eqn.~\eqref{eqn:dl_noInIE} into
    \begin{equation}\label{eqn:dl_case4}
    \begin{split}
       & \expt  \norm{\avgvect{u}_{t+1} - \vect{w}^*}^2 
       \leq (1-\eta_t \mu) \expt \norm{\avgvect{u}_t - \vect{w}^*}^2 + (1-\eta_t \mu)d \cdot q_{t}^2 M^2 \\
        & + \eta_t^2 \squab{\rev{\sum_{k=1}^{N}\frac{\sigma_k^2}{N^2}} + 6L \Gamma + 8(E-1)^2 H^2 + \frac{N-K}{N-1} \frac{4}{K} E^2 H^2}.
    \end{split}
    \end{equation}
        
   In summary, Eqn.~\eqref{eqn:dl_case4} holds for all cases. Let $\Delta_t = \expt \norm{\avgvect{u}_{t}- \vect{w}^*}^2$. If we increase the quantization level $B_{t}$ following $$B_{t} = \log_2{\left ( 1 + \frac{\sqrt{1-\eta_t \mu}}{\eta_t} \right )}$$ to make $$q_{t} = 1/(2^{B_{t}} -1 ) = \frac{\eta_t}{\sqrt{1-\eta_t \mu}}.$$ Then we have $ (1-\eta_t \mu) q_t^2 = \eta_t^2$ and we also have $$ \Delta_{t+1} \leq (1-\eta_t\mu) \Delta_{t} + \eta_t^2 D$$ where 
   \begin{equation*}
           D = \rev{\sum_{k=1}^{N}\frac{\sigma_k^2}{N^2}} + 6L \Gamma
            + 8(E-1)^2 H^2 + \frac{N-K}{N-1} \frac{4}{K} E^2 H^2 + dM^2.
   \end{equation*}
   Applying the same induction method in Appendix \ref{sec:compl_proof1} proves the theorem.

\bibliographystyle{IEEEtran}
\bibliography{FedLearn,wireless}

\end{document}